\documentclass[aps,prd,preprint,superscriptaddress]{revtex4}
\usepackage{graphicx}
\usepackage{amsmath,amssymb}
\usepackage{bm}% bold math
\usepackage[caption=false]{subfig}
\usepackage{color}
\usepackage{hyperref}

\def\p{\partial}

\def\=:{=\hspace{-.7em}\raisebox{1.1ex}{.}\hspace{.1em}\raisebox{-0.2ex}{.}}
\newcommand{\tr}{{\rm tr}\,}

\newcommand {\beq}{\begin{eqnarray}}
\newcommand {\eeq}{\end{eqnarray}}
\newcommand {\non}{\nonumber\\}

\newcommand {\1}[1]{\frac{1}{#1}}

\begin{document}

\preprint{NORDITA-2015-24}

\title{Fractional Skyrmions and their molecules}


\author{Sven Bjarke Gudnason}
\email{bjarke(at)impcas.ac.cn}
\affiliation{Institute of Modern Physics, Chinese Academy of Sciences,
  Lanzhou 730000, China}
\affiliation{Nordita, KTH Royal Institute of Technology and Stockholm
  University, Roslagstullsbacken 23, SE-106 91 Stockholm, Sweden
}
\author{Muneto Nitta}
\email{nitta(at)phys-h.keio.ac.jp}
\affiliation{Department of Physics, and Research and Education Center for Natural 
Sciences, Keio University, Hiyoshi 4-1-1, Yokohama, Kanagawa 223-8521, Japan
}
%\thanks{}



\date{\today}
\begin{abstract}
We study a Skyrme-type model with 
a quadratic potential for a field with $S^2$ vacua.
We consider two flavors of the model, the first is the Skyrme model
and the second has a sixth-order derivative term instead of the Skyrme
term; both with the added quadratic potential.
The model contains molecules of half Skyrmions, 
each of them is a global (anti-)monopole 
with baryon number $1/2$.
We numerically construct solutions with 
baryon numbers one through six, 
and find stable solutions which look like beads on rings. 
We also construct a molecule with 
fractional Skyrmions having the baryon numbers $1/3 + 2/3$,
by adding a linear potential term.
\end{abstract}

%\keywords{}

\maketitle

\section{Introduction}

Half a century ago Skyrme made a proposal \cite{Skyrme:1962vh} that
Skyrmions, i.e.~topological objects characterized by the third
homotopy group $\pi_3$, could describe nucleons in the pion effective
field theory 
\cite{Adkins:1983ya} if augmented by a higher-order derivative term.
This term was then later denoted as the Skyrme term. 
The Skyrme term is needed for stabilizing the Skyrmions against
shrinkage. 
Later nucleons are known to be bound states of quarks, described by
QCD. However, QCD at low energies is strongly coupled and hard to
tackle and hence the Skyrme model, with its small number of
parameters, has remained an attractive model at low energies. 
In the limit of a large number of colors, the Skyrmion is exactly the
baryon \cite{Witten:1983tw}. 
It has also recently been used in holographic QCD
\cite{Sakai:2004cn,Hata:2007mb}.

It has recently been found that 
half Skyrmions stably appear 
as constituents of a lattice
at finite baryon density 
\cite{Ma:2013ooa}. 
It is therefore important to understand 
the physical consequences of a half or non-integer 
topological charge of half (or fractional) Skyrmions, 
such as the statistics of them. 
It is, however, not easy to pick up an isolated half Skyrmion 
because it is not stable in isolation.

The fractionality of 
topological charge, however, has been better understood in
lower dimensions. 
As a lower-dimensional analog of Skyrmions,
baby Skyrmions were proposed 
in an O(3) sigma model with a fourth-order derivative term
in 2+1 dimensions 
\cite{Piette:1994ug,Weidig:1998ii},
which topologically are lumps 
characterized by $\pi_2$ 
instead of $\pi_3$, i.e.~the case for usual Skyrmions.  
2+1 dimensional Skyrmions  
often appear in various condensed-matter systems 
such as ferromagnets and quantum-Hall systems. 
A baby-Skyrme model with an XY-type potential
$V = m^2 n_3^2$ (which is also called easy-plane in ferromagnets) 
admits half baby Skyrmions 
\cite{Jaykka:2010bq,Kobayashi:2013aja,Kobayashi:2013wra}.
Each of them can be regarded as a global vortex 
covering the northern or southern hemisphere of the target space 
and consequently having a half $\pi_2$ topological charge.  
Each of them has logarithmically divergent energy when (if) isolated,
because it is a global vortex. 
If one separates the two by an infinite distance, 
the energy diverges logarithmically and 
thus they are confined to
the form of a molecule. 
If a U(1) subgroup of the O(3) symmetry in the O(3) model is gauged,
each constituent baby Skyrmion becomes a local vortex 
(here the fourth-order derivative term is not needed for stability), 
(still) carrying half a unit of $\pi_2$ charge 
\cite{Schroers:1995he,Baptista:2004rk,Nitta:2011um,Alonso-Izquierdo:2014cza}. 
In this case, 
if we choose the gauge coupling and the scalar coupling to be the
same, the baby Skyrmions are BPS and can be embedded 
in a supersymmetric theory \cite{Nitta:2011um}.
An entire molecule can be separated from other molecules without any
cost of energy. In this phase the two half Skyrmions can be separated
by a finite distance at a finite cost of energy. In this sense, the
half Skyrmion can be isolated.

In this paper, we construct 
3+1 dimensional 
half Skyrmions and their molecules 
in two Skyrme models with a potential term
in the form of $V = m^2 n_4^2$,
where we use the notation of the 
O(4) sigma model $n_A(x)$ $(A=1,2,3,4)$ 
with the constraint $\sum_A (n_A)^2=1$.
This potential is a potential for the would-be $S^2$ vacuum moduli and
we will denote it a Heisenberg-type potential. 
The two models we consider in this paper are the conventional Skyrme
model and a Skyrme-like model with the fourth-order derivative term
replaced by a sixth-order derivative term. The latter is inspired by
the BPS Skyrme model \cite{Adam:2010fg}. The sixth-order derivative
term is the baryon-current density squared
\cite{Adam:2010fg,Gudnason:2013qba,Gudnason:2014gla,Gudnason:2014jga}
and this term 
alone with an adequate potential provides the basis of an integrable
subsector of the model. 
For a short-term notation, we call them the 2+4 and 2+6 model,
respectively. 
Each constituent of the molecule is a half Skyrmion 
carrying half a baryon number, i.e.~the topological charge of
$\pi_3$. 
This turns out to be the case for both the 2+4 and the 2+6 models.  
The constituents are more separated 
for the 2+6 model than for the 2+4 model.
This is in contrast with the conventional Skyrmions (i.e.~without our
potential) for which the configuration for $B=1$ is
spherically symmetric, that for $B=2$ is toroidal, and those for $B>2$ 
have energy distributions with some point symmetry
\cite{Battye:1997qq,Houghton:1997kg}. 
We take also the limit of the mass-parameters of our
potential going to zero and recover the spherically symmetric
Skyrmion for $B=1$ and the transition is smooth as expected.
We construct also higher baryon numbered solutions numerically;
specifically $B=2,3,4,5,6$ and find that the lowest-energy states are
found using an axially symmetric initial guess (basically a torus) and
the relaxation method finds the solutions which look like beads on 
rings. We confirm that the configurations of the form of beads on rings
are really the lowest-energy states, by checking different initial
conditions; i.e.~both those with an axial symmetric Ansatz and
others with a rational map Ansatz having an appropriate point symmetry
\cite{Battye:1997qq,Houghton:1997kg}.
The rings of higher $B$ are all stable against decaying into a sum of
smaller $B$ rings (for the $B$ we have found explicitly here). 
This is a three-dimensional analog
of baby Skyrmions as beads on a ring
in a baby-Skyrme model with the XY-potential
\cite{Kobayashi:2013aja,Kobayashi:2013wra}. 
We also construct fractional Skyrmions 
with an arbitrary baryon number by using the potential 
$V = m^2 (n_4-c)^2$, for which each unit Skyrmion charge  
is split into two fractional Skyrmions with 
baryon number $(1+c)/2$ and $(1-c)/2$, respectively. 
We call them unequal fractional molecules. 
We also note that fractional Skyrmions
can be identified as global monopoles 
having divergent energy in an infinite system.

If one gauges an SO(3) subgroup 
acting on $n_1,n_2,n_3$ 
(a diagonal subgroup of the chiral symmetry), 
a monopole becomes the local 't Hooft-Polyakov monopole
having finite energy 
\cite{Brihaye:1998,Brihaye:1998vr,Kleihaus:1999ea,Brihaye:2000ku,Brihaye:2001je,Grigoriev:2002qc,Brihaye:2004pz,Brihaye:2007gk}, 
similarly to a vortex as a half lump in 2+1 dimensions
\cite{Schroers:1995he,Nitta:2011um,Alonso-Izquierdo:2014cza}. 
Our case is a global analog of this case.

This paper is organized as follows. 
In Sec.~\ref{sec:model}, we present our two models and their numerical
solutions in Sec.~\ref{sec:molecule}, i.e.~of fractional-Skyrmion
molecules.
In Sec.~\ref{sec:monopole}, 
we construct an isolated fractional Skyrmion as a global monopole.
In Sec.~\ref{sec:higher}, numerical solutions with 
higher baryon numbers $B=2,3,4,5,6$ are presented. 
Finally, in Sec.~\ref{sec:fractional}, we construct a molecule where
the weight of the two constituents of the molecule is altered,
i.e.~an unequal fractional molecule.
Sec.~\ref{sec:summary} is devoted to a summary and discussions. 
Two investigations are delegated to the appendices. In
App.~\ref{app:rational_map_2+4} we try different initial guesses for
the numerical relaxation and find only metastable solutions. Finally, 
App.~\ref{app:low_mass_limit} shows how the molecule turns into a
spherical Skyrmion by turning off the potential under study.

%%%%%%%%%%%%%%%%%%%%%%%%%%
\section{A Skyrme-like model with a Heisenberg-type
  potential \label{sec:model}} 

We consider the SU(2) principal chiral model with the addition of the
Skyrme term and a sixth-order derivative term in $d=3+1$ dimensions. 
In terms of the SU(2)-valued field $U(x)\in$ SU(2), the Lagrangian
which we are considering is given by 
\beq
\mathcal{L} = -\frac{c_2}{4} 
\tr (\p_{\mu}U^{\dagger} \p^{\mu} U) 
+ c_4 \mathcal{L}_4
+ c_6 \mathcal{L}_6
- V(U),
\eeq
where we use the mostly-positive metric and the higher-derivative
terms are given by
\begin{align}
\mathcal{L}_4 &= -\frac{1}{32} 
\tr (\big[U^\dag \p_{\mu} U, U^\dag \p_{\nu} U\big]^2),\\ 
\mathcal{L}_6 &= \frac{1}{144} 
\left(\epsilon^{\mu\nu\rho\sigma}\tr\big[U^\dag\p_\nu U U^\dag \p_\rho
  U U^\dag \p_\sigma U\big]\right)^2,
\end{align}
where we are using Skyrme units in which lengths are measured in units
of $2/(e f_\pi)$ and energy is measured in units of $f_\pi/(2e)$. 
$\mathcal{L}_4$ is the Skyrme term and $\mathcal{L}_6$ is the baryon
current density squared, which is inspired by the BPS Skyrme model
\cite{Adam:2010fg}. 
The symmetry of the Lagrangian for $V=0$ is 
$\tilde G =$ SU(2)$_{\rm L} \times $SU(2)$_{\rm R}$ acting on $U$ as 
$U \to U'= g_{\rm L} U g_{\rm R}^\dag$.
In the vacuum, this symmetry is spontaneously broken
down to $\tilde H \simeq$
SU(2)$_{\rm L+R}$, which in turn acts on $U$ as 
$U \to U'= g U g^\dag$, giving rise to the target space 
$\tilde G/\tilde H \simeq$ SU(2)$_{\rm L-R}$.
The conventional potential term in the Skyrme model, viz.~the pion
mass term, is $V = m_{\pi}^2\tr (2{\bf 1}_2 - U - U^\dag)$,
and it \emph{explicitly} breaks the symmetry $\tilde G$ to 
SU(2)$_{\rm L+R}$. 

In this paper, it will prove convenient to use the following notation
of an O(4) nonlinear sigma model where we express the field $U$ in
terms of four real scalar fields $n_A(x)$ ($A=1,2,3,4$)
with the constraint $\sum_A n_A^2 =1$:
\beq
U = i n_a \sigma^a + n_4 \mathbf{1}_2 \equiv 
\mathbf{n} \cdot \mathbf{t} \, , 
\eeq
where $a=1,2,3$ is summed over, $\sigma^a$ are the Pauli matrices and
$U^\dag U = \mathbf{1}_2$ is equivalent to
$\mathbf{n}\cdot\mathbf{n}=1$.
We thus obtain the O(4) sigma model with the
Skyrme and six-derivative terms
\begin{align}
\mathcal{L} &= 
-\frac{c_2}{2}\p_\mu\mathbf{n}\cdot\p^\mu\mathbf{n}
+ c_4 \mathcal{L}_4 
+ c_6 \mathcal{L}_6  - V(\mathbf{n})  , \label{eq:LO4} \\
\mathcal{L}_4 &=
-\frac{1}{4} \left[
\left(\p_\mu\mathbf{n}\cdot\p^\mu\mathbf{n}\right)^2 
-
\left(\p_\mu\mathbf{n}\cdot\p_\nu\mathbf{n}\right)^2
\right]
\, ,  \\
\mathcal{L}_6 
&= \frac{1}{36} \left(\epsilon^{A B C D} \epsilon^{\mu\nu\rho\sigma} 
n_A \partial_\nu n_B \partial_\rho n_C \partial_\sigma n_D\right)^2
\\
&= -\frac{1}{6}\left[
\left(\p_\mu\mathbf{n}\cdot\p^\mu\mathbf{n}\right)^3
-3\left(\p_\mu\mathbf{n}\cdot\p^\mu\mathbf{n}\right)
\left(\p_\nu\mathbf{n}\cdot\p_\rho\mathbf{n}\right)^2
+2\left(\p_\mu\mathbf{n}\cdot\p^\nu\mathbf{n}\right)
\left(\p_\nu\mathbf{n}\cdot\p^\rho\mathbf{n}\right)
\left(\p_\rho\mathbf{n}\cdot\p^\mu\mathbf{n}\right)
\right].
\nonumber
\end{align}
The symmetry SO(4) $\sim$ SU(2) $\times$ SU(2) 
for $V=0$ is thus manifest.

The target space (the vacuum manifold with $V=0$) 
$\mathcal{M}\simeq$ SU(2) $\simeq S^3$ has a nontrivial homotopy group  
\beq
\pi_3(M) = \mathbb{Z}, 
\eeq
which admits Skyrmions. 
The baryon number (the Skyrme charge) of $B \in \pi_3(S^3)$ is defined
as 
\beq
B &=& -\1{24\pi^2} \int d^3x \; \epsilon^{ijk} 
\tr \left( U^\dag\p_i U U^\dag\p_j U U^\dag\p_k U\right) \non
&=& \1{24\pi^2} \int d^3x \; \epsilon^{ijk} 
\tr \left( U^\dag\p_i U\p_j U^\dag\p_k U\right) \non
%&=& \1{4\pi^2} \int d^3x \; \epsilon^{ijk} 
%\phi^\dag \p_i\phi \p_j\phi^\dag \p_k \phi .
%\non
&=& -\frac{1}{12 \pi^2} \int d^3x \; \epsilon^{ABCD} \epsilon^{ijk} 
\partial_i n_A \partial_j n_B \partial_k n_C n_D\non
 &=& -\frac{1}{2 \pi^2} \int d^3x \; \epsilon^{ABCD} 
\partial_1 n_A \partial_2 n_B \partial_3 n_C n_D.
\eeq

Instead of the conventional potential term, we consider here the
following potential term 
\beq
V = m^2 n_4^2. 
\label{eq:pot}
\eeq
We call this potential the Heisenberg type.
The vacua of this potential is determined by $n_4=0$, 
and thus the vacuum manifold is a sphere
\beq 
\mathcal{M} \simeq S^2,
\label{eq:vac_manifold}
\eeq 
parametrized by $n_1,n_2,n_3$ 
with a constraint $\sum_{a=1,2,3} n_a^2=1$.
From the homotopy group 
\beq
\pi_2({\cal M}) \simeq {\mathbb Z},
\eeq 
this model admits a monopole, viz.~a global monopole. 

In our previous papers
\cite{Nitta:2012wi,Gudnason:2013qba,Gudnason:2014nba,Gudnason:2014hsa,Gudnason:2014jga}, 
we considered instead the potential term 
$V = m^2 (1-n_4^2)$, i.e.~a modified mass term.
This admits two discrete vacua and a domain wall
interpolating between them.

In this paper, the potential \eqref{eq:pot} differs seemingly only by
the overall sign. The physics described by this potential is, however,
vastly different. The boundary condition at spatial infinity is
\beq
n_4 = 0, \qquad \mathrm{for}\ \ r\to\infty, 
\eeq
which has the vacuum manifold \eqref{eq:vac_manifold}. 
The vacuum state is thus any point on the two-sphere, given
by $n_1^2+n_2^2+n_3^2=1$. 
The potential \eqref{eq:pot} breaks the symmetry 
$\tilde G =$ SU(2)$_{\rm L} \times $SU(2)$_{\rm R}$ down to
SU(2)$_{L+R}$, \emph{explicitly}.
This SU(2) or O(3) symmetry is, however, spontaneously broken down to
O(2).
The spontaneous breaking O(3)/O(2) gives rise to 2 Nambu-Goldstone
bosons which remain massless as well as 1 massive pion.
Abusing notation a bit, we can say that there are 2 massless pions and 
1 massive pion. 
The phase of QCD we are trying to mimic, is high-density QCD at
sufficiently high density; above the critical density for the
formation of half-Skyrmions \cite{Ma:2013ooa} and below the chiral
restoration density. 
The potential \eqref{eq:pot} is an effective potential that can model
this phase in the sense that it allows for half Skyrmions. It may or
may not be the complete potential describing real QCD at mentioned
densities, but in this paper we will study the above presented
Skyrme-like models in the presence of said potential.

%%%%%%%%%%%%%%%%%%%%%%%%%%
\section{Fractional Skyrmion molecules \label{sec:molecule}} 

In this section we will calculate numerical solutions of fractional
molecules in the $B=1$ sector, which are bound states of two half
baryons. The only free parameter is the mass $m$, i.e.~the coefficient
in front of the Heisenberg-type potential. We will carry out the study
in both the 2+4 and the 2+6 models. Instead of varying the mass, we
will keep the mass fixed and vary the parameters $(c_2,c_4)$ and
$(c_2,c_6)$ for the 2+4 and the 2+6 models, respectively. This
includes the possibility of having $c_2 = 0$ which is not possible in
the rescaled system where only the mass is varied. The $c_2=0$ region
of parameter space is especially interesting due to the BPS properties 
that are present in that limit, see \cite{Adam:2010fg} for the 2+6
model and \cite{Harland:2013rxa} for a possibility in the 2+4 model
(although this requires a particular potential). 

As an Ansatz for the initial guess fed to the relaxation procedure, we
will simply use the hedgehog Ansatz, suitable for the $B=1$ sector
\beq
\mathbf{n} = \left(
-\cos f(r),
\hat{\mathbf{x}}\sin f(r)
\right), \label{eq:hedgehog}
\eeq
where $\hat{\mathbf{x}}$ is the 3-dimensional spatial unit vector and
$r^2=x^2+y^2+z^2$ is the radial coordinate. The relaxation method will
then deform the initial guess away from the spherical initial guess to
the correct minimal-energy state of molecular shape (due to the
Heisenberg-type potential). 
Note that we have chosen a particular value on the vacuum manifold,
namely $n_1=-1$, which of course by O(3) symmetry is equivalent to any
other choice. 

We are now ready to perform the numerical calculation and we use the
finite difference method to discretize the fields $\mathbf{n}$, feed
the initial guess to the relaxation algorithm and simply evolve a
linear time operator until the numerical precision of the solution
satisfies our criteria. 

\begin{figure}[!ptbh]
\begin{center}
\captionsetup[subfloat]{labelformat=empty}
\mbox{
\subfloat[$c_2=0,c_4=\tfrac{1}{4}$]{\includegraphics[width=0.33\linewidth]{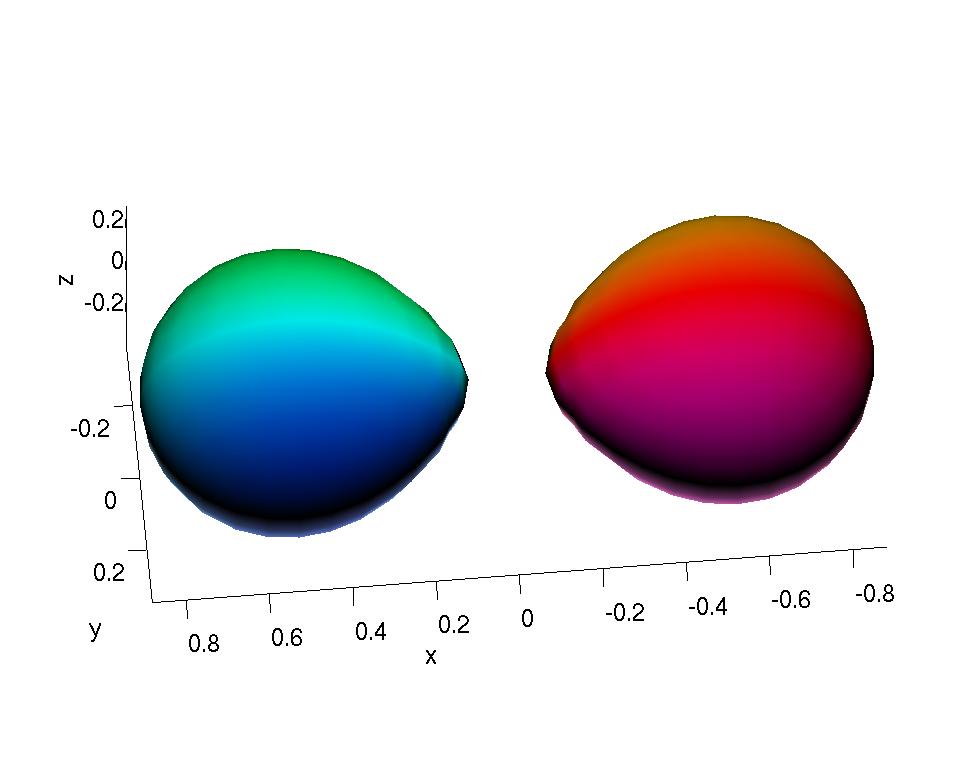}}
\subfloat[$c_2=0,c_4=1$]{\includegraphics[width=0.33\linewidth]{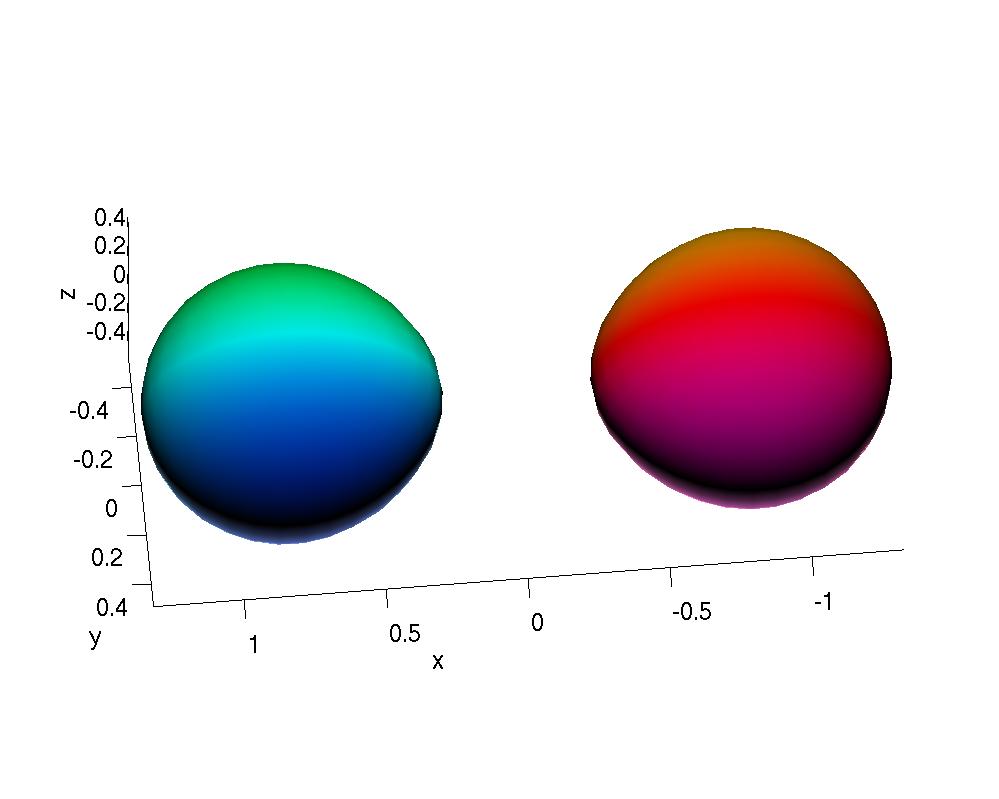}}
\subfloat[$c_2=0,c_4=4$]{\includegraphics[width=0.33\linewidth]{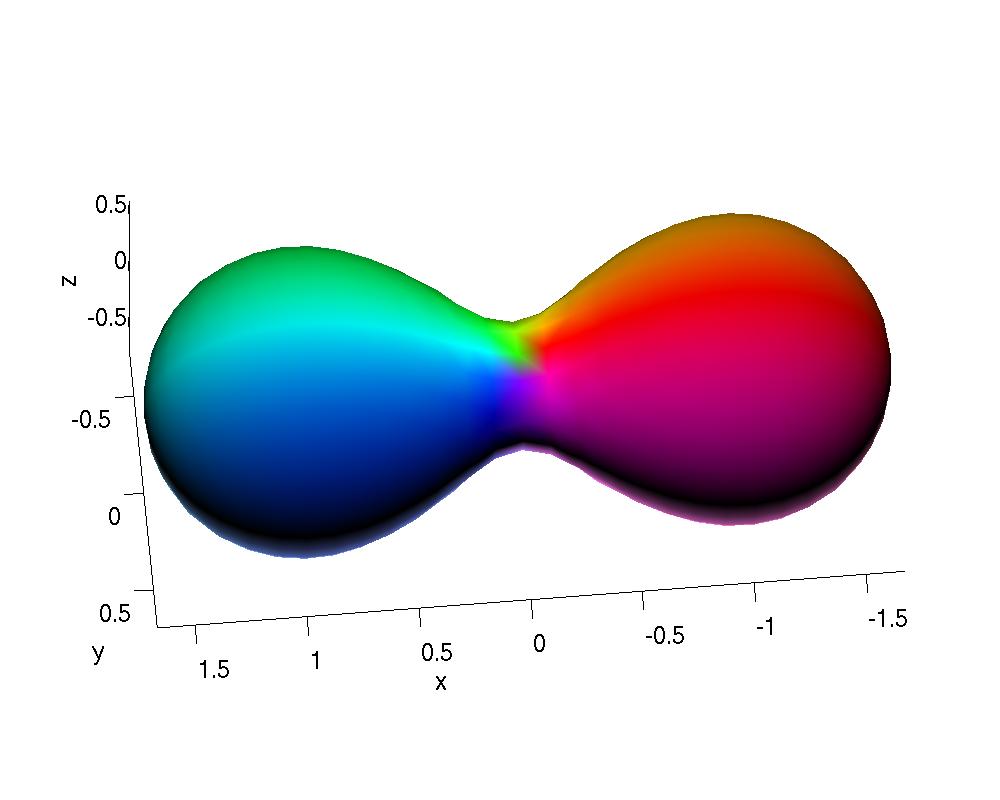}}}
\mbox{
\subfloat[$c_2=\tfrac{1}{4},c_4=\tfrac{1}{4}$]{\includegraphics[width=0.33\linewidth]{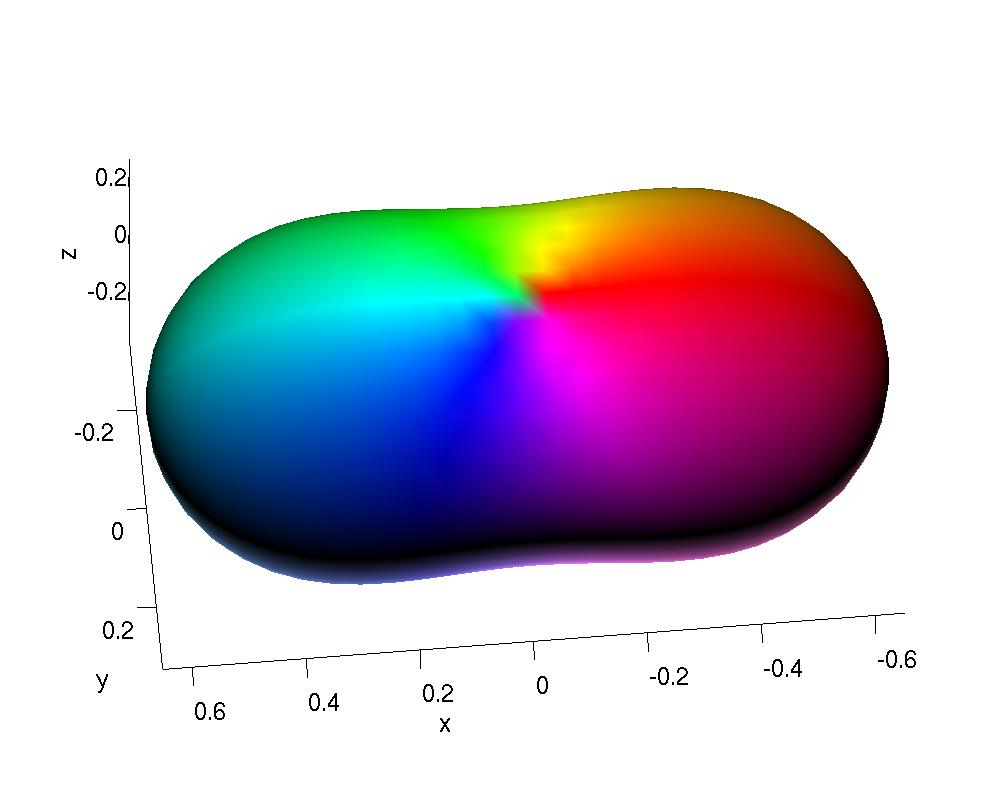}}
\subfloat[$c_2=\tfrac{1}{4},c_4=1$]{\includegraphics[width=0.33\linewidth]{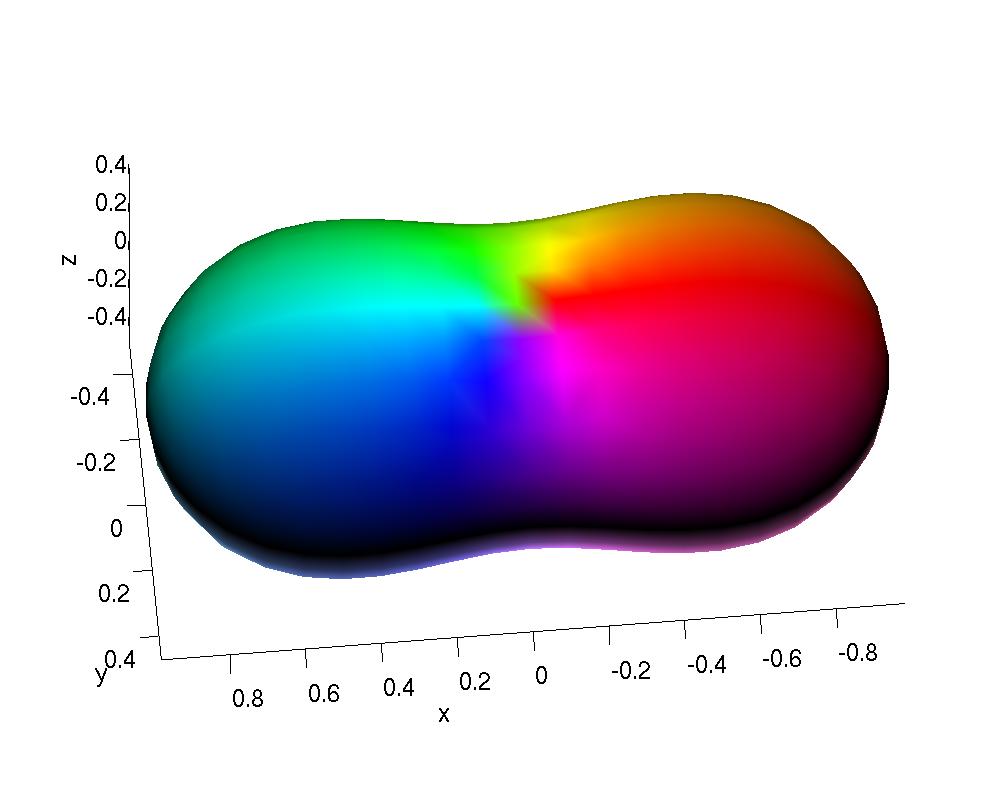}}
\subfloat[$c_2=\tfrac{1}{4},c_4=4$]{\includegraphics[width=0.33\linewidth]{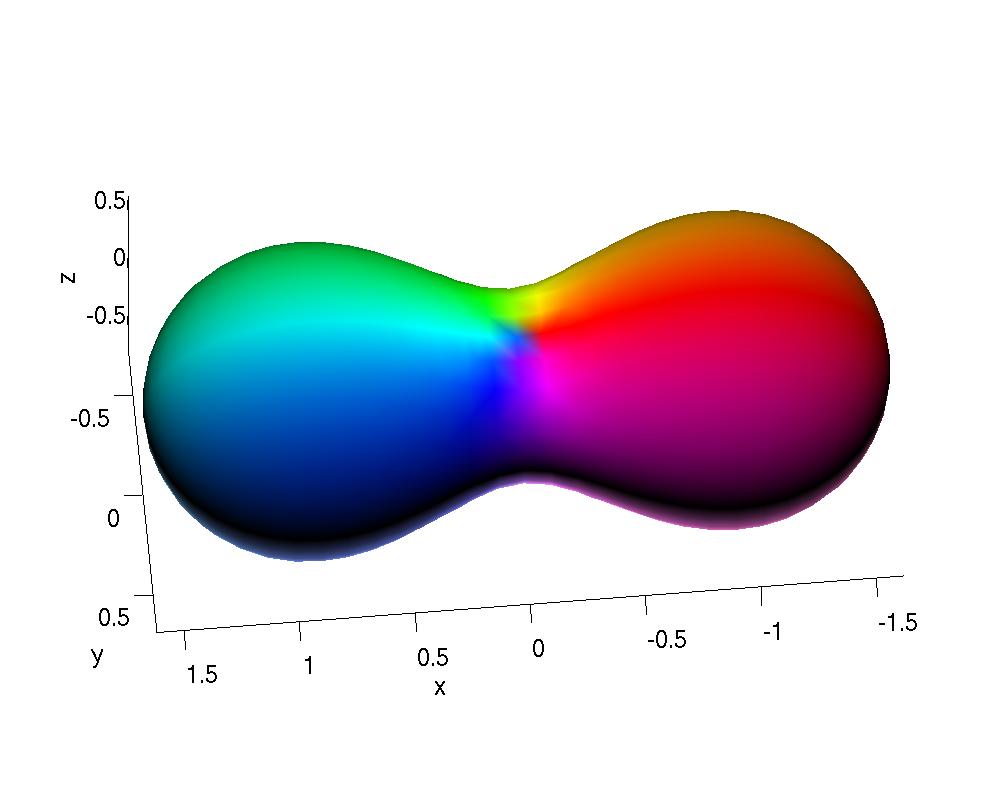}}}
\mbox{
\subfloat[$c_2=1,c_4=\tfrac{1}{4}$]{\includegraphics[width=0.33\linewidth]{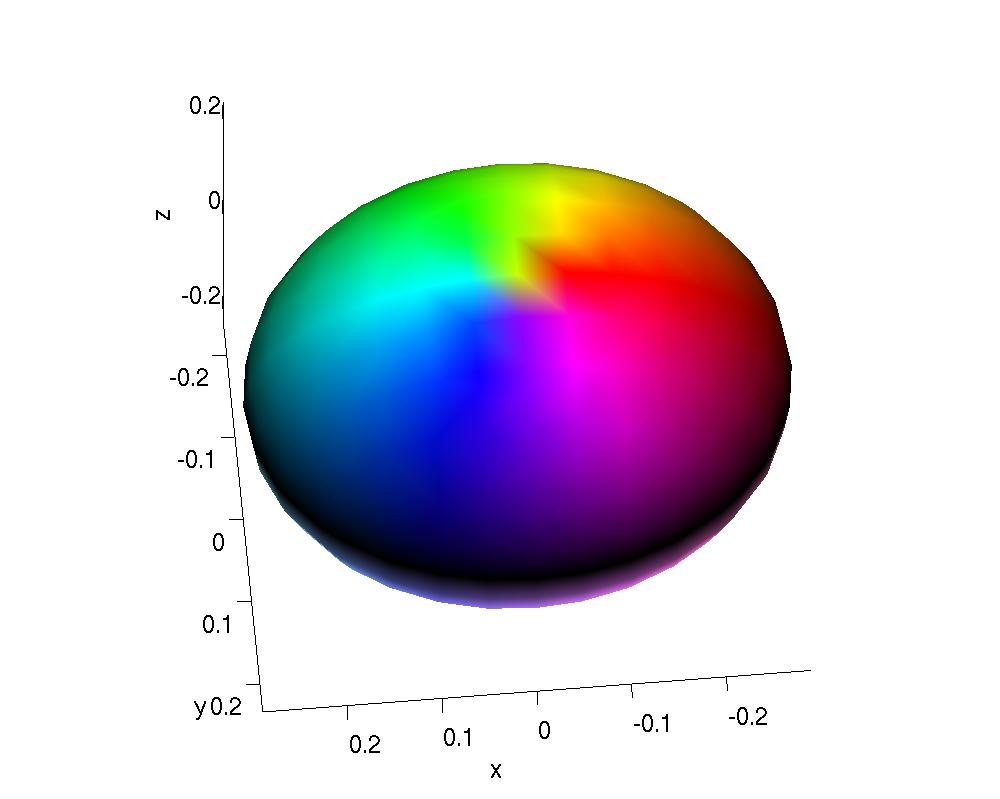}}
\subfloat[$c_2=1,c_4=1$]{\includegraphics[width=0.33\linewidth]{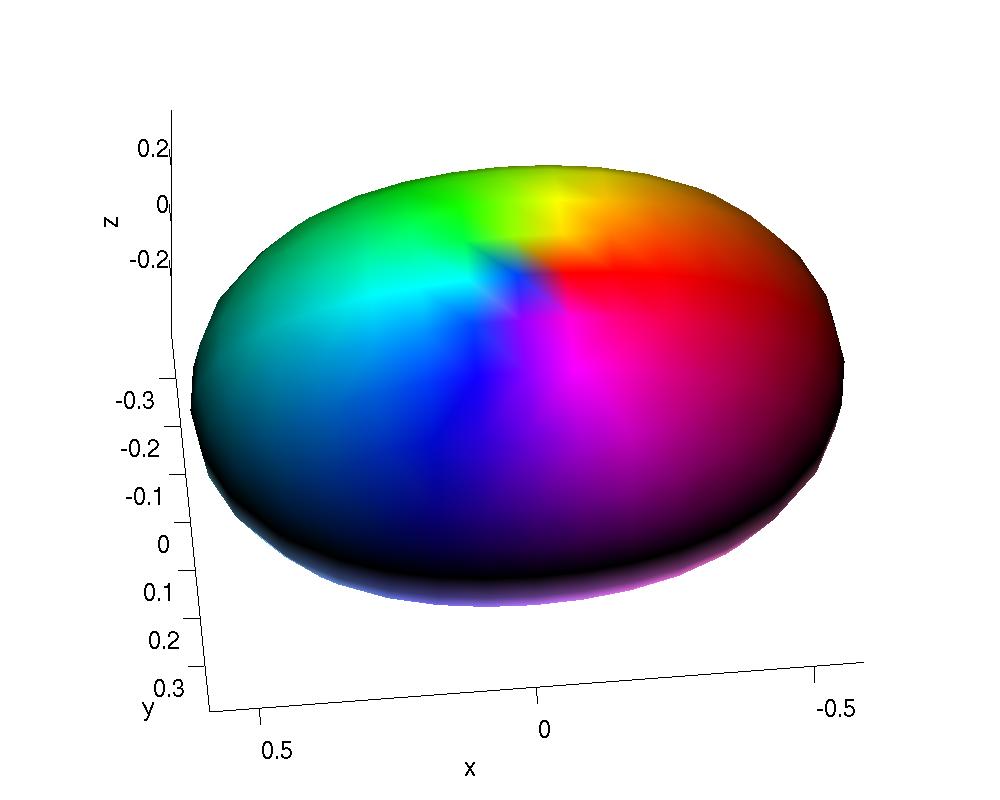}}
\subfloat[$c_2=1,c_4=4$]{\includegraphics[width=0.33\linewidth]{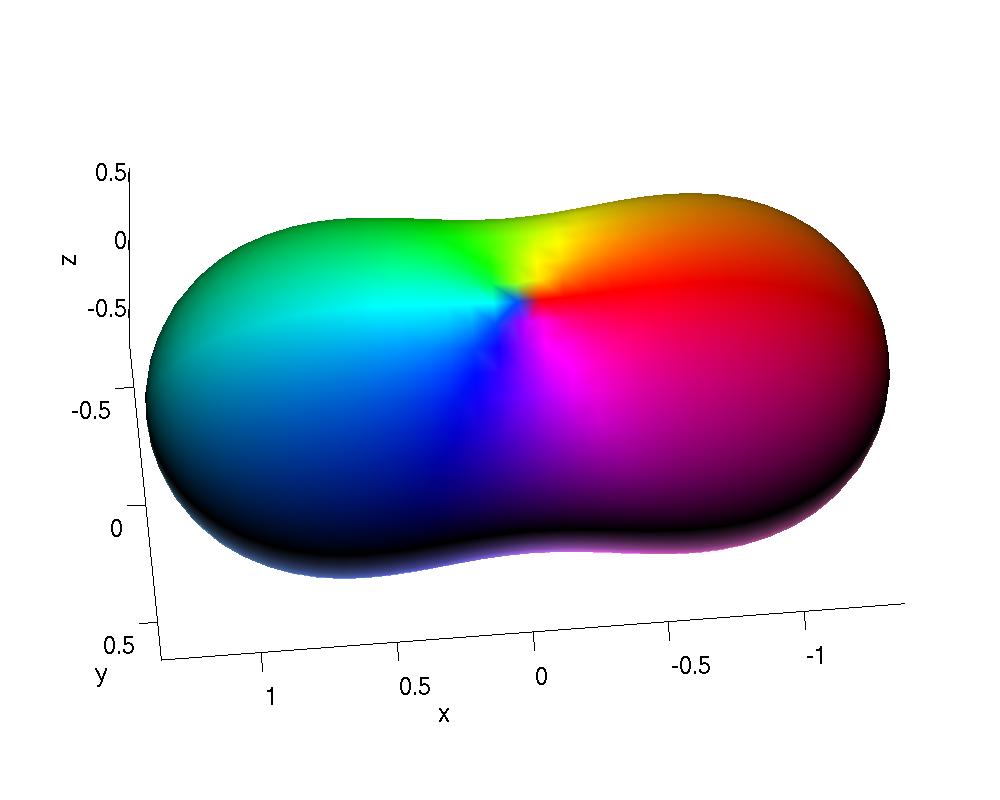}}}
\caption{Isosurfaces showing the half-maximum of the baryon charge
  density in the 2+4 model for various choices of $(c_2,c_4)$ for
  fixed mass $m=4$. The color scheme represents the normalized
  3-vector $\vec\pi=\frac{\vec n}{|\vec n|}$, where ${\rm
    arg}(\pi_2+i\pi_3)$ is the hue and the lightness is given by
  $|\pi_1|$. 
}
\label{fig:M4B1}
\end{center}
\end{figure}

We consider first the 2+4 model, i.e.~$c_2\geq 0,c_4>0$ and
$c_6=0$. 
In Fig.~\ref{fig:M4B1} is shown an array of molecule baryon charge
density isosurfaces for various choices of the coefficients
$(c_2,c_4)$ for fixed mass $m=4$. The coloring scheme used is based on
the hue-saturation-lightness (HSL) parameters and the hue is 
given by the phase  
${\rm arg}(\pi_2+i\pi_3)$, the lightness is given by $|\pi_1|$ and
$\vec\pi\equiv \vec n/|\vec n|$ is a normalized 3-vector. 
All the numerical calculations throughout the paper are
  carried out on an $81^3$ cubic lattice using the relaxation method. 
We observe
that the molecular shape is most pronounced when the coefficients are
small (which is clear, because that corresponds to a large mass) and
even more so when $c_2\ll c_4$.

\begin{figure}[!ptbh]
\begin{center}
\captionsetup[subfloat]{labelformat=empty}
\mbox{
\subfloat[$c_2=0,c_4=\tfrac{1}{4}$]{\includegraphics[width=0.33\linewidth]{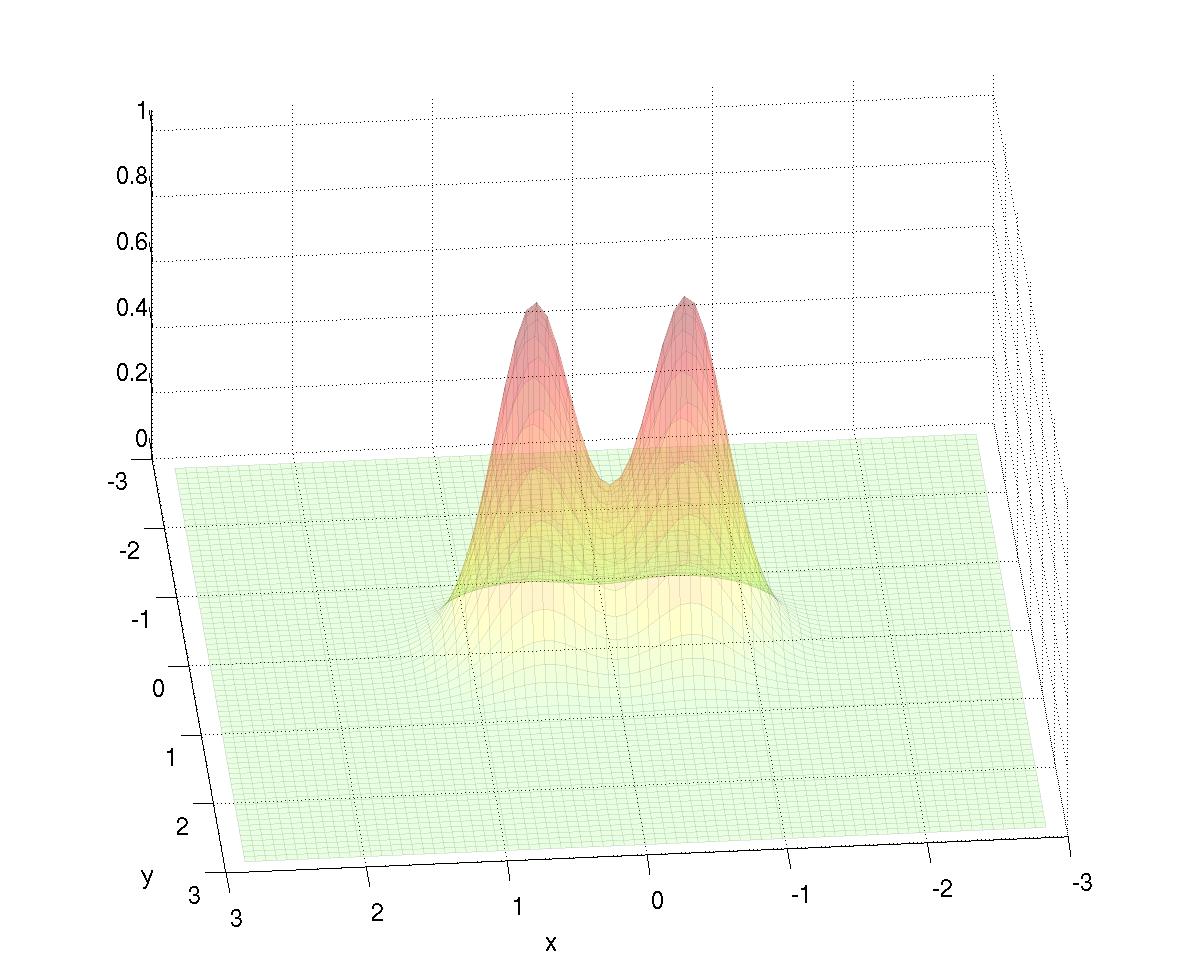}}
\subfloat[$c_2=0,c_4=1$]{\includegraphics[width=0.33\linewidth]{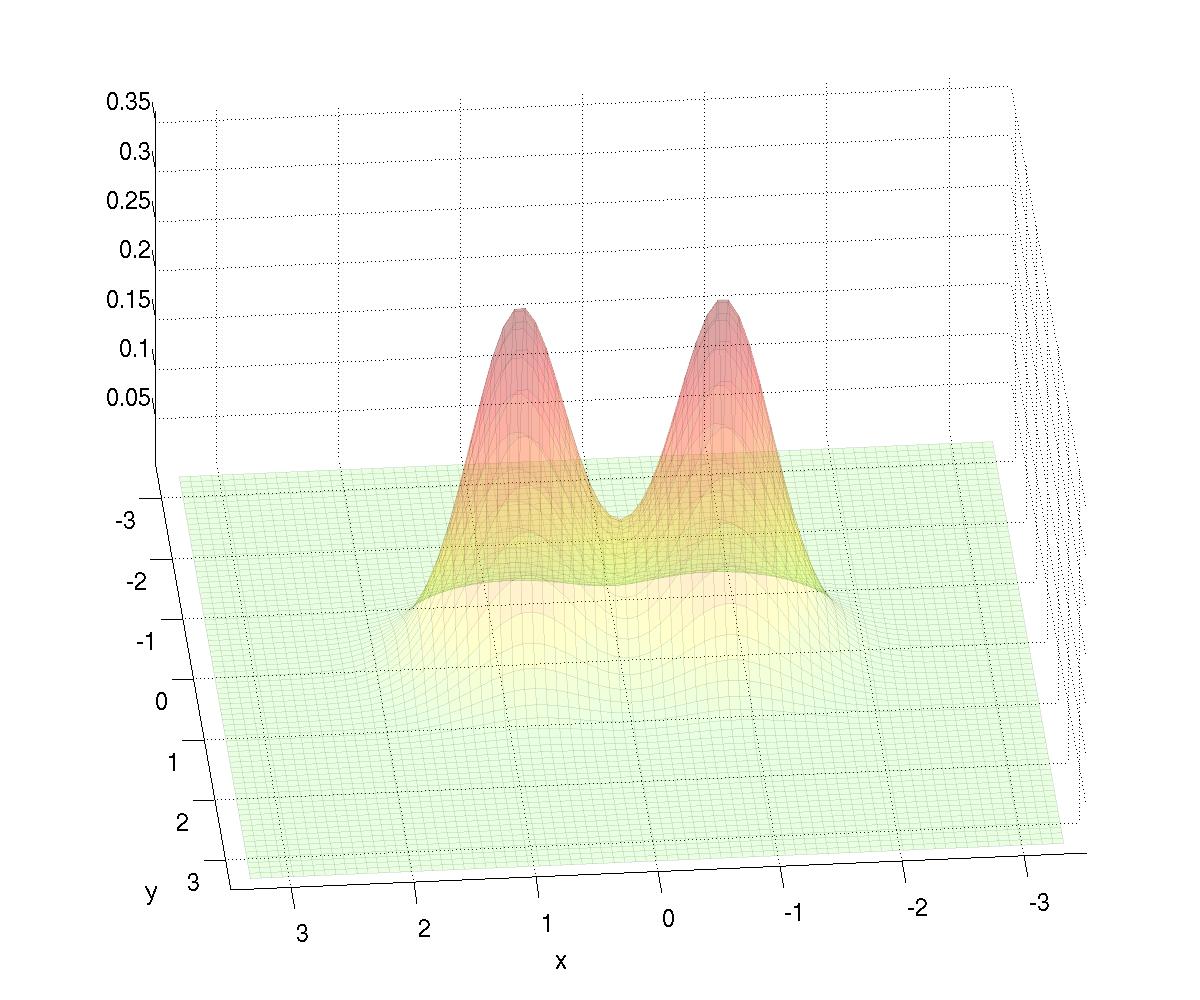}}
\subfloat[$c_2=0,c_4=4$]{\includegraphics[width=0.33\linewidth]{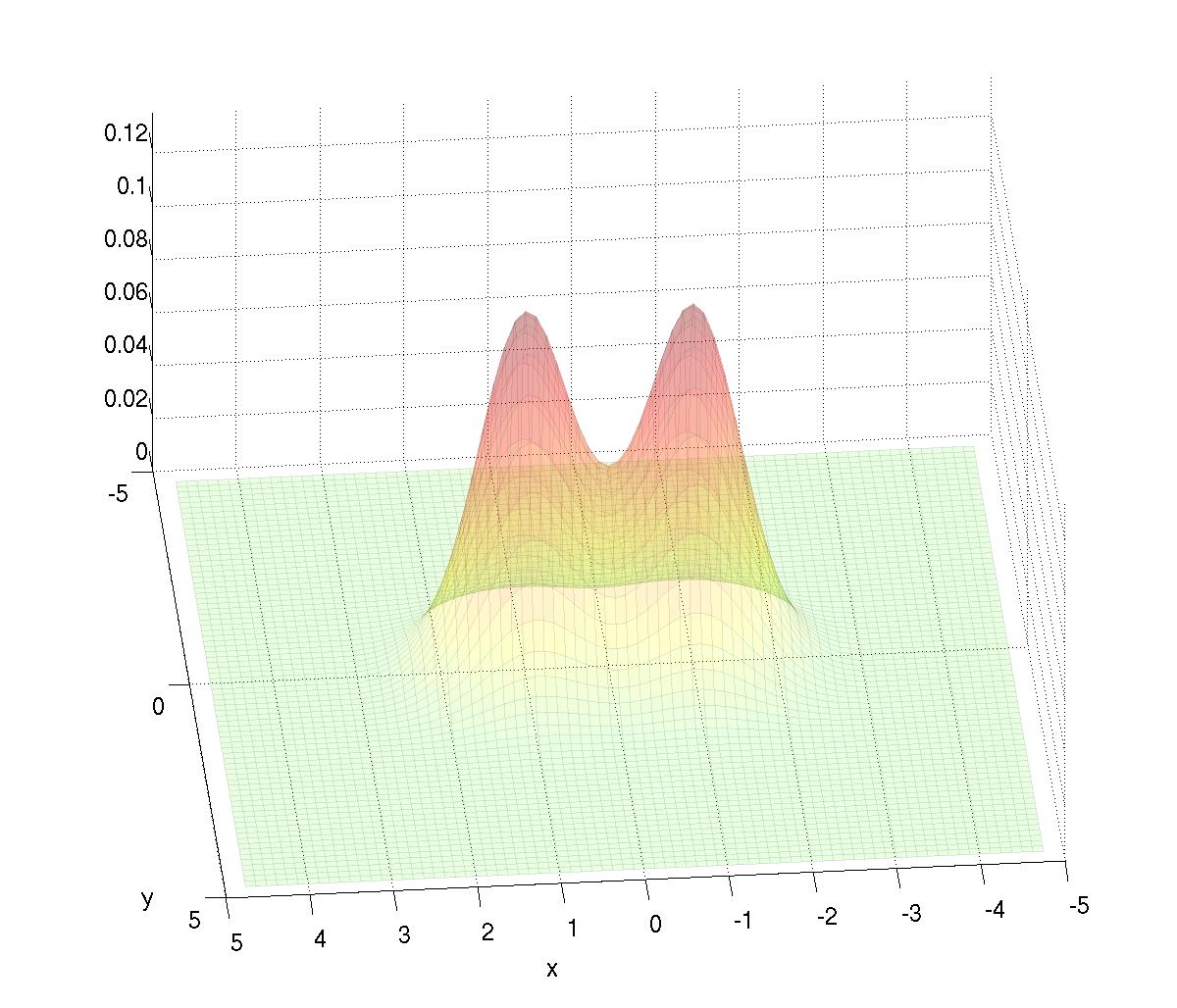}}}
\mbox{
\subfloat[$c_2=\tfrac{1}{4},c_4=\tfrac{1}{4}$]{\includegraphics[width=0.33\linewidth]{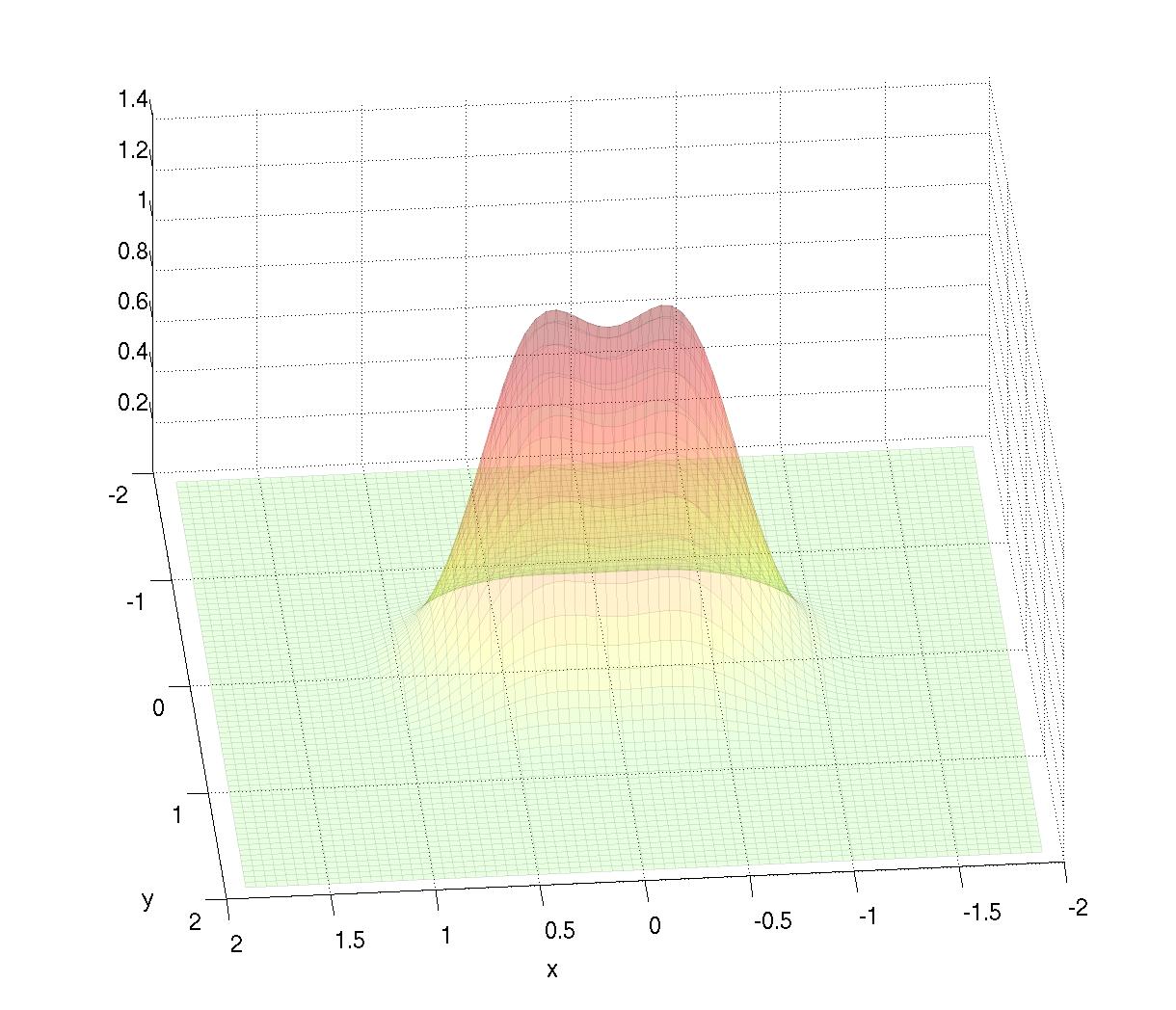}}
\subfloat[$c_2=\tfrac{1}{4},c_4=1$]{\includegraphics[width=0.33\linewidth]{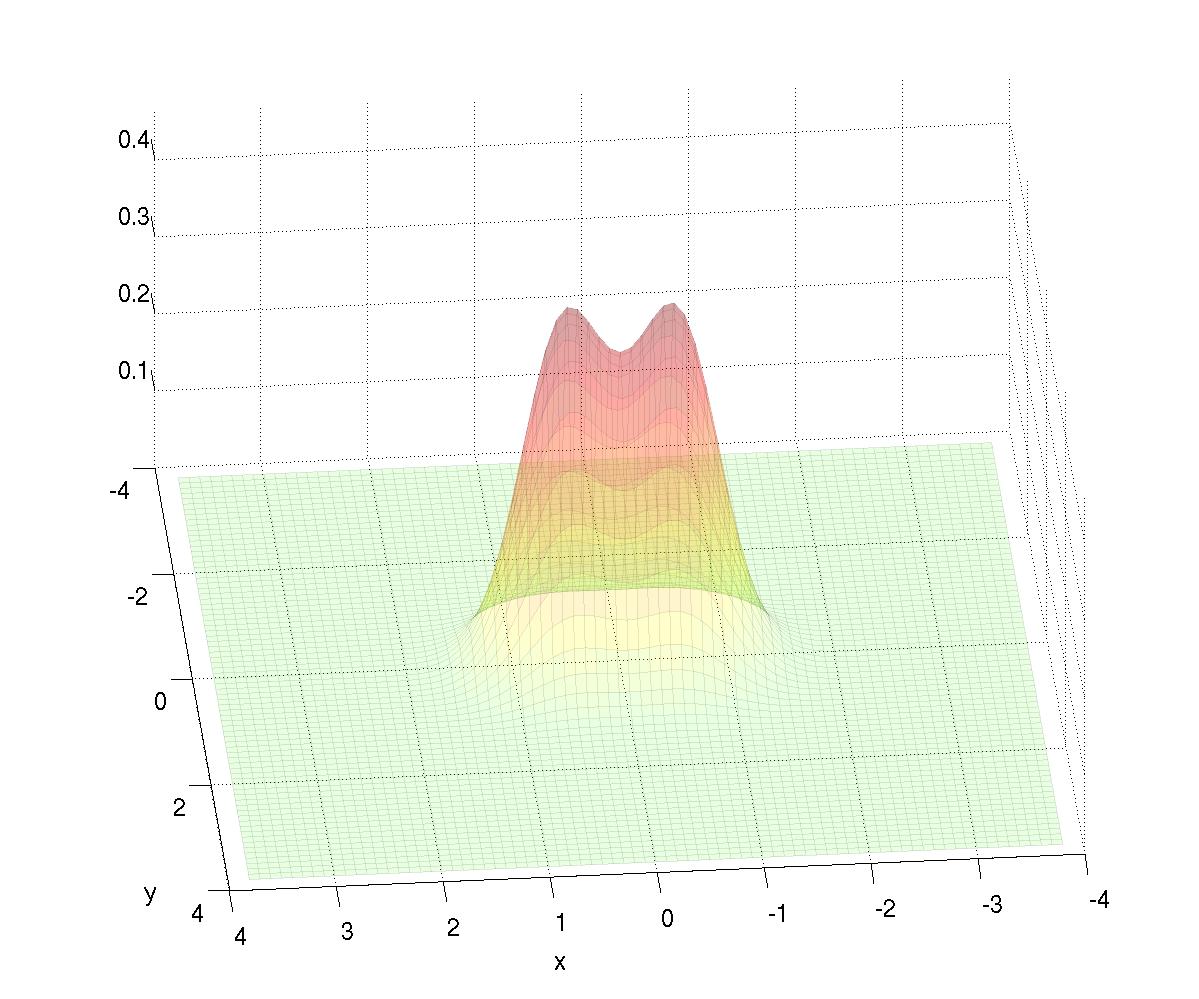}}
\subfloat[$c_2=\tfrac{1}{4},c_4=4$]{\includegraphics[width=0.33\linewidth]{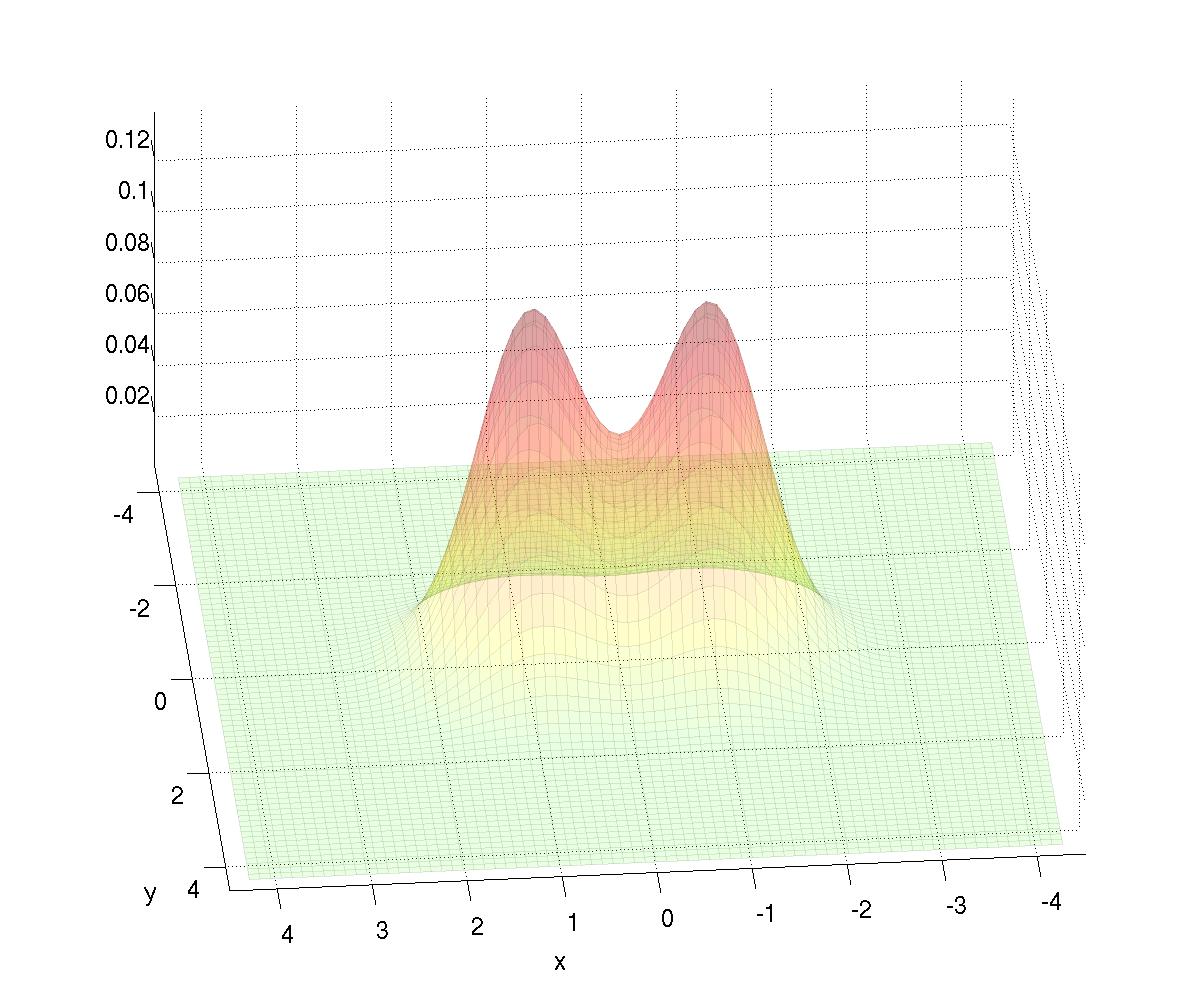}}}
\mbox{
\subfloat[$c_2=1,c_4=\tfrac{1}{4}$]{\includegraphics[width=0.33\linewidth]{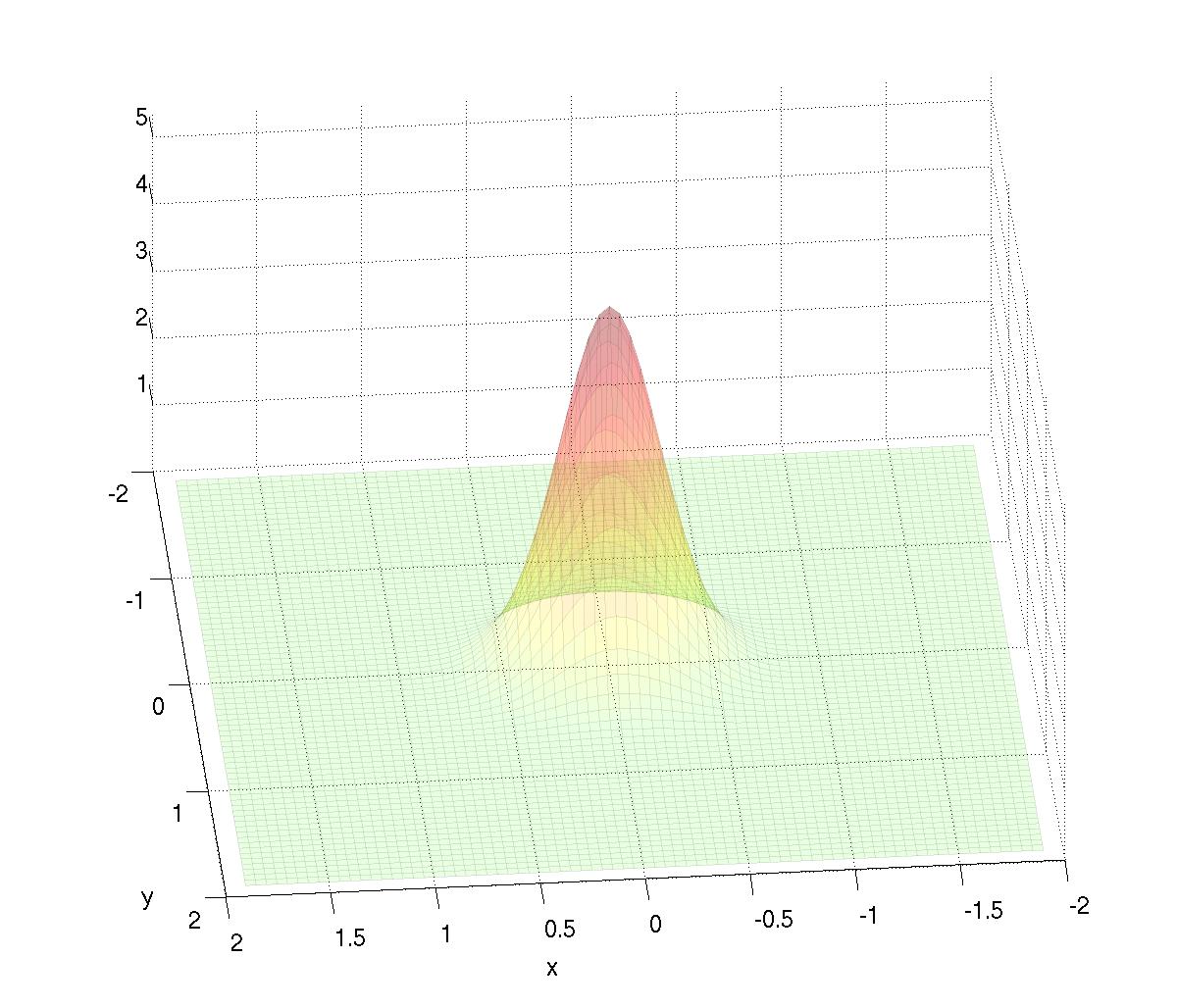}}
\subfloat[$c_2=1,c_4=1$]{\includegraphics[width=0.33\linewidth]{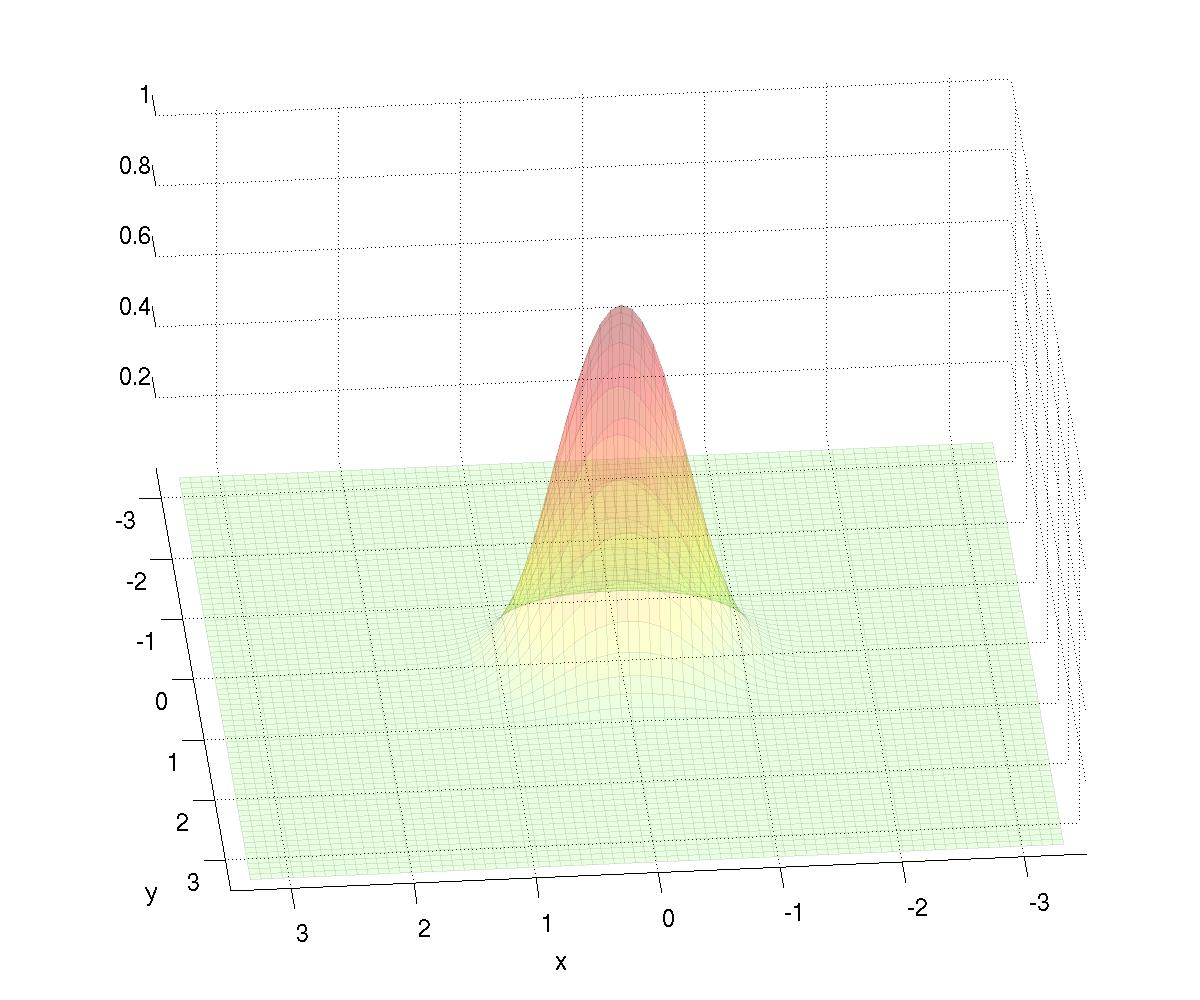}}
\subfloat[$c_2=1,c_4=4$]{\includegraphics[width=0.33\linewidth]{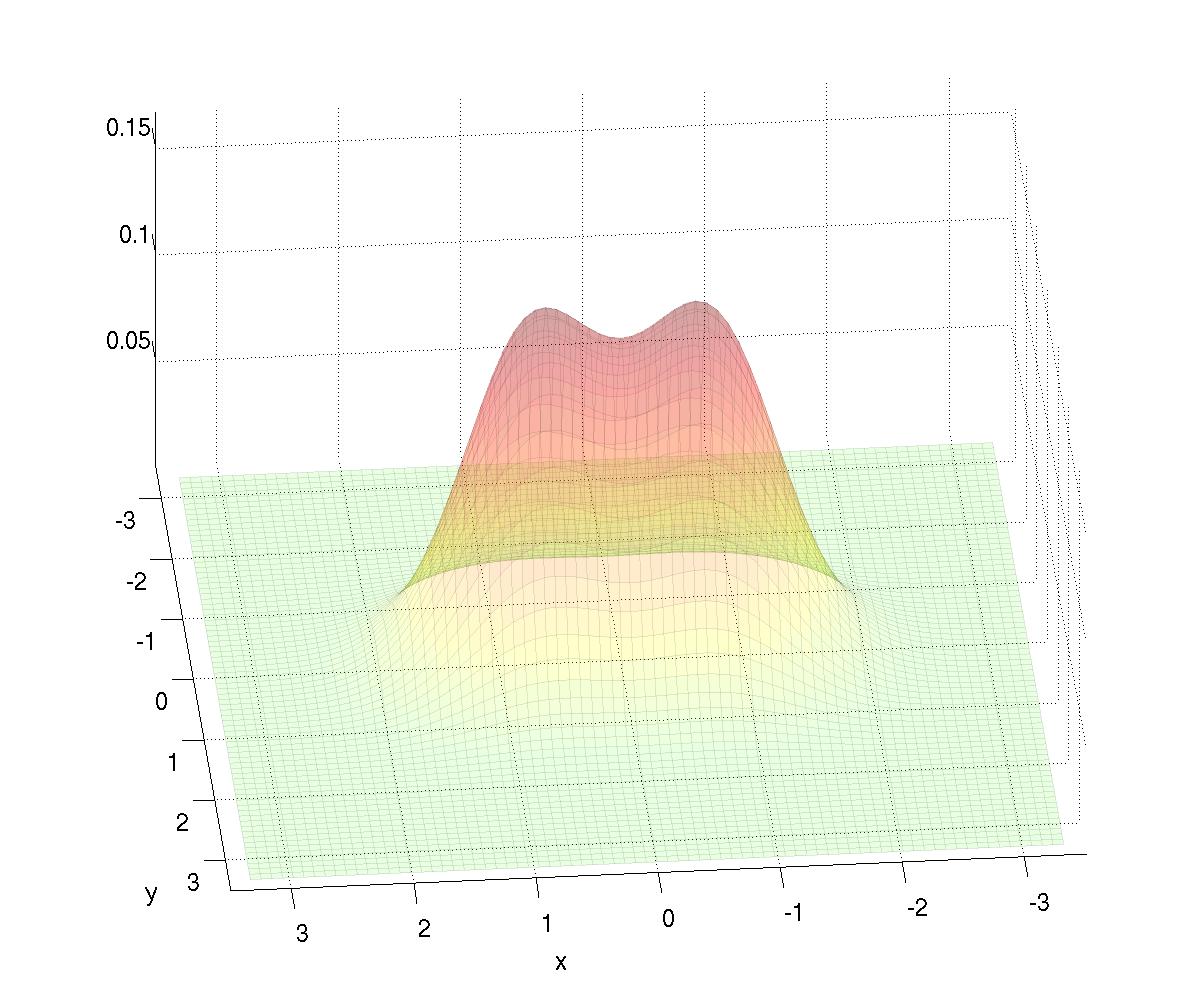}}}
\caption{Baryon charge density at a spatial slice through the molecule
  at $z=0$ in the 2+4 model for various choices of $(c_2,c_4)$ and for
  fixed mass $m=4$. 
}
\label{fig:M4B1_baryonslice}
\end{center}
\end{figure}

\begin{figure}[!ptbh]
\begin{center}
\captionsetup[subfloat]{labelformat=empty}
\mbox{
\subfloat[$c_2=0,c_4=\tfrac{1}{4}$]{\includegraphics[width=0.33\linewidth]{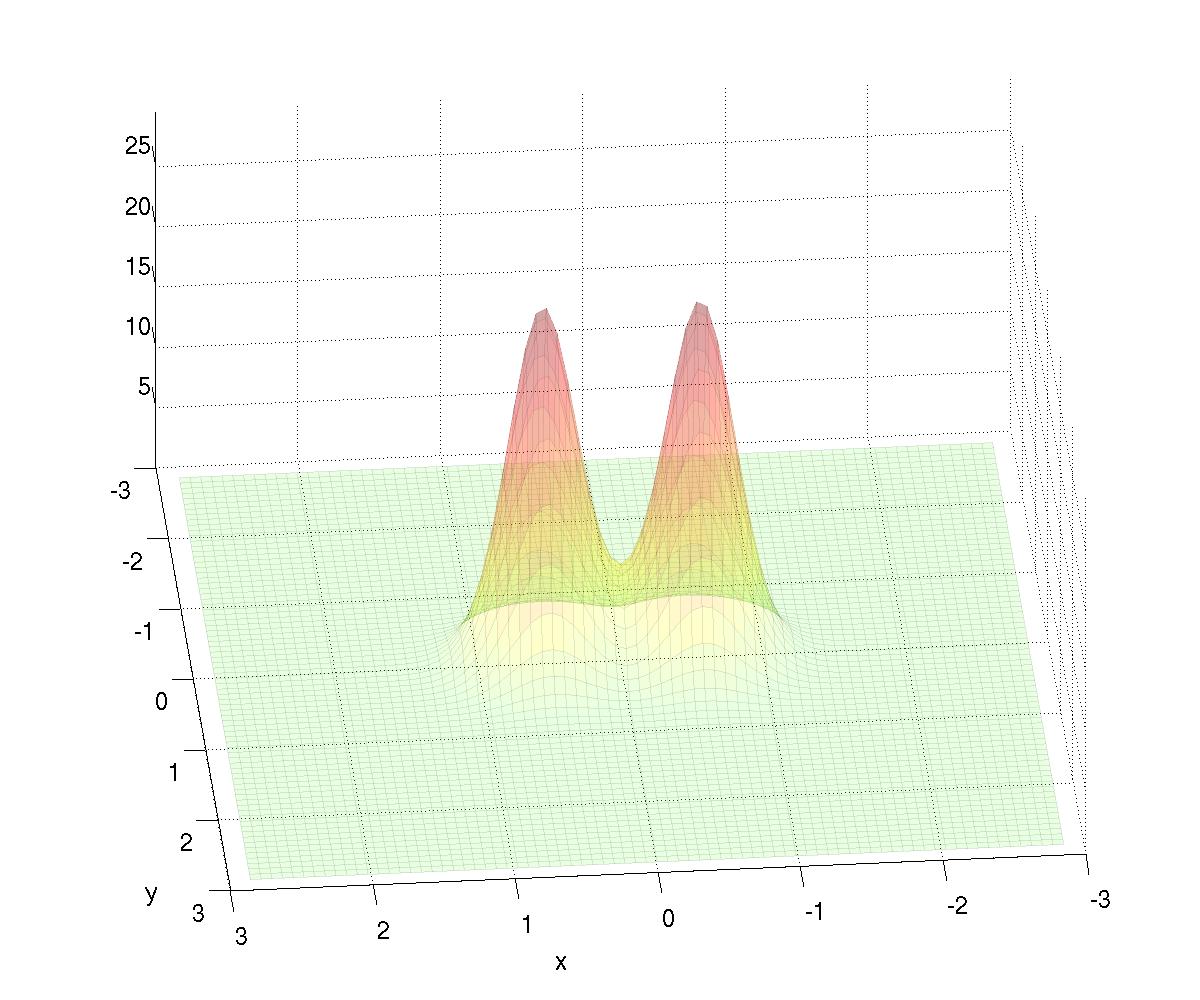}}
\subfloat[$c_2=0,c_4=1$]{\includegraphics[width=0.33\linewidth]{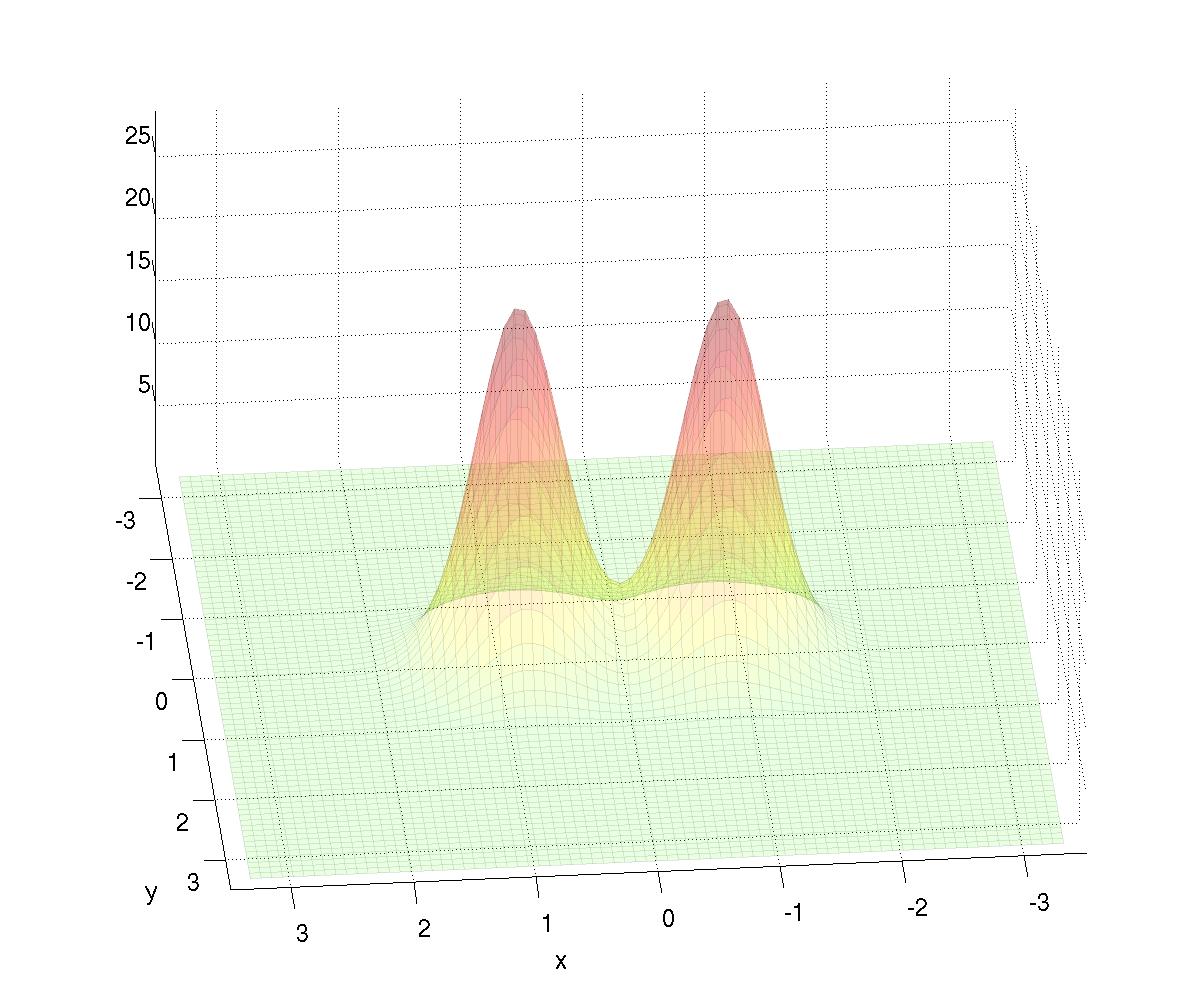}}
\subfloat[$c_2=0,c_4=4$]{\includegraphics[width=0.33\linewidth]{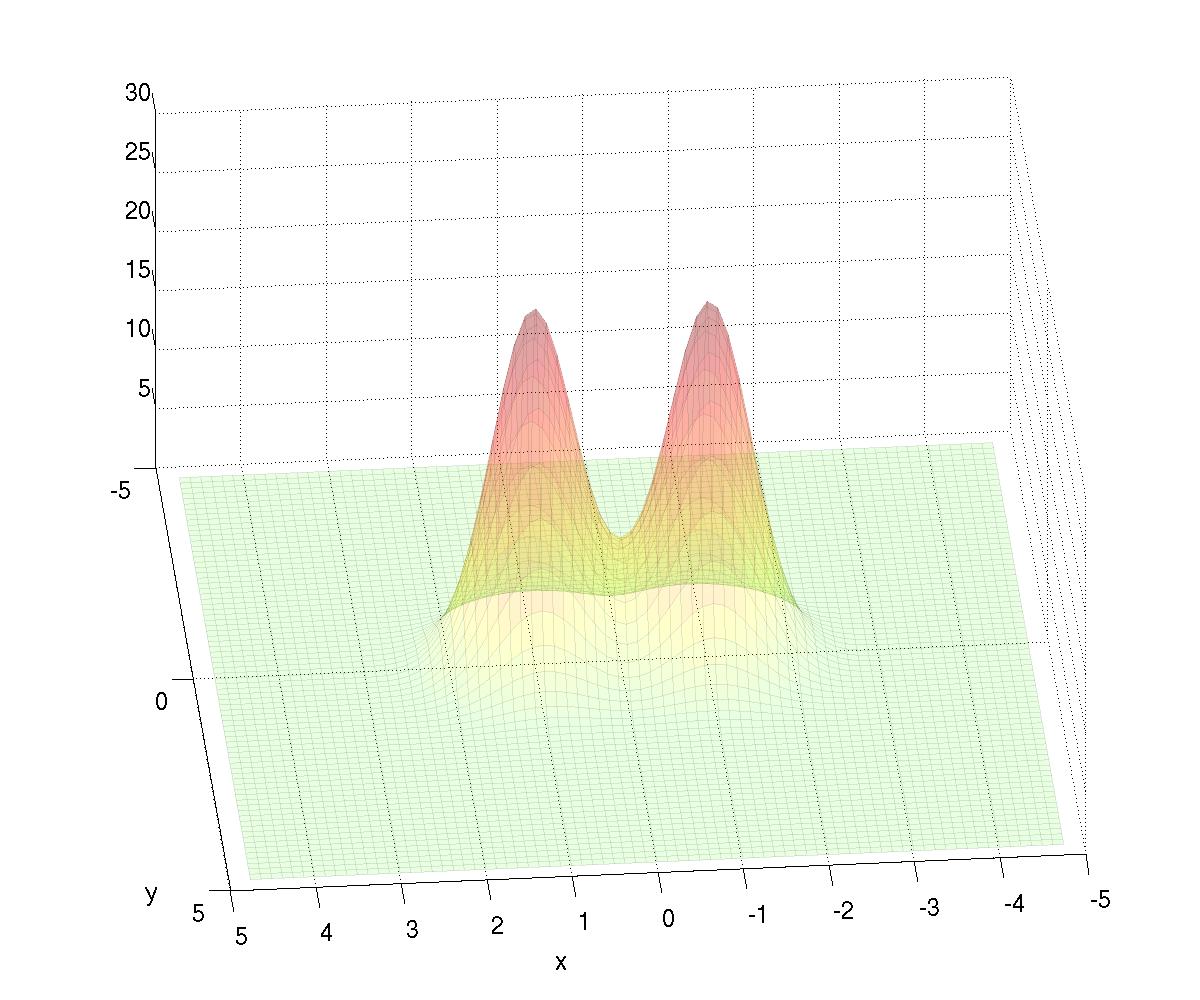}}}
\mbox{
\subfloat[$c_2=\tfrac{1}{4},c_4=\tfrac{1}{4}$]{\includegraphics[width=0.33\linewidth]{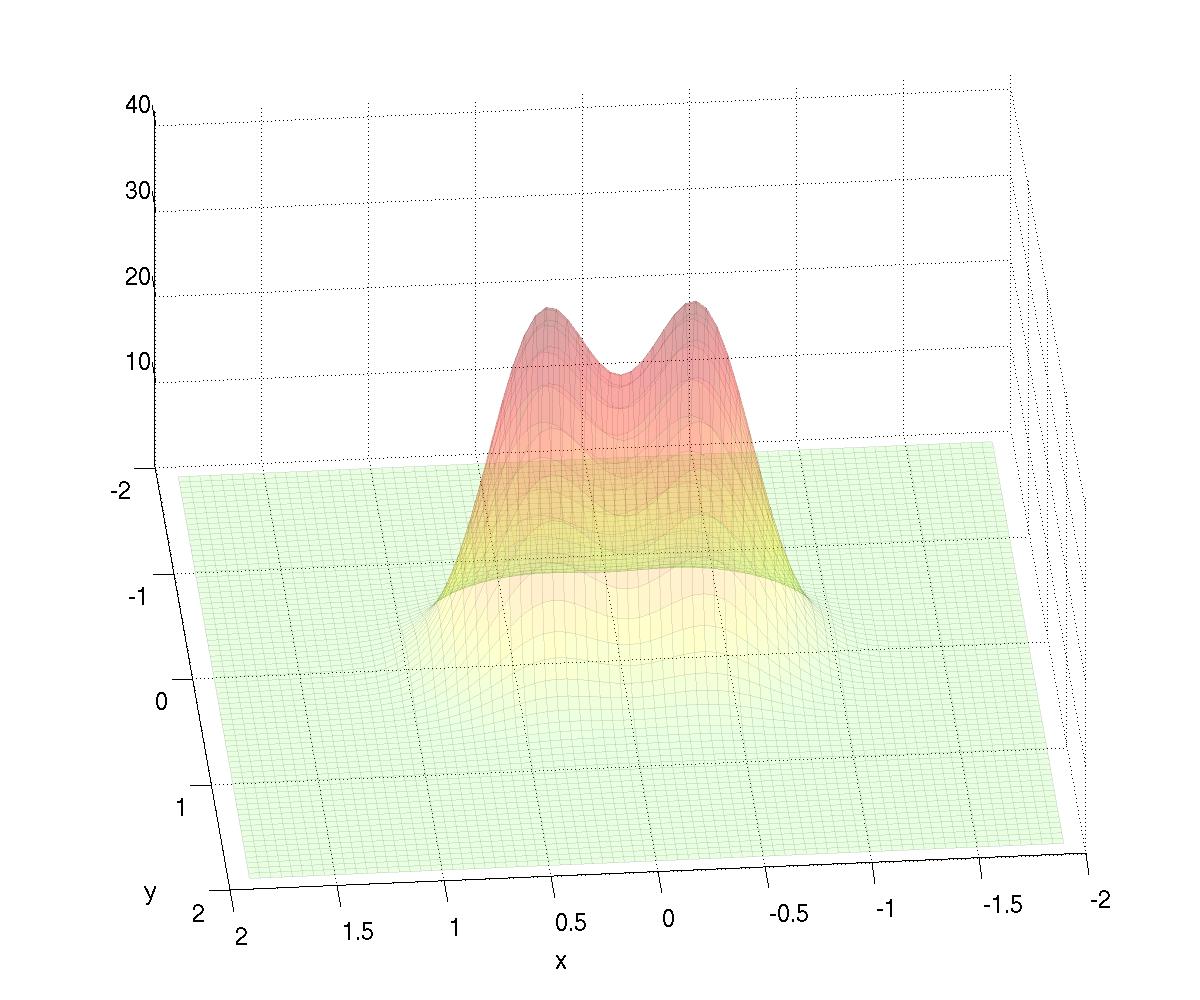}}
\subfloat[$c_2=\tfrac{1}{4},c_4=1$]{\includegraphics[width=0.33\linewidth]{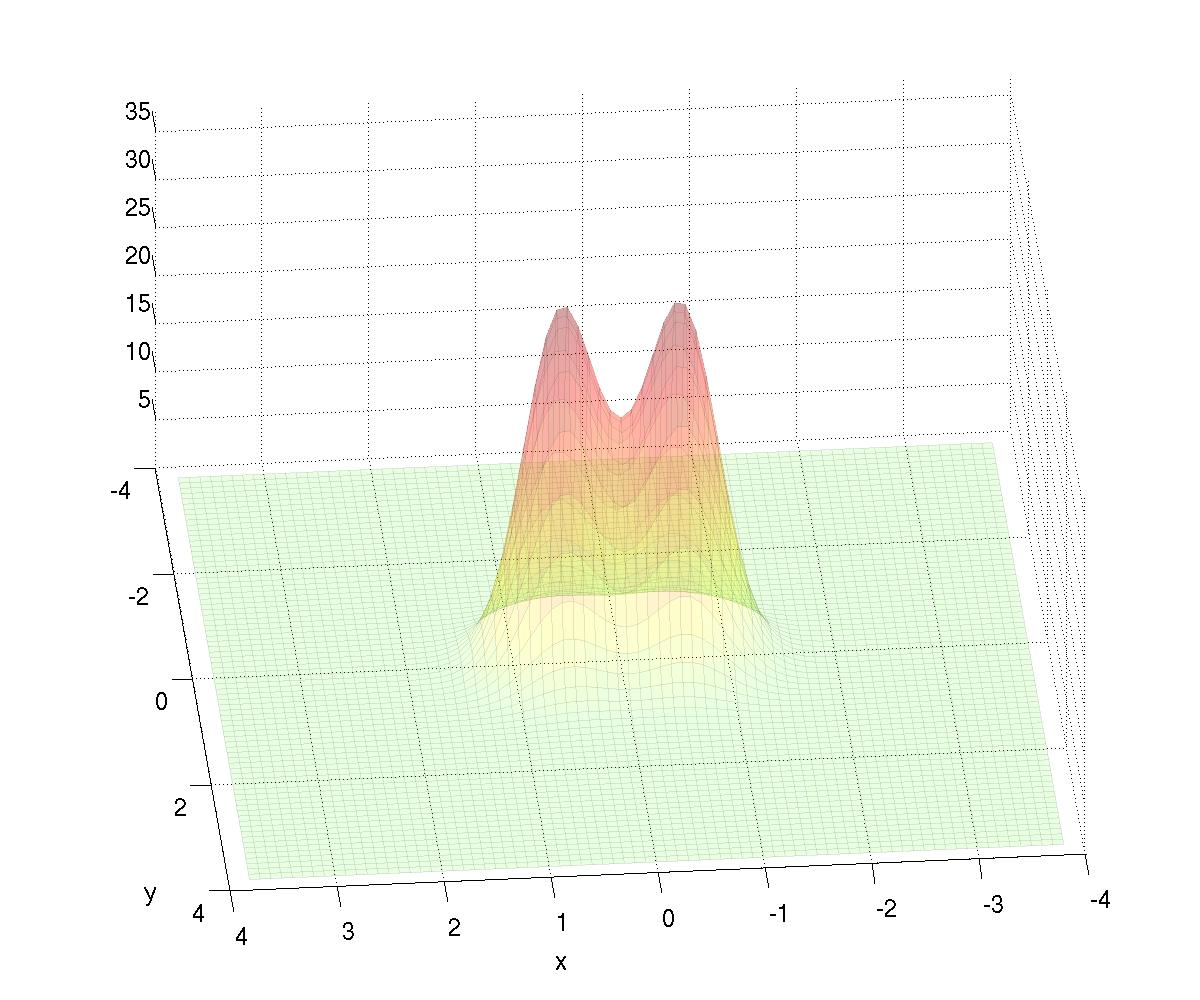}}
\subfloat[$c_2=\tfrac{1}{4},c_4=4$]{\includegraphics[width=0.33\linewidth]{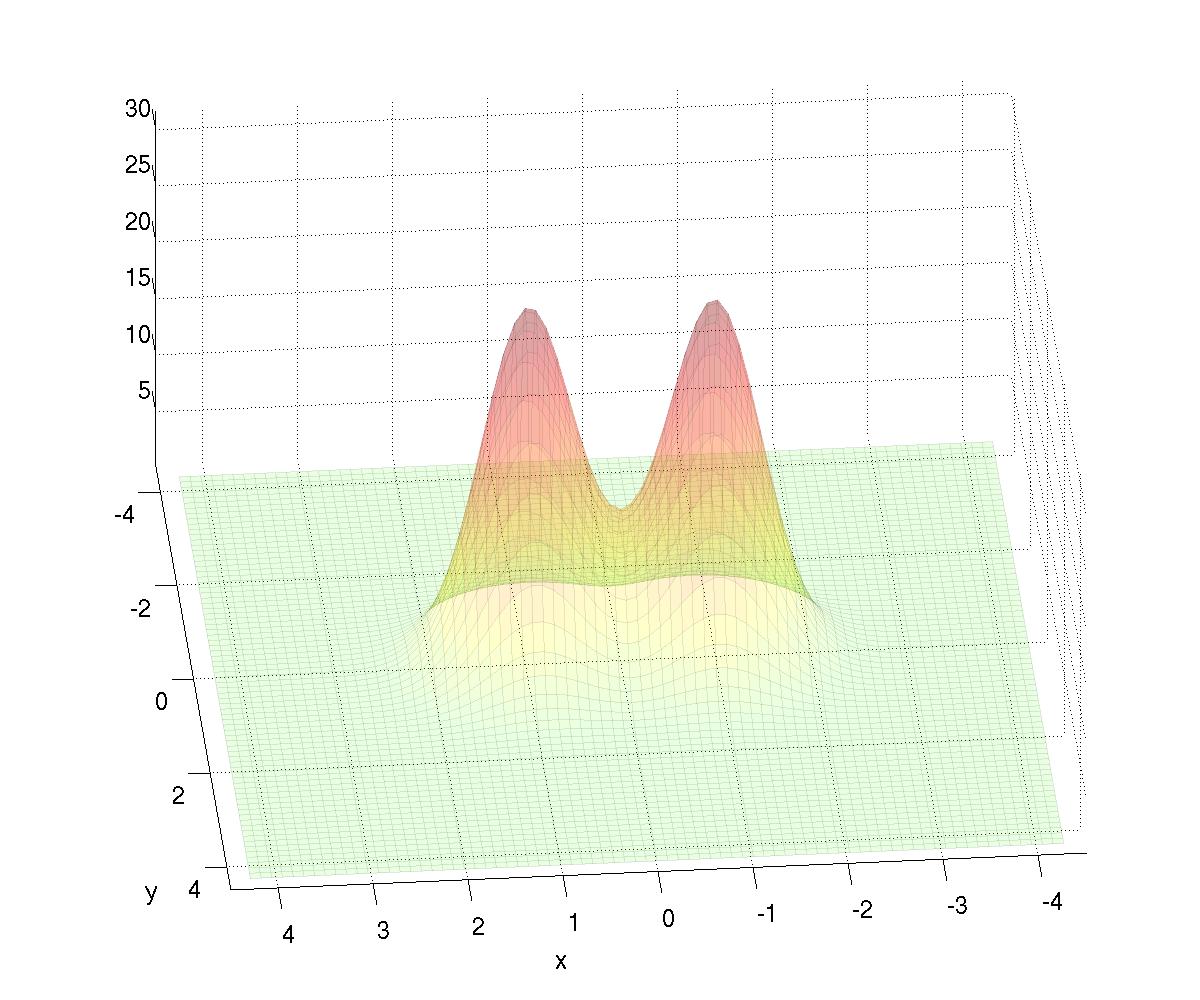}}}
\mbox{
\subfloat[$c_2=1,c_4=\tfrac{1}{4}$]{\includegraphics[width=0.33\linewidth]{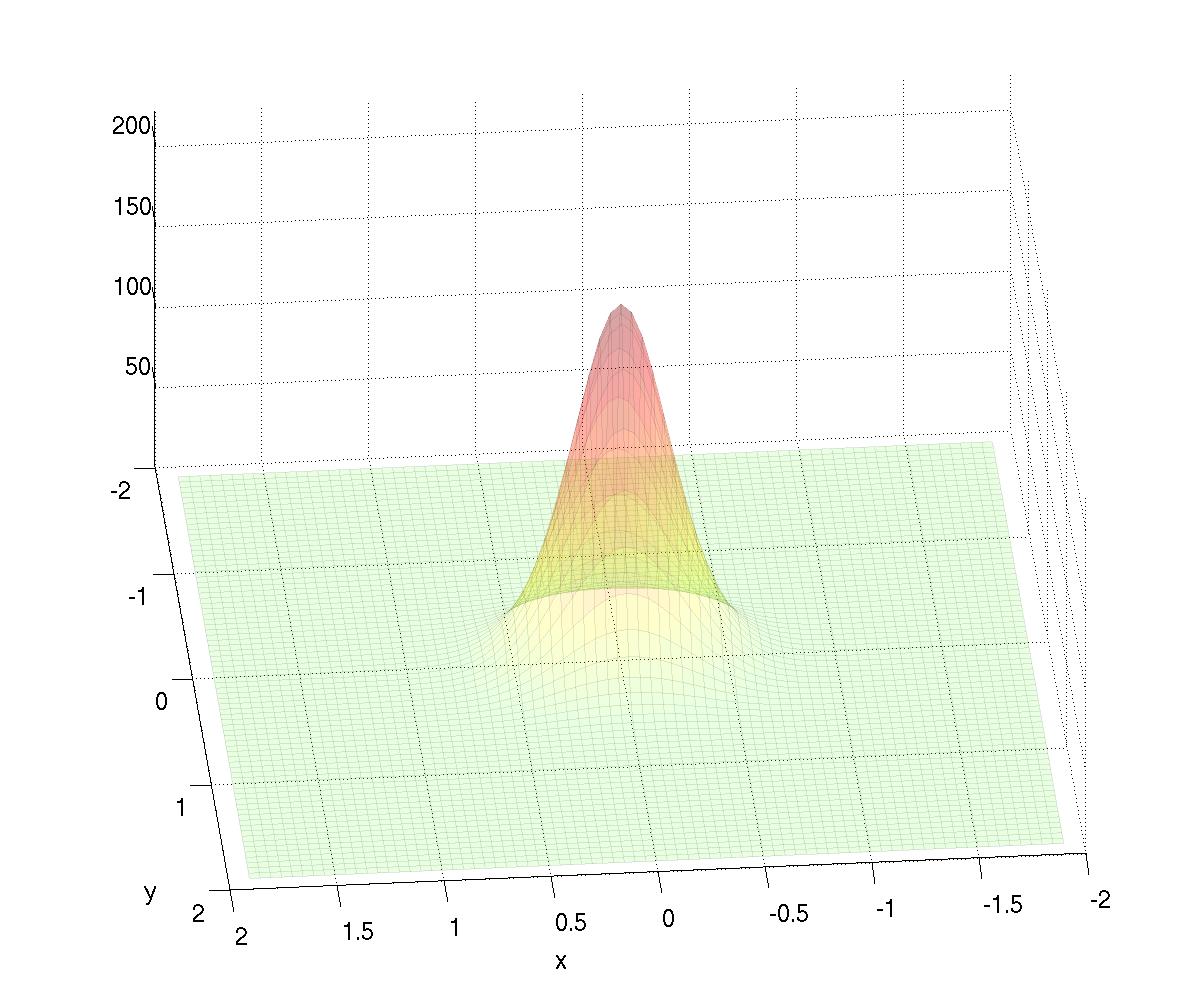}}
\subfloat[$c_2=1,c_4=1$]{\includegraphics[width=0.33\linewidth]{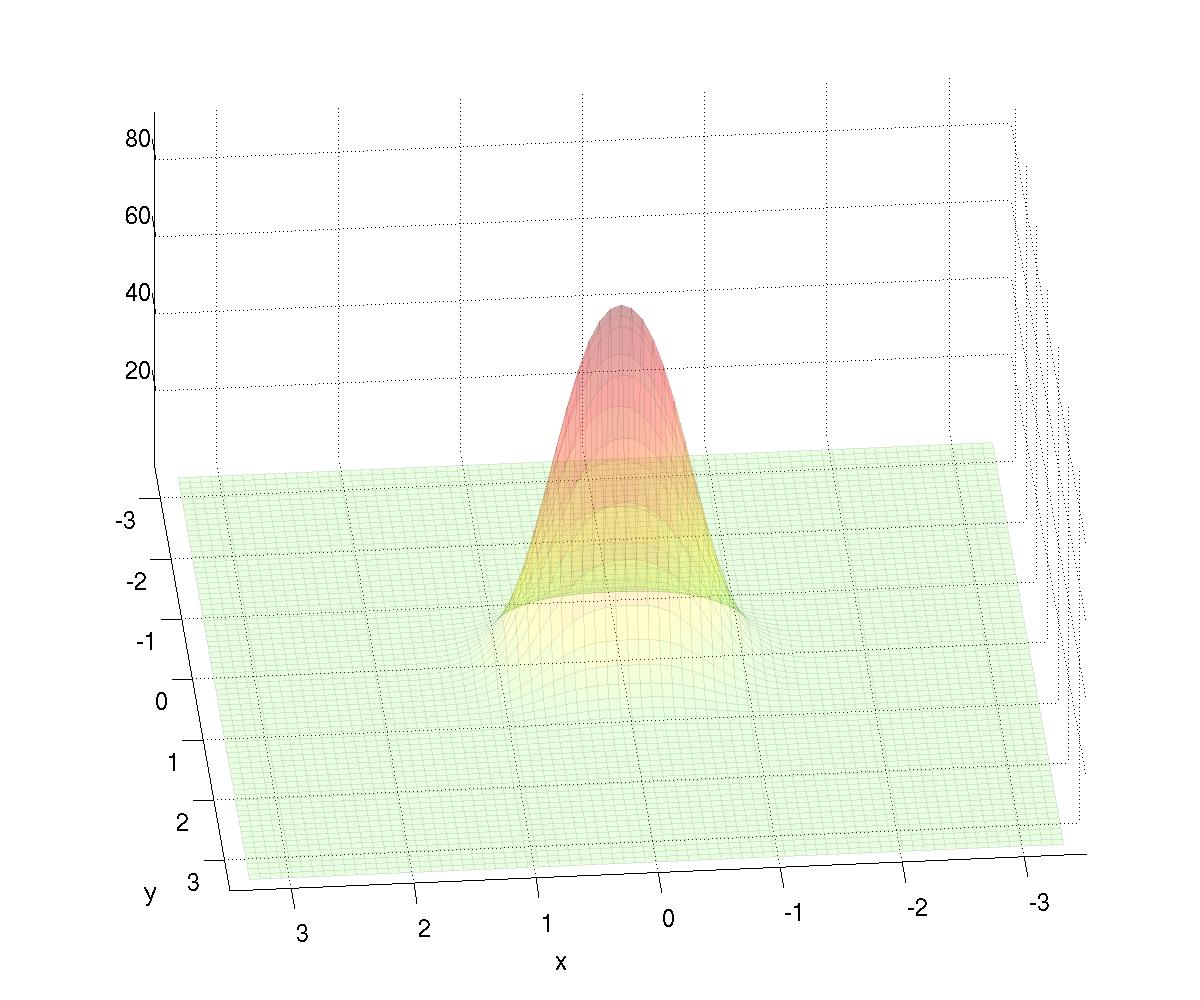}}
\subfloat[$c_2=1,c_4=4$]{\includegraphics[width=0.33\linewidth]{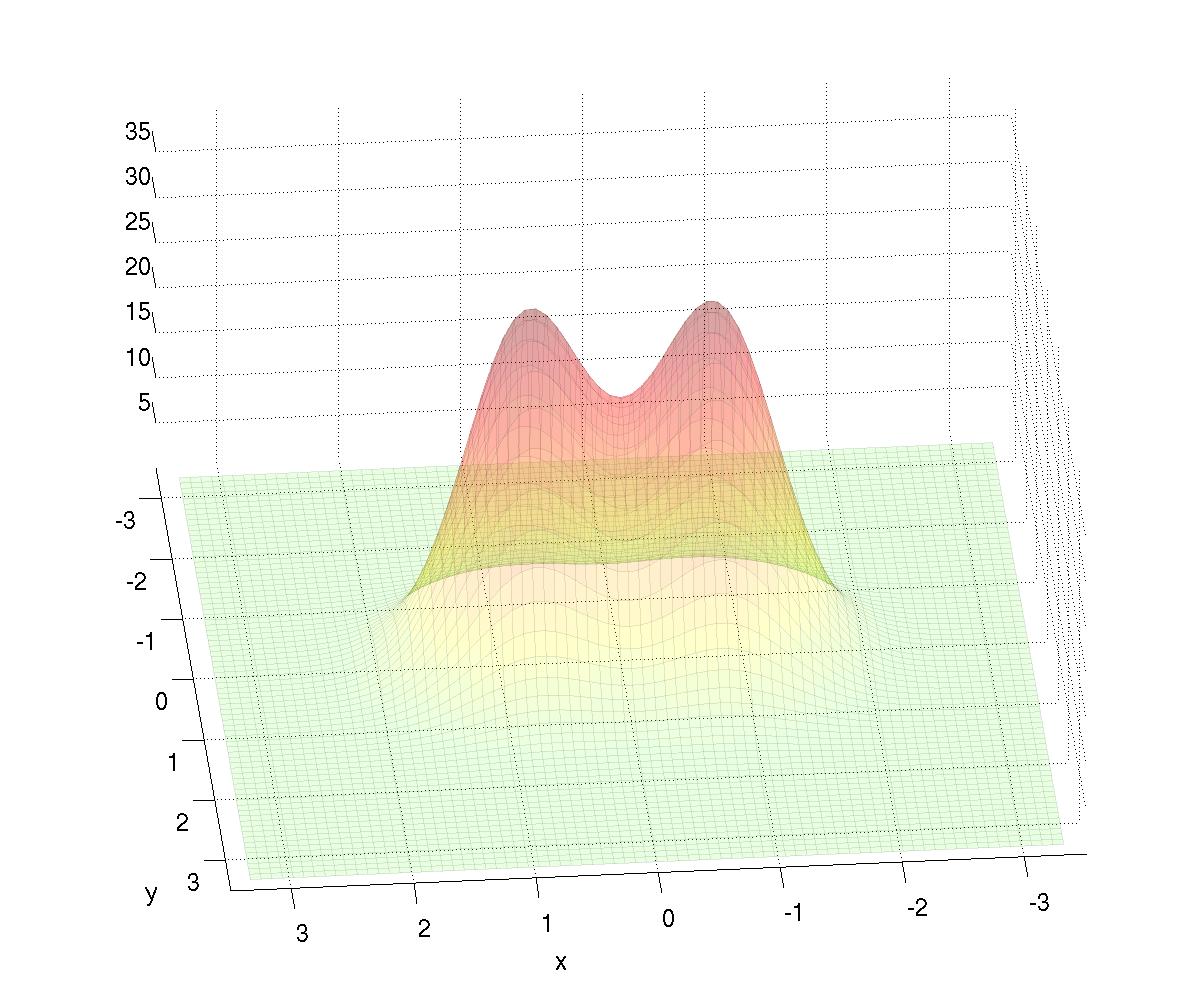}}}
\caption{Energy density at a spatial slice through the molecule
  at $z=0$ in the 2+4 model for various choices of $(c_2,c_4)$ and for
  fixed mass $m=4$. 
}
\label{fig:M4B1_energyslice}
\end{center}
\end{figure}

In Figs.~\ref{fig:M4B1_baryonslice} and \ref{fig:M4B1_energyslice} are
shown cross sections at $z=0$ of the baryon charge density and energy
density, respectively. 
The numerically integrated baryon charge density, denoted by 
$B^{\rm numerical}$ gives a handle on the precision of the numerical
solution, see Tab.~\ref{tab:M4B1}.
Note that the molecular shape is slightly more pronounced in the
energy density than in the baryon charge density, viz.~the depth of
the valley between the two peaks is deeper. In the cases
$c_2=1,c_4=\tfrac{1}{4}$ and $c_2=c_4=1$, there is no valley between
the two peaks; there is however still some amount of ``polarization;''
or better, there is a dipole moment. In order to quantify the amount
to which the Skyrmion is moleculized, we define the following quantity 
\beq
\mathfrak{p}^B \equiv 2\int d^3 x\; \left(
\mathop{\rm sign}(n_4) x \mathcal{B} - {\rm sign}(n_3) y \mathcal{B}
\right),
\eeq
which we call the baryonic dipole moment and 
\beq
\mathcal{B} \equiv -\frac{1}{2\pi^2}
\epsilon^{ABCD} \p_1 n_A \p_2 n_B \p_3 n_C n_D,
\eeq
is the baryon charge density. The physical meaning of $\mathfrak{p}^B$,
which has units of length, is to which degree the baryon charge
corresponding to the northern and southern hemisphere (distinguished
by the sign of $n_4$) is separated compared to how the charge is
distributed in the transverse direction. This quantity is absolute and
should be compared to the size of the Skyrmion (see the figures). For
a Skyrmion without the potential \eqref{eq:pot}, the baryonic dipole
moment vanishes: $\mathfrak{p}^B=0$. For convenience we define the
size of the Skyrmion as 
\beq
\mathfrak{s}^B \equiv \sqrt{\int d^3 x\; r^2 \mathcal{B}}.
\eeq

We mentioned already that the two coefficients $c_2,c_4$ are not
independent of each other and they can be scaled to unity giving a
mass which we will denote $m^{\rm canonical}$ which can be calculated
as
\beq
m^{\rm canonical} = \frac{\sqrt{c_4}}{c_2} m,
\eeq 
where $m$ is the mass in the noncanonically normalized Lagrangian,
i.e.~with $c_2\neq 1$ and $c_4\neq 1$. The values of 
$m^{\rm canonical}$ for the various configurations shown in
Figs.~\ref{fig:M4B1}, \ref{fig:M4B1_baryonslice} and
\ref{fig:M4B1_energyslice} are given in Tab.~\ref{tab:M4B1}.

\begin{table}[!htb]
\caption{The numerically integrated baryon charge, the numerically
  integrated energy and the numerically integrated baryonic dipole
  moment for the various configurations in the 2+4 model, shown in
  Figs.~\ref{fig:M4B1}, \ref{fig:M4B1_baryonslice} and
  \ref{fig:M4B1_energyslice}. } 
\label{tab:M4B1}
\begin{tabular}{ccccccc}
$c_2$ & $c_4$ & $m^{\rm canonical}$ & $B^{\rm numerical}$ 
& $E^{\rm numerical}$ & $\mathfrak{p}^B$ & $\mathfrak{s}^B$\\
\hline
\hline
$0$ & $\tfrac{1}{4}$ & $\infty$ & $0.99969$ & $18.930$ & $0.575$ & $0.772$\\
$0$ & $1$ & $\infty$ & $0.99986$ & $52.720$ & $0.970$ & $1.151$\\
$0$ & $4$ & $\infty$ & $0.99989$ & $154.74$ & $0.998$ & $1.472$\\
$\tfrac{1}{4}$ & $\tfrac{1}{4}$ & $2^3$ & $0.99989$ & $25.832$ &
$0.288$ & $0.620$\\
$\tfrac{1}{4}$ & $1$ & $2^4$ & $0.99967$ & $65.527$ & $0.478$ & $0.901$\\
$\tfrac{1}{4}$ & $4$ & $2^5$ & $0.99989$ & $167.24$ & $0.920$ & $1.437$\\
$1$ & $\tfrac{1}{4}$ & $2^1$ & $0.99962$ & $40.078$ & $0.087$ & $0.426$\\
$1$ & $1$ & $2^2$ & $0.99958$ & $89.365$ & $0.185$ & $0.713$\\
$1$ & $4$ & $2^3$ & $0.99972$ & $203.93$ & $0.635$ & $1.282$
\end{tabular}
\end{table}

\begin{figure}[!ptb]
\begin{center}
\captionsetup[subfloat]{labelformat=empty}
\mbox{
\subfloat[$c_2=\tfrac{1}{4},c_6=\tfrac{1}{4}$]{\includegraphics[width=0.33\linewidth]{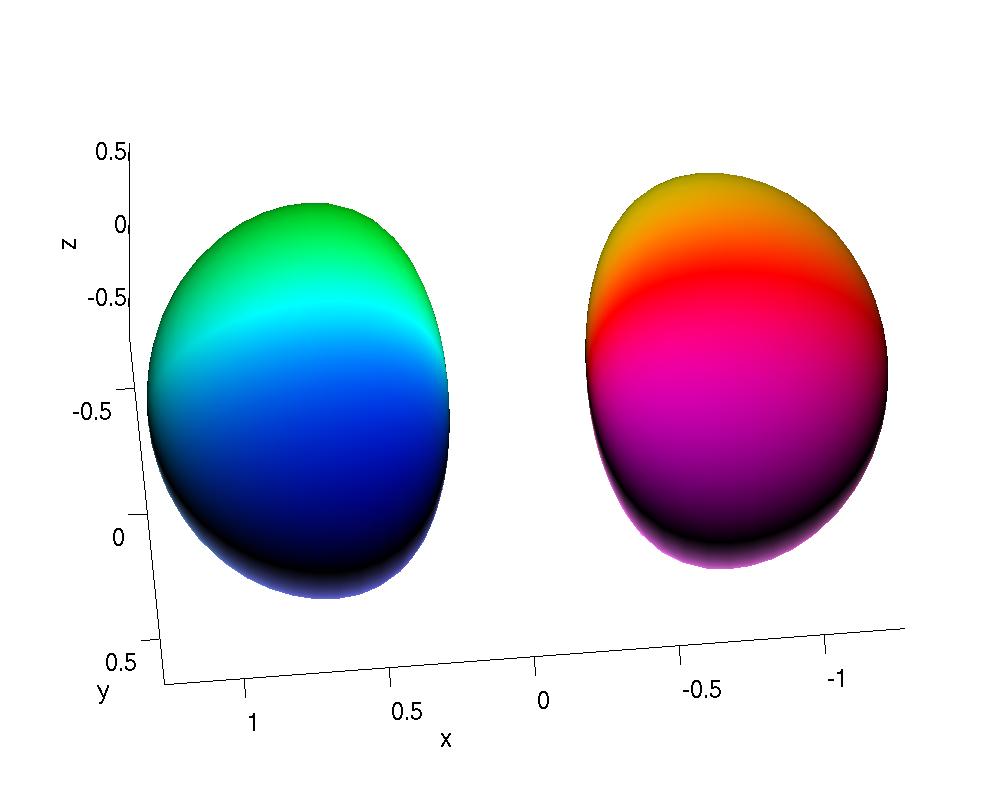}}
\subfloat[$c_2=\tfrac{1}{4},c_6=1$]{\includegraphics[width=0.33\linewidth]{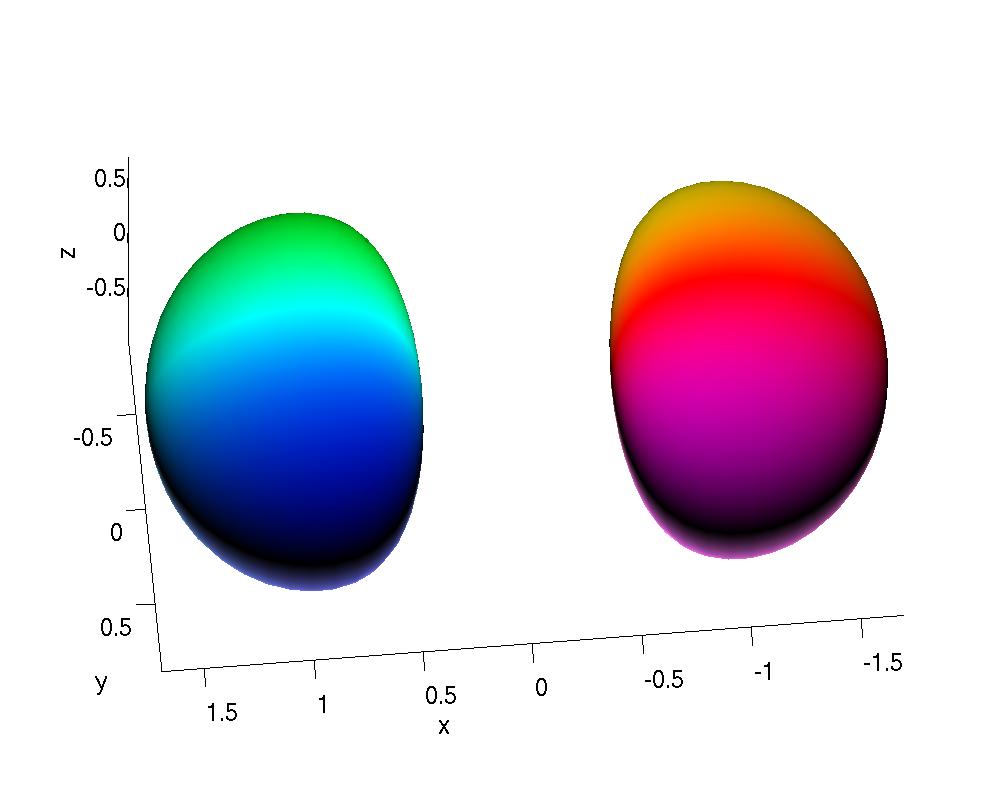}}
\subfloat[$c_2=\tfrac{1}{4},c_6=4$]{\includegraphics[width=0.33\linewidth]{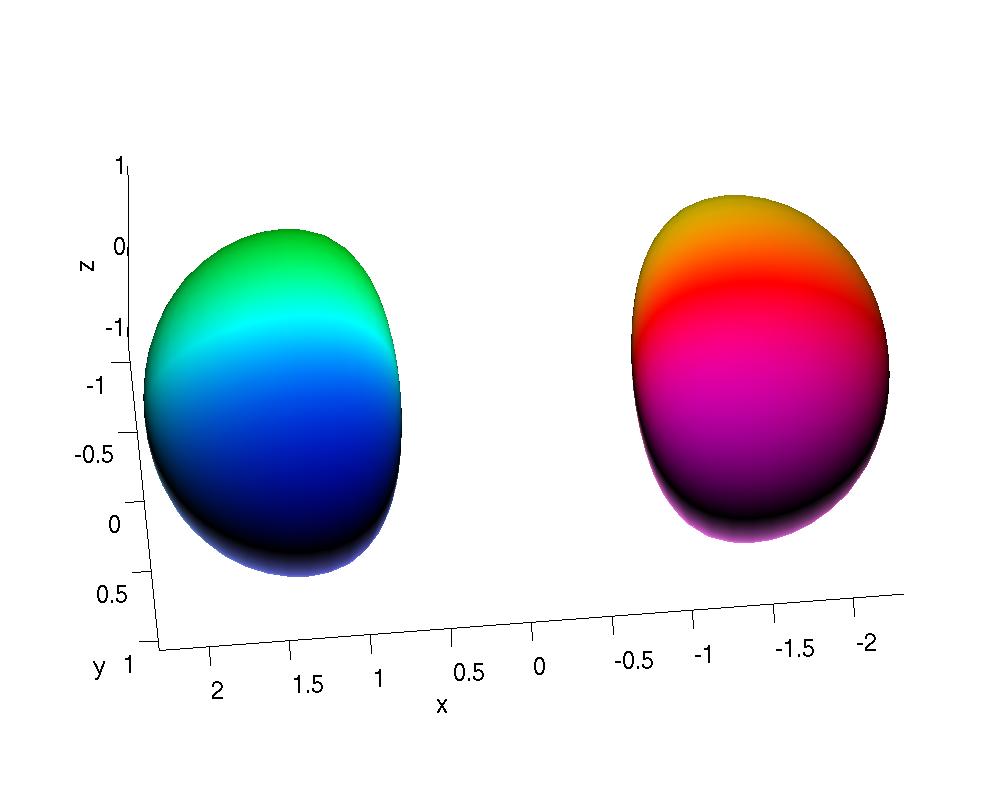}}}
\mbox{
\subfloat[$c_2=1,c_6=\tfrac{1}{4}$]{\includegraphics[width=0.33\linewidth]{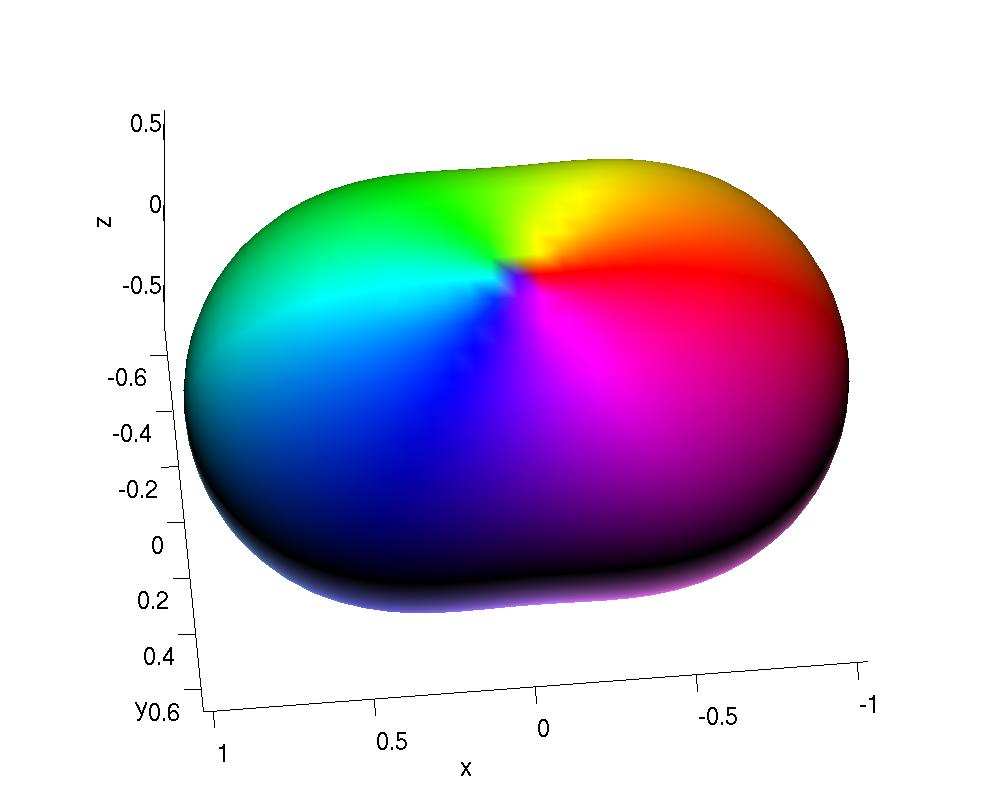}}
\subfloat[$c_2=1,c_6=1$]{\includegraphics[width=0.33\linewidth]{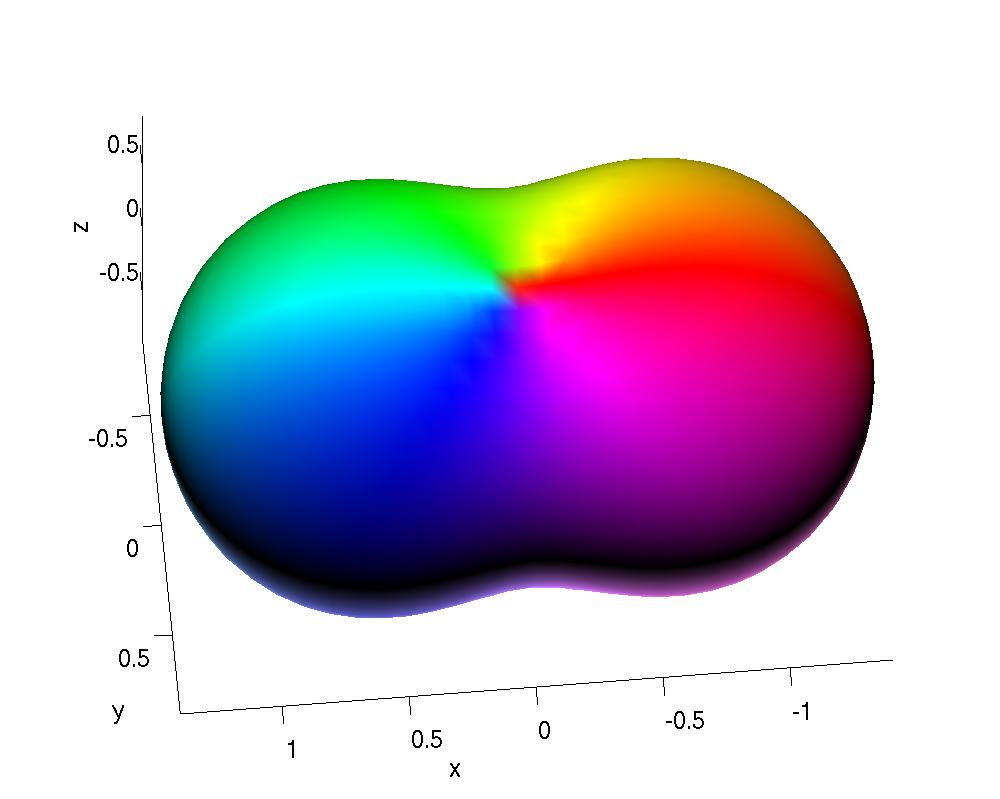}}
\subfloat[$c_2=1,c_6=4$]{\includegraphics[width=0.33\linewidth]{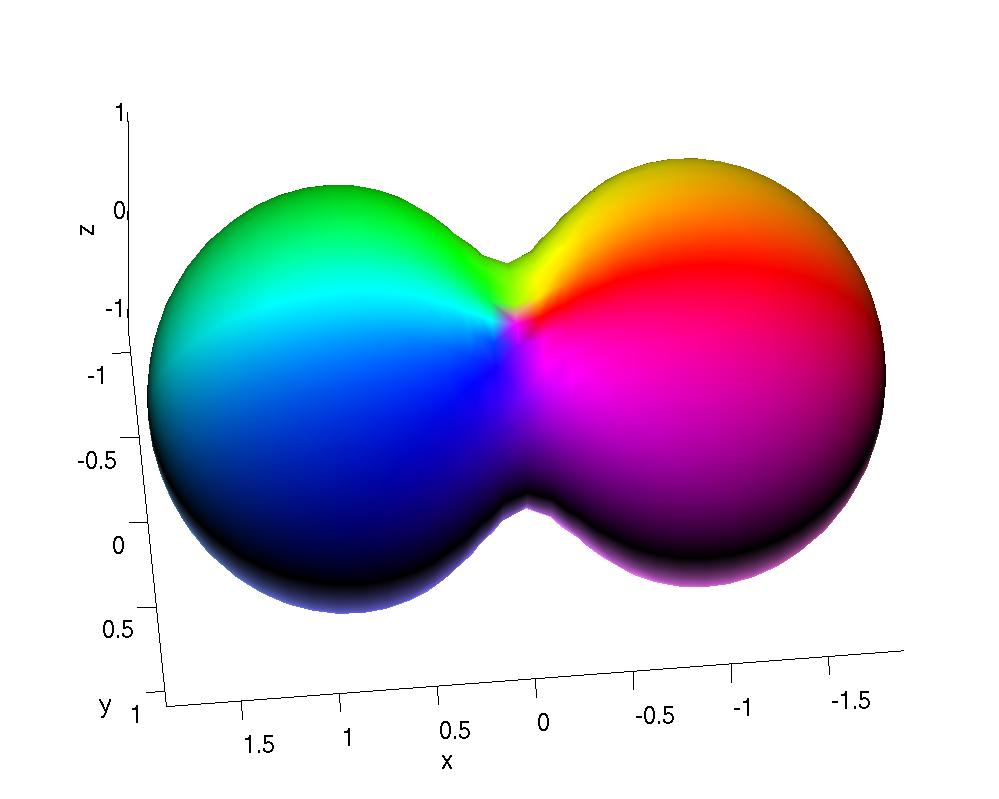}}}
\caption{Isosurfaces showing the half-maximum of the baryon charge
  density in the 2+6 model for various choices of $(c_2,c_6)$ for
  fixed mass $m=4$. 
  The color scheme is the same as that in Fig.~\ref{fig:M4B1}.
}
\label{fig:M6B1}
\end{center}
\end{figure}

We will now consider the 2+6 model, i.e.~$c_2\geq 0,c_6>0$ and
$c_4=0$. 
In Fig.~\ref{fig:M6B1} is shown an array of molecule baryon charge
density isosurfaces for similar values of the coefficients $(c_2,c_6)$
as in the previous case and for fixed mass $m=4$. The coloring scheme
is the same as used above. Again the molecular shape is mostly
pronounced when the coefficients are small and $c_2\ll c_6$. We also
calculate the canonical mass, 
\beq
m^{\rm canonical} = \sqrt[4]{\frac{c_6}{c_2^3}} m,
\eeq
i.e.~the mass corresponding to a rescaled equation of motion with
$c_2=c_6=1$ and show those corresponding values in
Tab.~\ref{tab:M6B1}.

\begin{table}[!htb]
\caption{The numerically integrated baryon charge, the numerically
  integrated energy and the numerically integrated baryonic dipole
  moment for the various configurations in the 2+6 model, shown in
  Figs.~\ref{fig:M6B1}, \ref{fig:M6B1_baryonslice} and
  \ref{fig:M6B1_energyslice}. } 
\label{tab:M6B1}
\begin{tabular}{ccccccc}
$c_2$ & $c_6$ & $m^{\rm canonical}$ & $B^{\rm numerical}$ 
& $E^{\rm numerical}$ & $\mathfrak{p}^B$ & $\mathfrak{s}^B$\\
\hline
\hline
$\tfrac{1}{4}$ & $\tfrac{1}{4}$ & $2^3$ & $0.99989$ & $34.440$ &
$0.595$ & $0.978$\\
$\tfrac{1}{4}$ & $1$ & $2^{\frac{7}{2}}$ & $0.99986$ & $61.728$ &
$0.863$ & $1.278$\\
$\tfrac{1}{4}$ & $4$ & $2^4$ & $0.99986$ & $113.89$ & $1.353$ & $1.707$\\
$1$ & $\tfrac{1}{4}$ & $2^{\frac{3}{2}}$ & $0.99986$ & $60.705$ &
$0.281$ & $0.822$\\
$1$ & $1$ & $2^2$ & $0.99984$ & $97.104$ & $0.440$ & $1.091$\\
$1$ & $4$ & $2^{\frac{5}{2}}$ & $0.99985$ & $161.15$ & $0.708$ & $1.442$
\end{tabular}
\end{table}

\begin{figure}[!ptb]
\begin{center}
\captionsetup[subfloat]{labelformat=empty}
\mbox{
\subfloat[$c_2=\tfrac{1}{4},c_6=\tfrac{1}{4}$]{\includegraphics[width=0.33\linewidth]{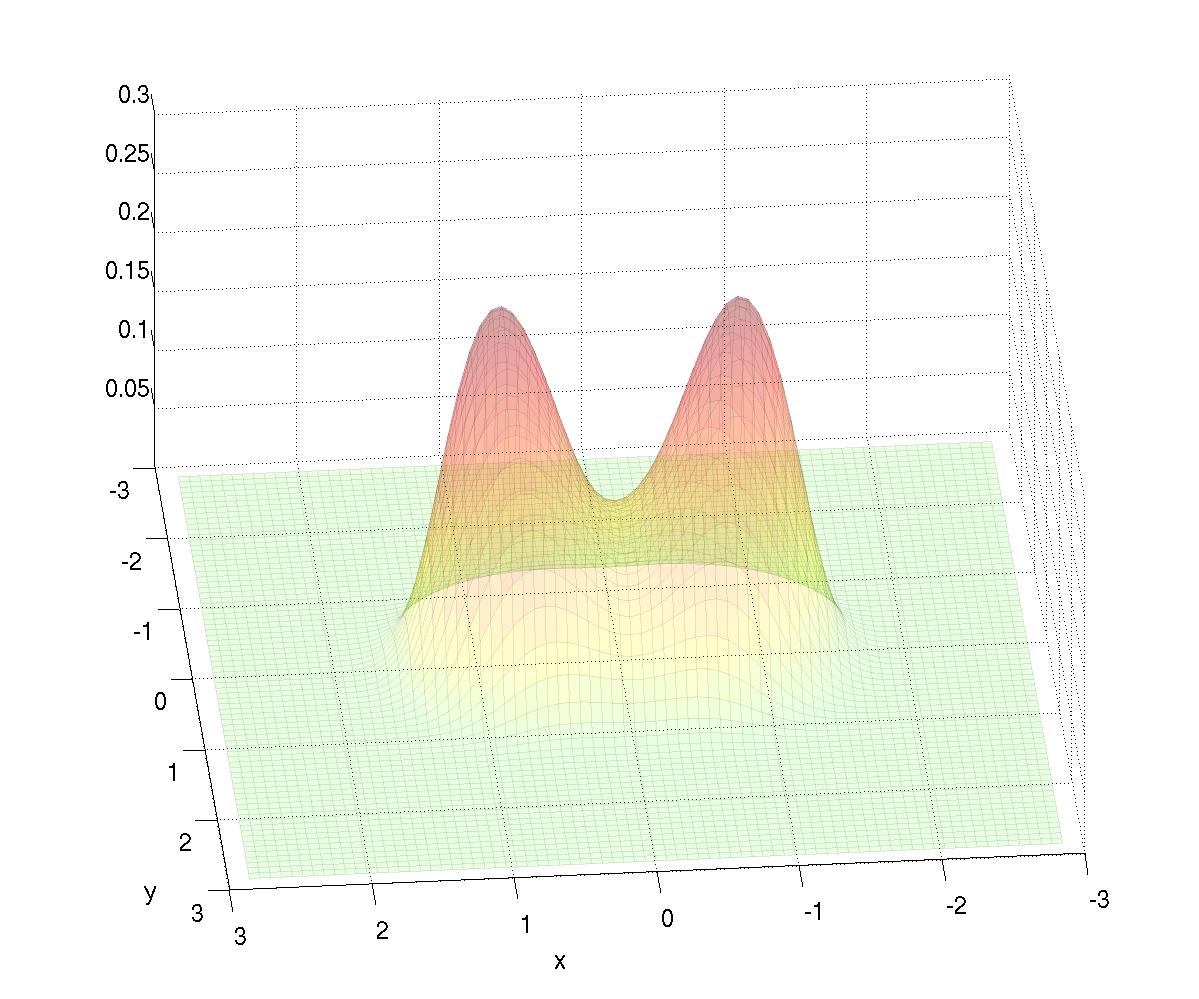}}
\subfloat[$c_2=\tfrac{1}{4},c_6=1$]{\includegraphics[width=0.33\linewidth]{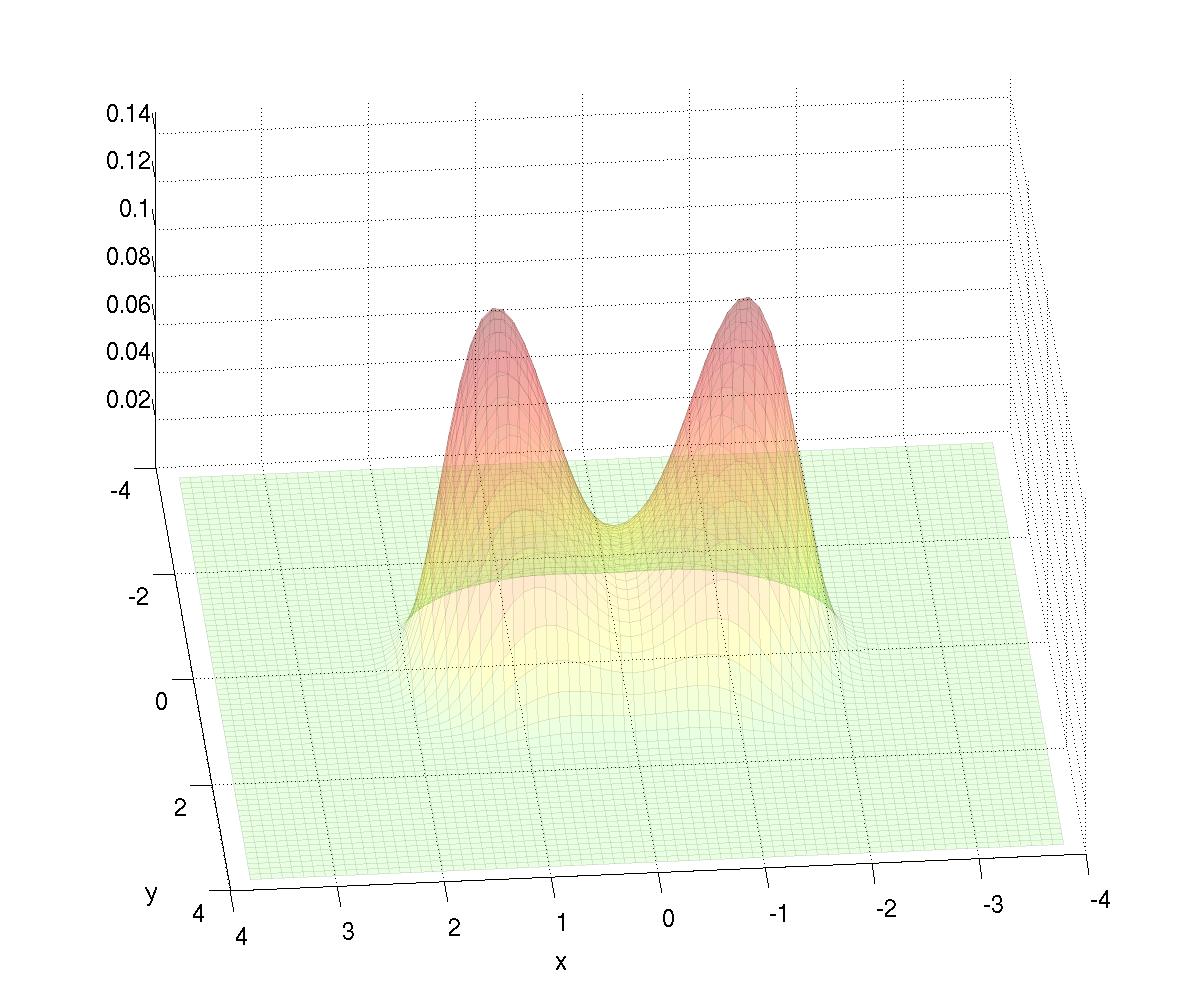}}
\subfloat[$c_2=\tfrac{1}{4},c_6=4$]{\includegraphics[width=0.33\linewidth]{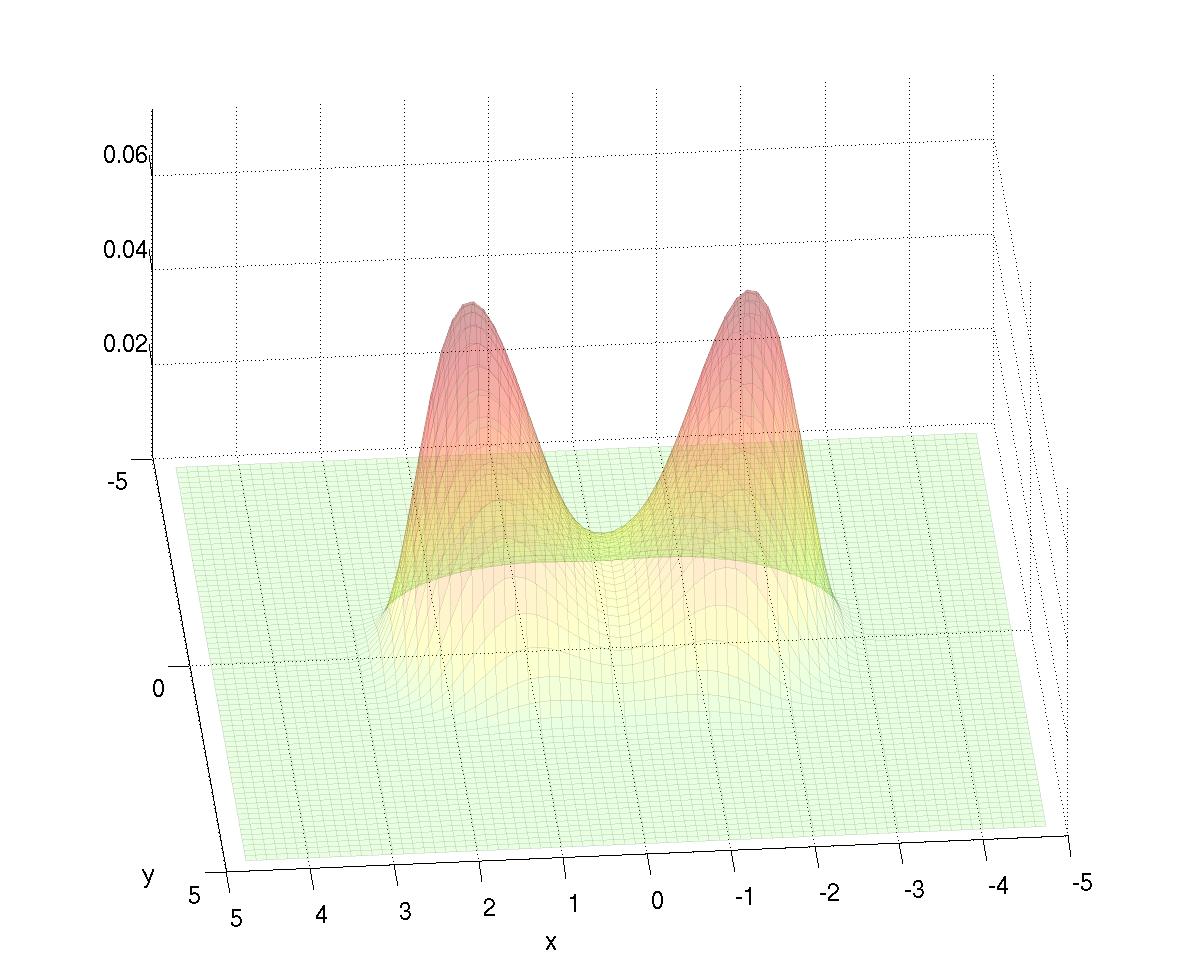}}}
\mbox{
\subfloat[$c_2=1,c_6=\tfrac{1}{4}$]{\includegraphics[width=0.33\linewidth]{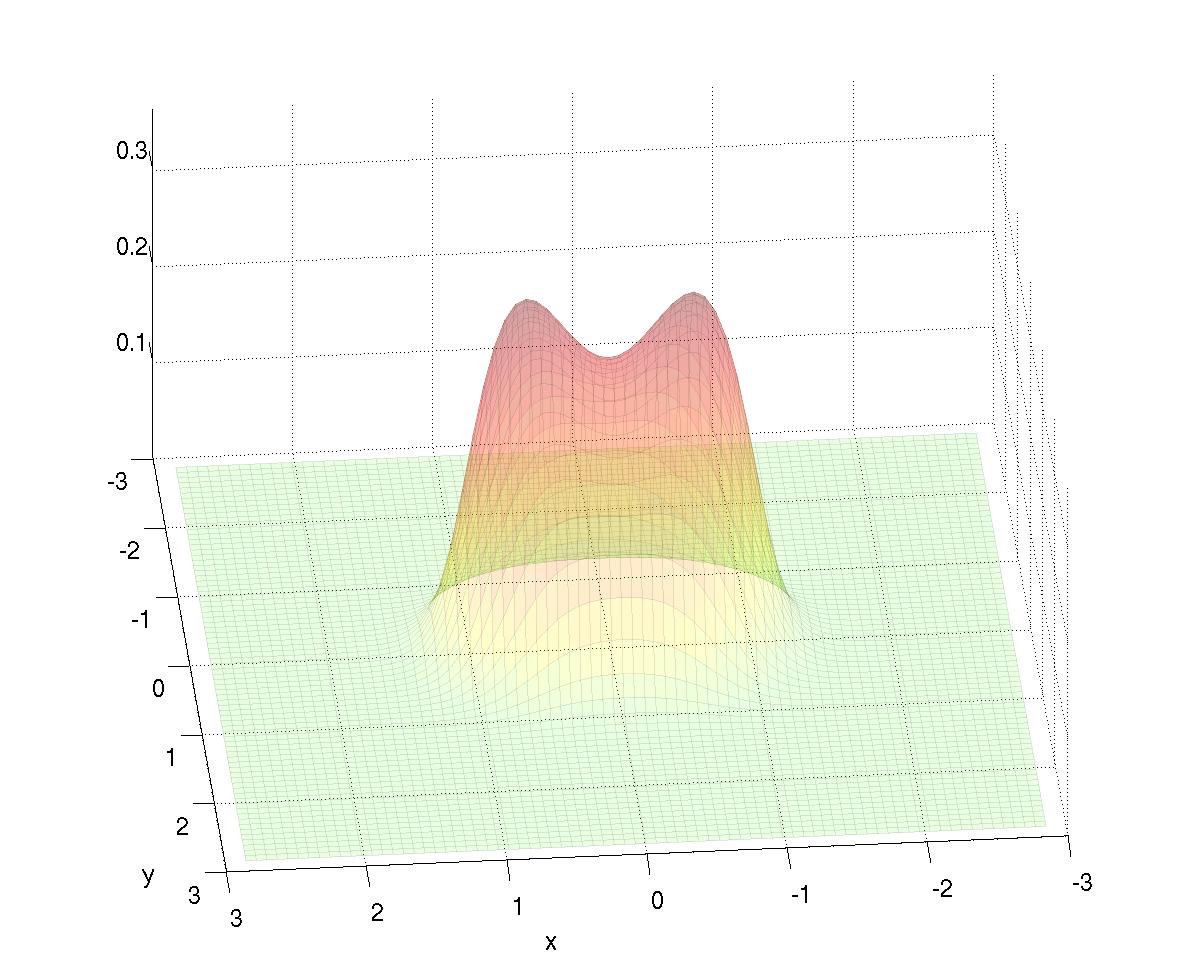}}
\subfloat[$c_2=1,c_6=1$]{\includegraphics[width=0.33\linewidth]{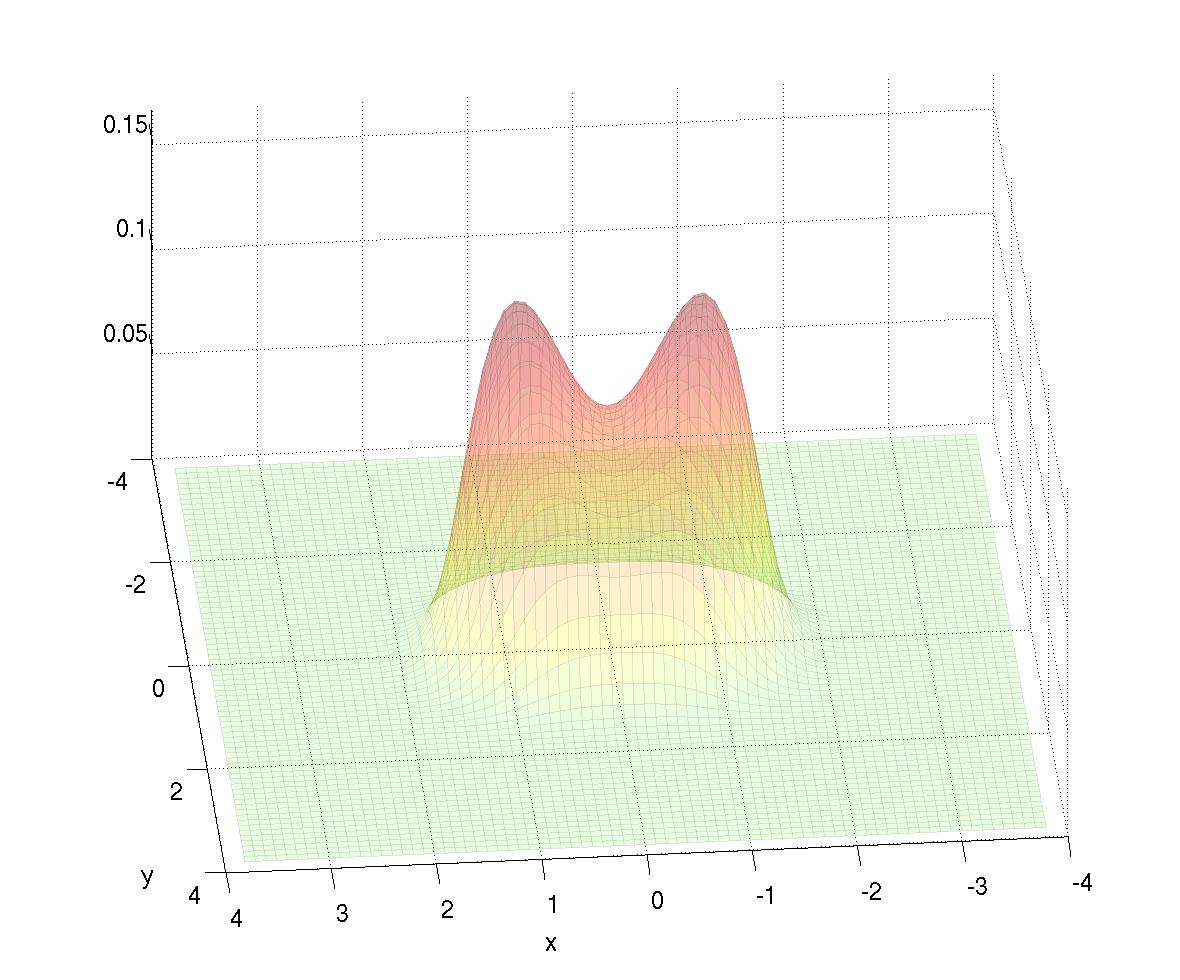}}
\subfloat[$c_2=1,c_6=4$]{\includegraphics[width=0.33\linewidth]{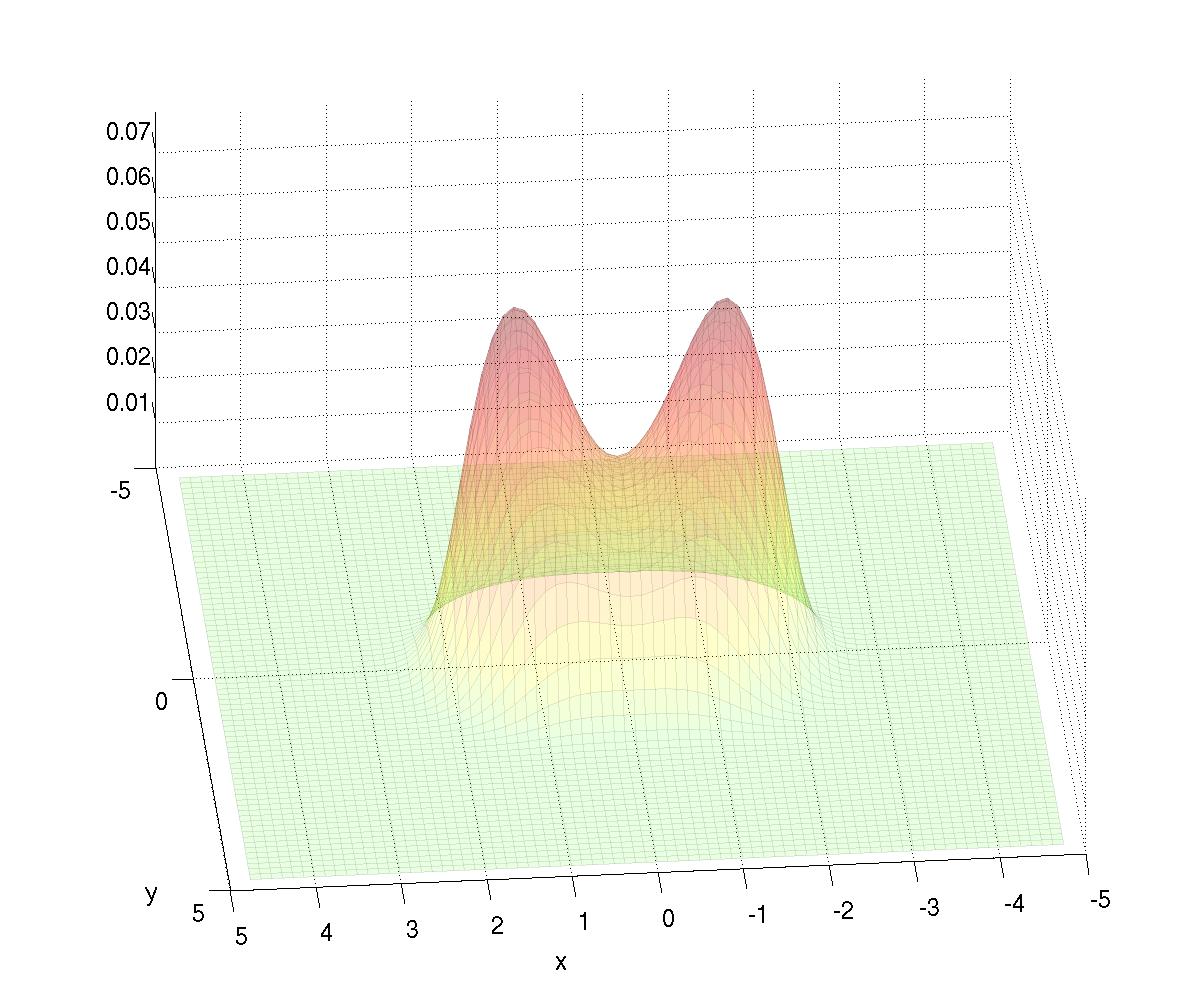}}}
\caption{Baryon charge density at a spatial slice through the molecule
  at $z=0$ in the 2+6 model for various choices of $(c_2,c_6)$ and for
  fixed mass $m=4$. 
}
\label{fig:M6B1_baryonslice}
\end{center}
\end{figure}

\begin{figure}[!ptb]
\begin{center}
\captionsetup[subfloat]{labelformat=empty}
\mbox{
\subfloat[$c_2=\tfrac{1}{4},c_6=\tfrac{1}{4}$]{\includegraphics[width=0.33\linewidth]{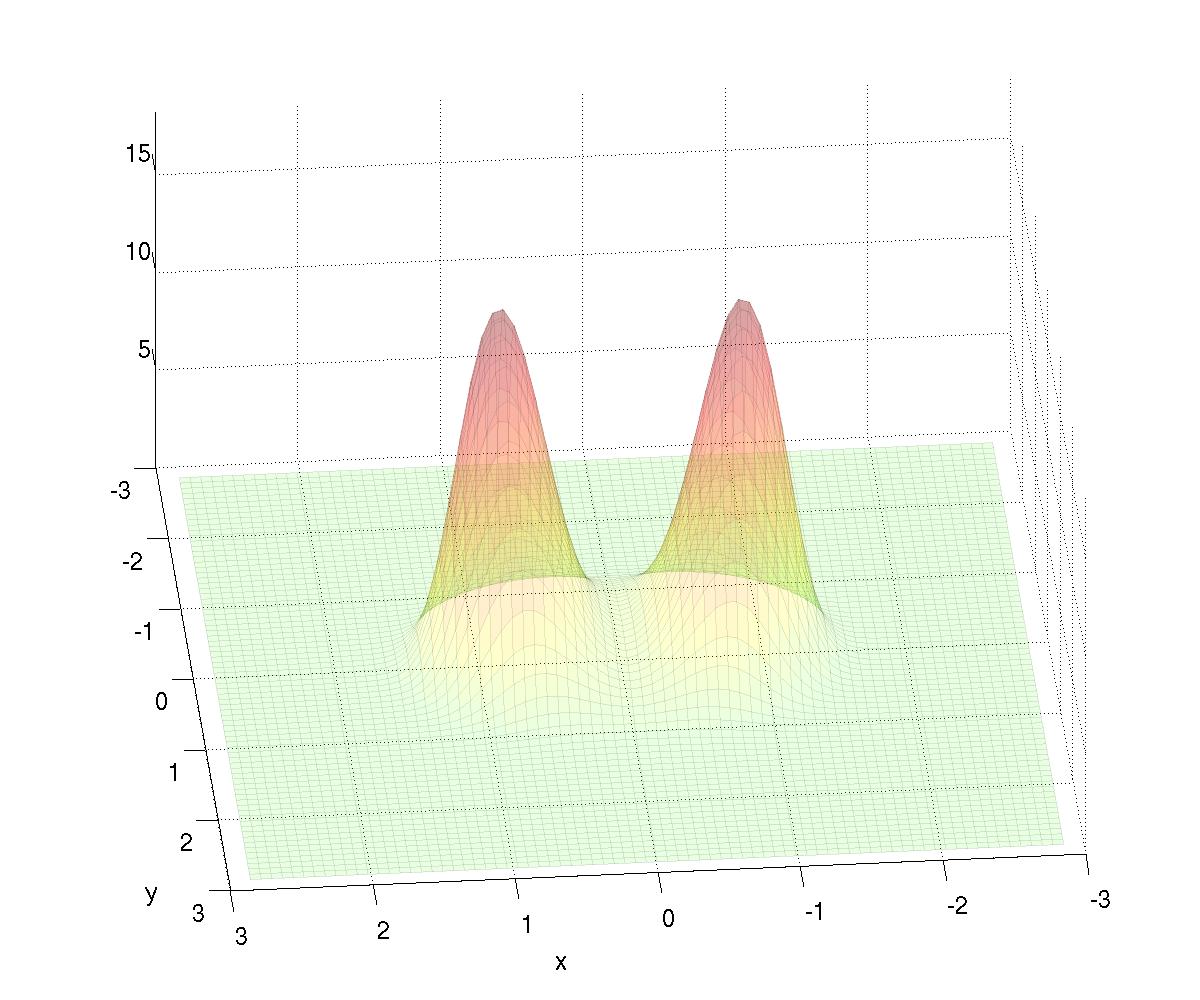}}
\subfloat[$c_2=\tfrac{1}{4},c_6=1$]{\includegraphics[width=0.33\linewidth]{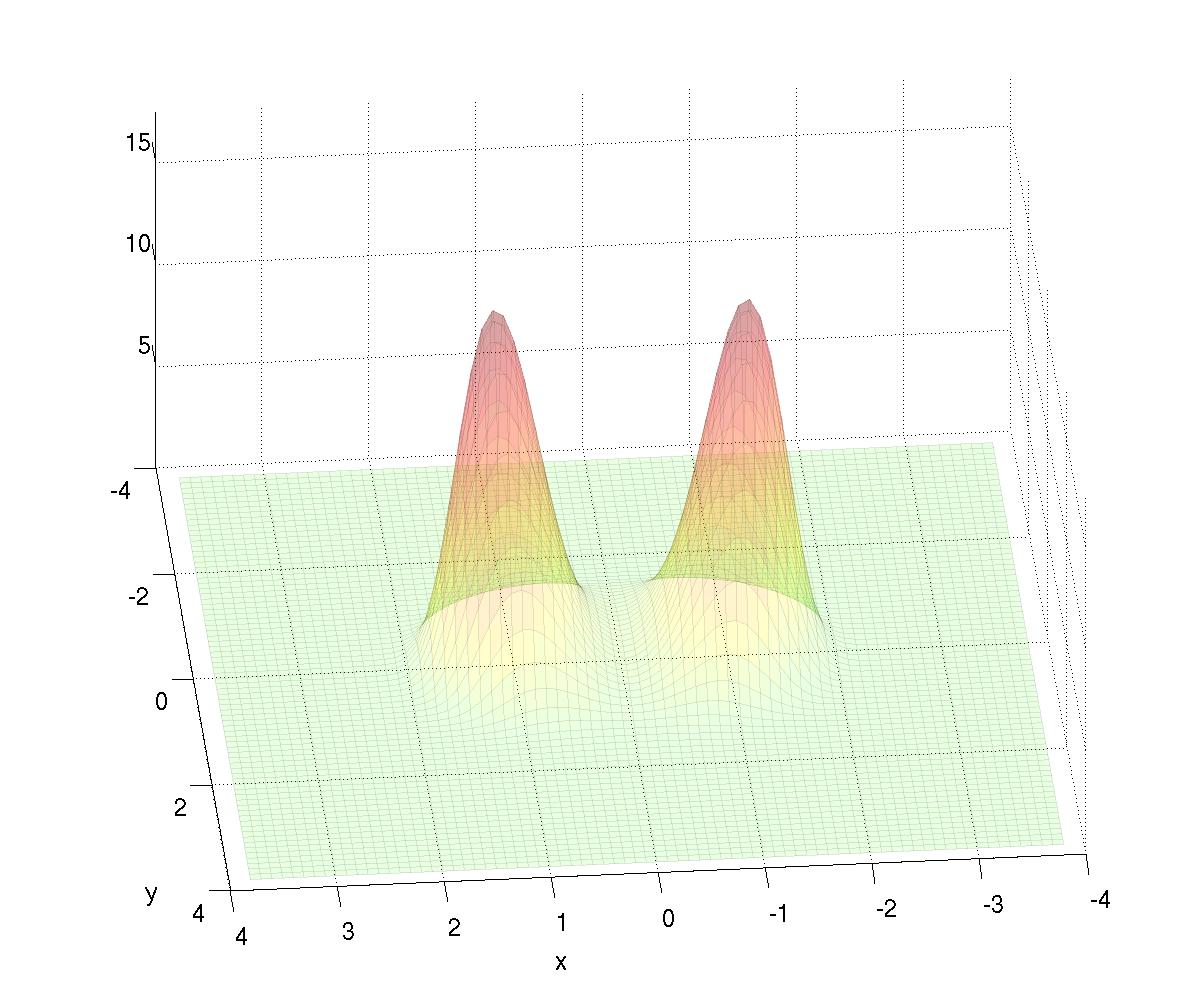}}
\subfloat[$c_2=\tfrac{1}{4},c_6=4$]{\includegraphics[width=0.33\linewidth]{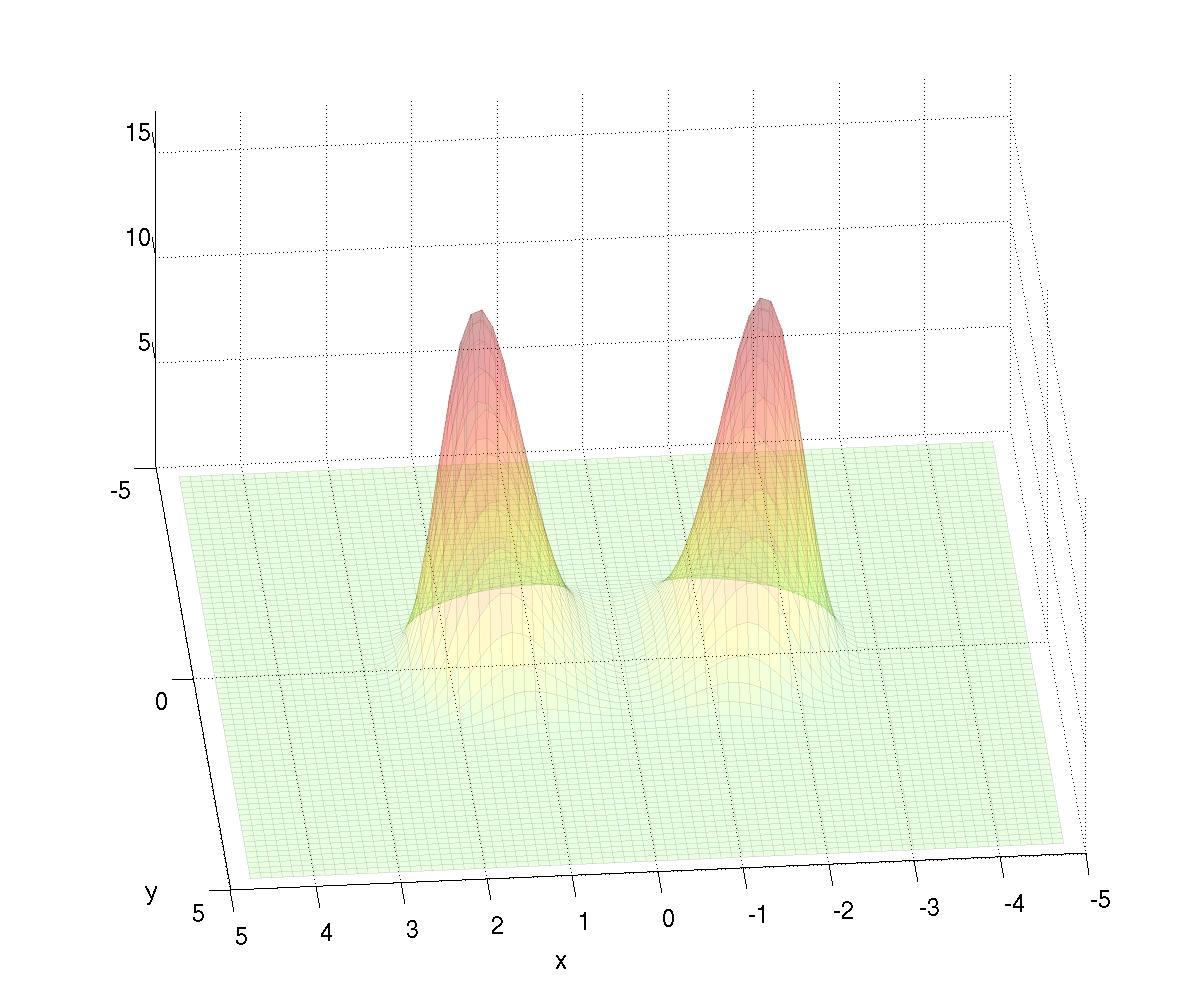}}}
\mbox{
\subfloat[$c_2=1,c_6=\tfrac{1}{4}$]{\includegraphics[width=0.33\linewidth]{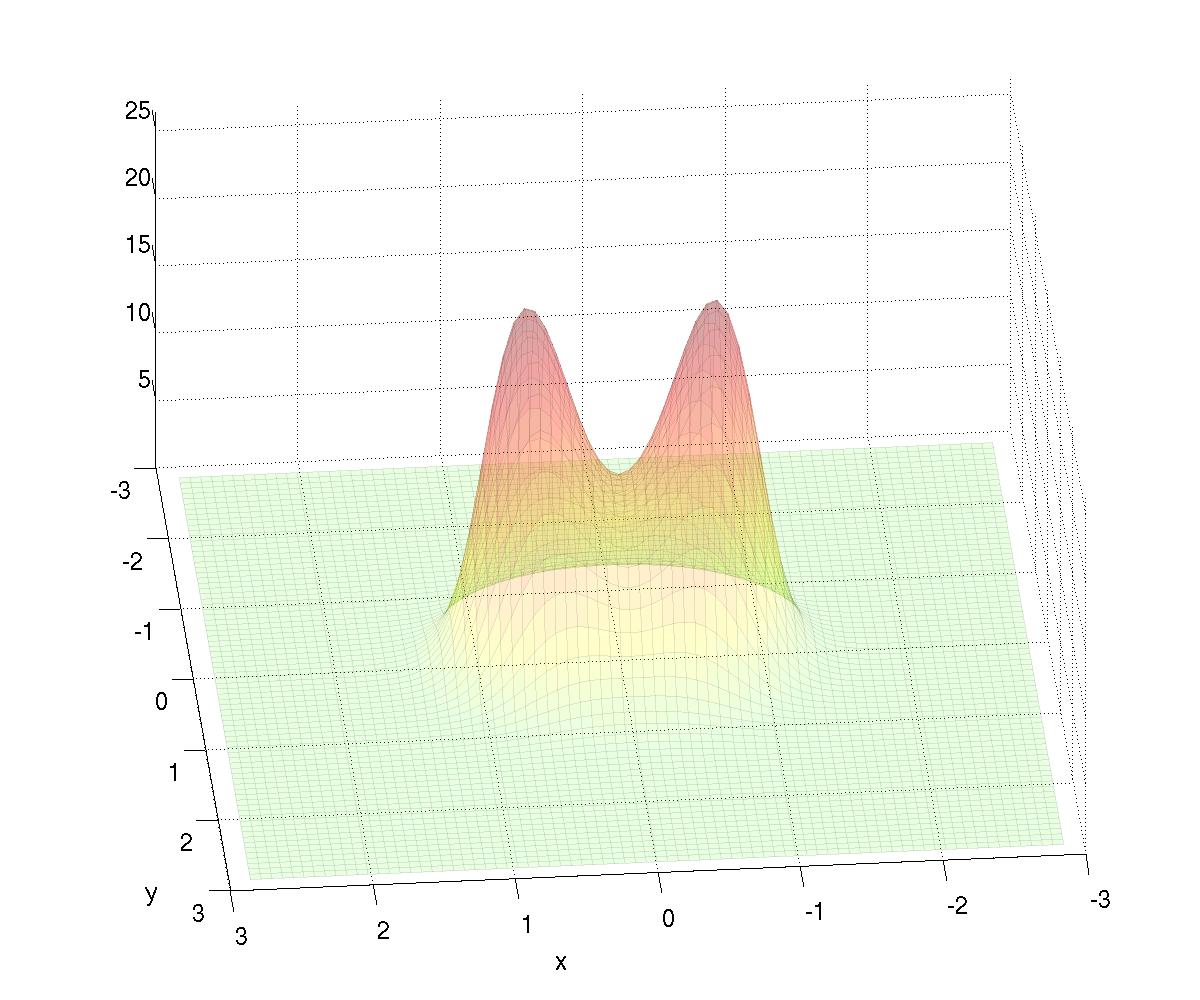}}
\subfloat[$c_2=1,c_6=1$]{\includegraphics[width=0.33\linewidth]{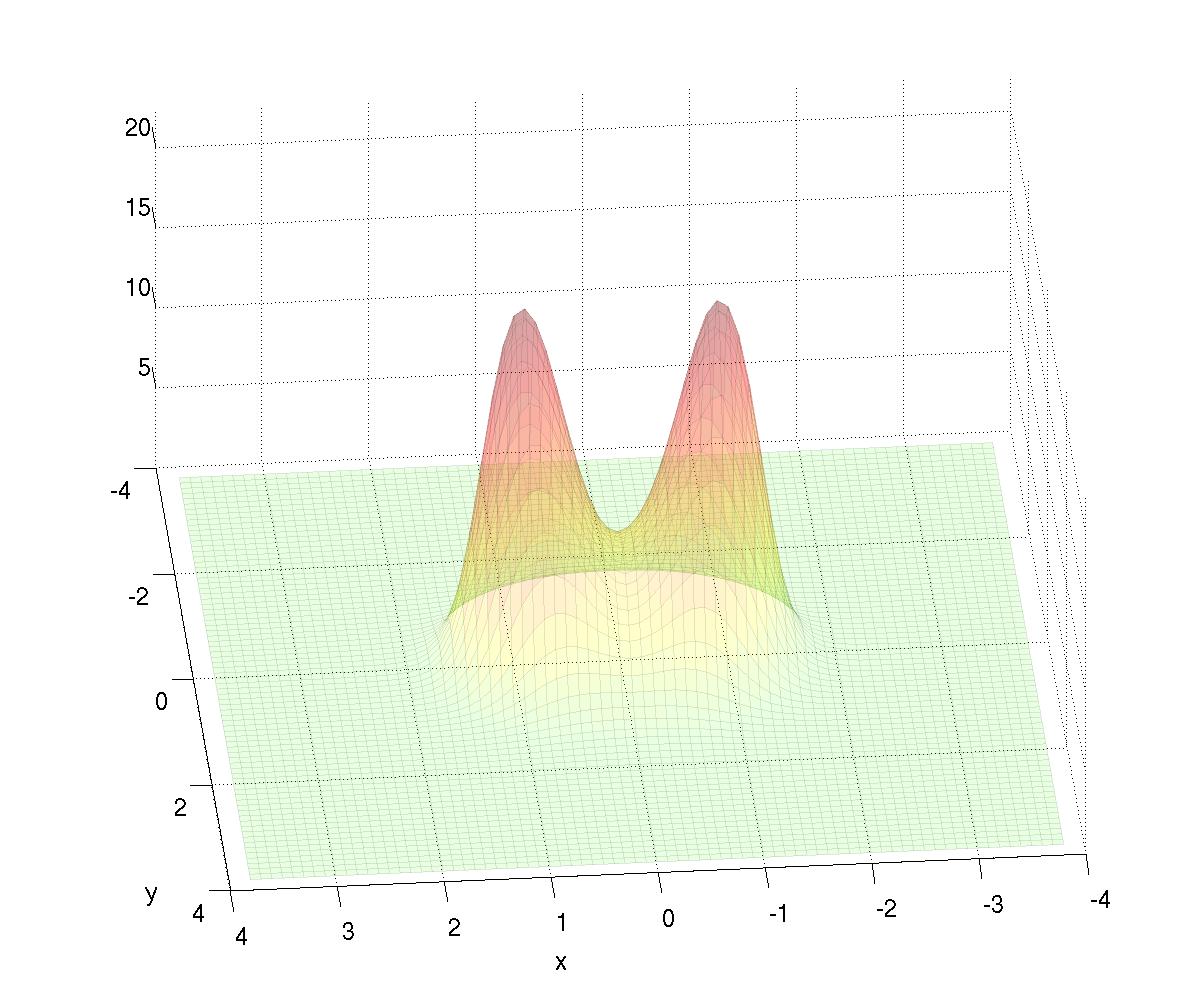}}
\subfloat[$c_2=1,c_6=4$]{\includegraphics[width=0.33\linewidth]{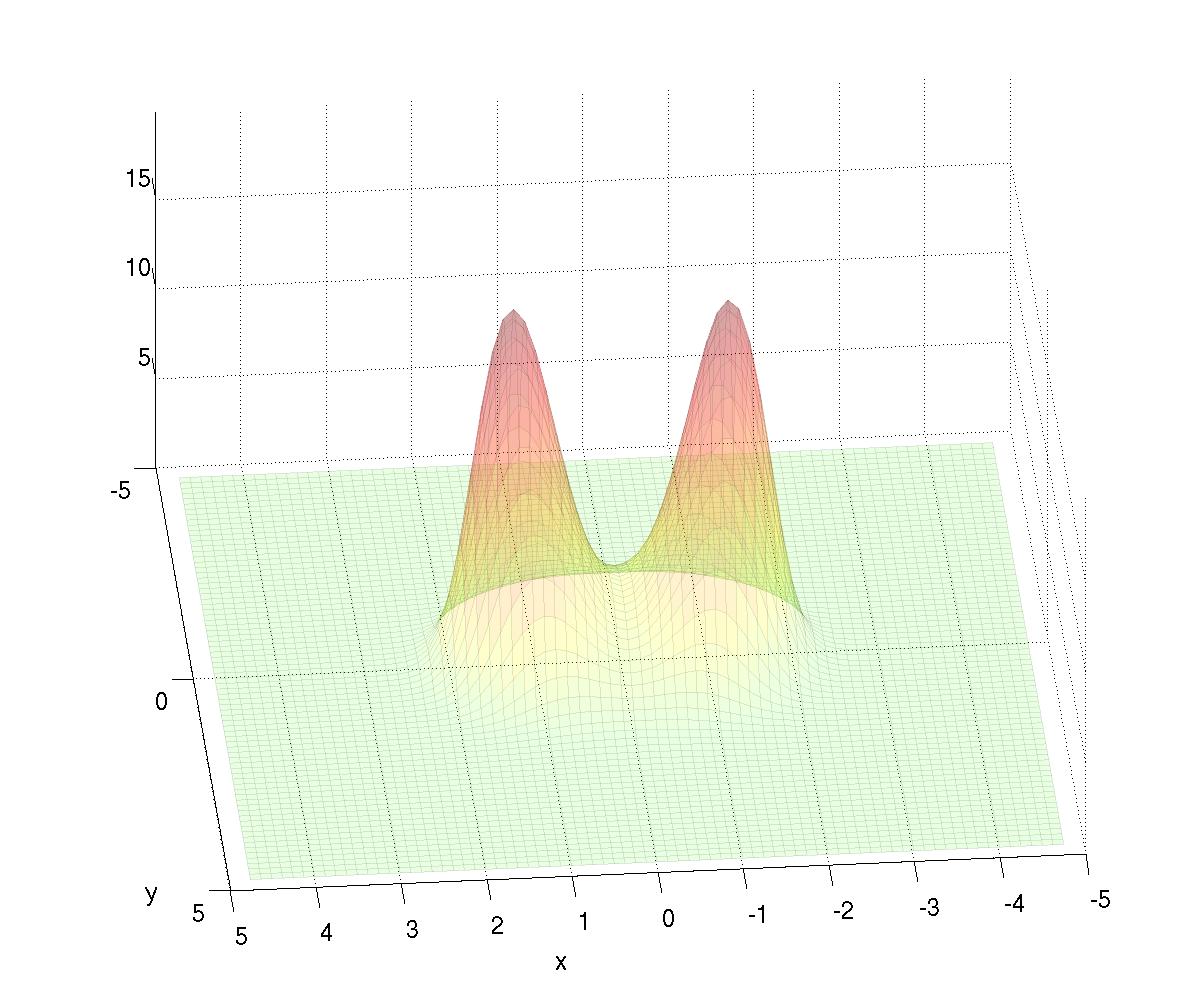}}}
\caption{Energy density at a spatial slice through the molecule
  at $z=0$ in the 2+6 model for various choices of $(c_2,c_6)$ and for
  fixed mass $m=4$. 
}
\label{fig:M6B1_energyslice}
\end{center}
\end{figure}

In Figs.~\ref{fig:M6B1_baryonslice} and \ref{fig:M6B1_energyslice} are
shown cross sections at $z=0$ of the baryon charge density and energy
density, respectively. 
The numerically integrated baryon charge density, denoted 
$B^{\rm numerical}$ gives again a handle on the precision of the
numerical solution, see Tab.~\ref{tab:M6B1}.
We observe that the energy density is more spiky than the baryon
charge density, hence more of a molecular shape.

The case of $c_2=0$ and $c_6>0$ is particularly interesting because of
its BPS and integrable properties. It is by now called the BPS Skyrme
model \cite{Adam:2010fg}. The integrable property is very appealing
since analytic solutions can readily be calculated for a large class
of potentials. Numerically, however, it is a rather difficult problem,
because the Skyrmion in the BPS limit turns into a compacton
\cite{Adam:2010fg}, giving the soliton a finite size (hence no
exponentially damped tail) and so a cusp in the fields at a finite
distance. There is no cusp in the energy density, which is smooth, but
the cusp in the fields requires a special technique in order to be
studied numerically. We will not pursue this problem further in this
paper.

\begin{figure}[!ptb]
\begin{center}
\includegraphics[width=0.5\linewidth]{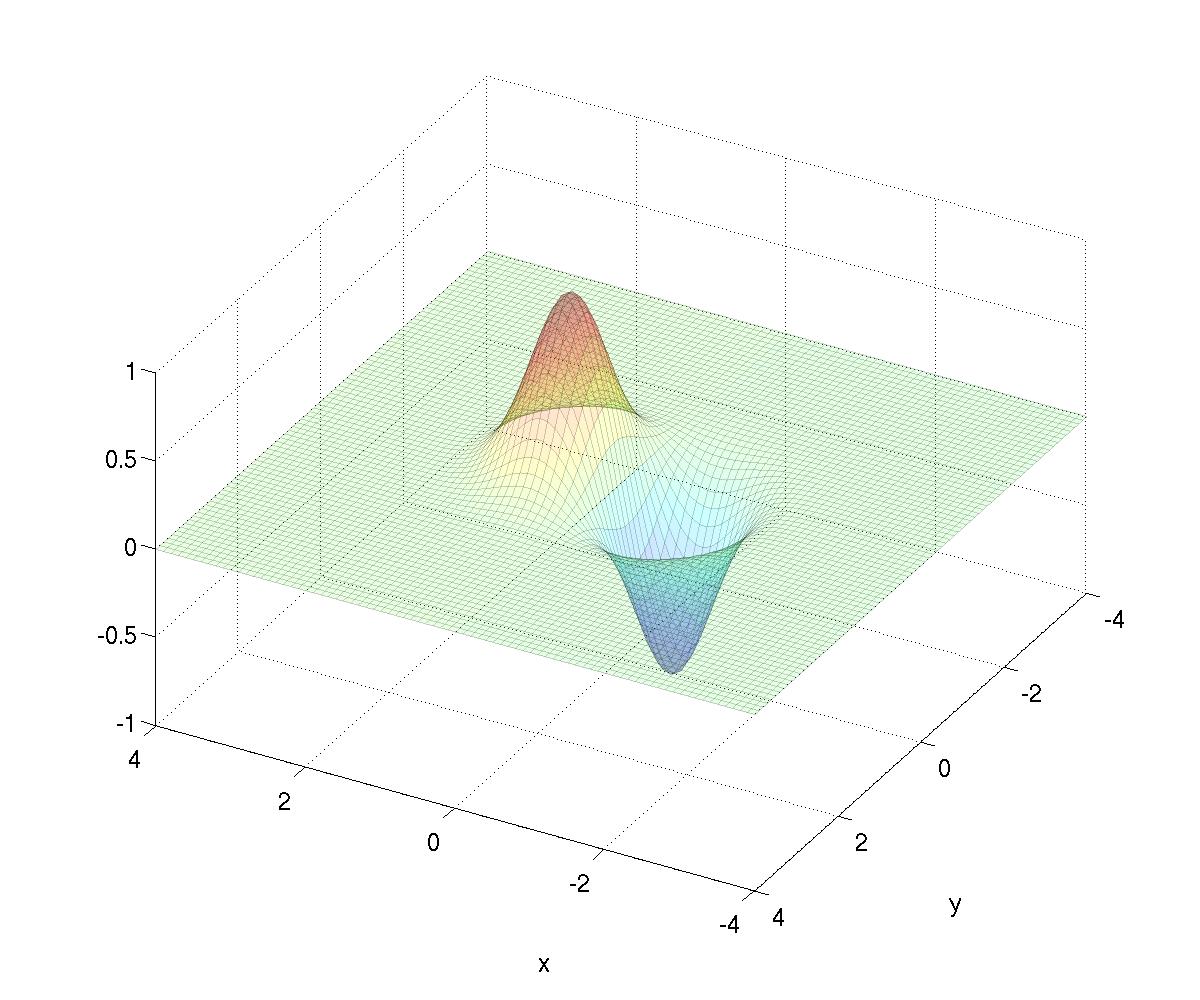}
\caption{The $n_4$ component at a spatial slice through a molecule at
  $z=0$ in the 2+4 model with $c_2=\tfrac{1}{4}$ and $c_4=1$. }
\label{fig:M4B1_025_1_n4}
\end{center}
\end{figure}
The half-Skyrmion molecule has two components which we can interpret
as global monopoles, both of baryon charge 1/2. In
Fig.~\ref{fig:M4B1_025_1_n4} is shown a cross section at $z=0$ of the
$n_4$ component of the molecule in the 2+4 model with
$c_2=\tfrac{1}{4}$ and $c_4=1$, as an example. 
The same characteristic holds for all the molecules that we found in
this section. The figure nicely shows that $n_4=0$, not only 
at infinity, but also on a plane separating the two constituents of the
molecule (which is seen as a line in the cross section shown in
Fig.~\ref{fig:M4B1_025_1_n4}). 
In the next section we will construct only one of the two constituents
isolated, however, at the cost of having an infinite total energy of
the configuration.

%%%%%%%%%%%%%%%%%%%%%%%%%%
\section{Fractional Skyrmions as global monopoles \label{sec:monopole}} 

In this section we will consider a limit in which one constituent of
the molecule has been drawn away to infinity, thus leaving the
molecule with only a half baryon number. The result is a half
Skyrmion. This is usually not possible. The loop hole is that the
configuration has a divergent total energy. 
By inserting the Ansatz 
\beq
\mathbf{n} = \left(
\hat{\mathbf{x}}\sin f(r), \cos f(r)
\right), 
\eeq
into the Lagrangian
\eqref{eq:LO4}, we get
\begin{align}
-\mathcal{L} &=
\frac{c_2}{2} f_r^2 + \frac{c_2}{r^2}\sin^2 f
+\frac{c_4}{r^2}\sin^2(f) f_r^2 + \frac{c_4}{2r^4}\sin^4f
+\frac{c_6}{r^4}\sin^4(f) f_r^2
+\frac{m^2}{2}\cos^2 f,
\end{align}
which has the corresponding equation of motion
\begin{align}
c_2\left(
  f_{rr} 
  + \frac{2}{r}f_r 
  - \frac{1}{r^2}\sin 2f
\right)
+ c_4\left(
  \frac{2}{r^2}\sin^2(f) f_{rr} 
  + \frac{1}{r^2}\sin(2f) f_r^2
  - \frac{1}{r^4}\sin^2 f\sin 2f
\right) \non
+ c_6\left(
  \frac{2}{r^4}\sin^4(f) f_{rr}
  - \frac{4}{r^5}\sin^4(f) f_r
  + \frac{2}{r^4}\sin^2 f\sin(2f) f_r^2
\right)
+ \frac{1}{2}m^2\sin 2f = 0,
\end{align}
where $f_r\equiv\p_r f$.
The difference between this Lagrangian with corresponding equation of
motion and the normal case with a mass term is that the potential
$\sin^2 f$ is replaced with $\cos^2 f$ and in turn $-\sin 2f$ in the
equation of motion becomes $+\sin 2f$. We therefore need to consider
the boundary conditions for this system in order to construct a half
Skyrmion. One possibility is to choose 
\beq
\textrm{southern molecule constituent:}\qquad
f(0) = \pi, \qquad
f(\infty) = \frac{\pi}{2},
\eeq
giving a half Skyrmion winding on only the southern hemisphere of the
target space. Alternatively, we could have chosen
\beq
\textrm{northern molecule constituent:}\qquad
f(0) = 0, \qquad
f(\infty) = \frac{\pi}{2},
\eeq
giving instead a half anti-Skyrmion which winds only on the northern
hemisphere of the target space. 

\begin{figure}[!htb]
\begin{center}
\mbox{
\subfloat[$n_4=\cos f$]{\includegraphics[width=0.49\linewidth]{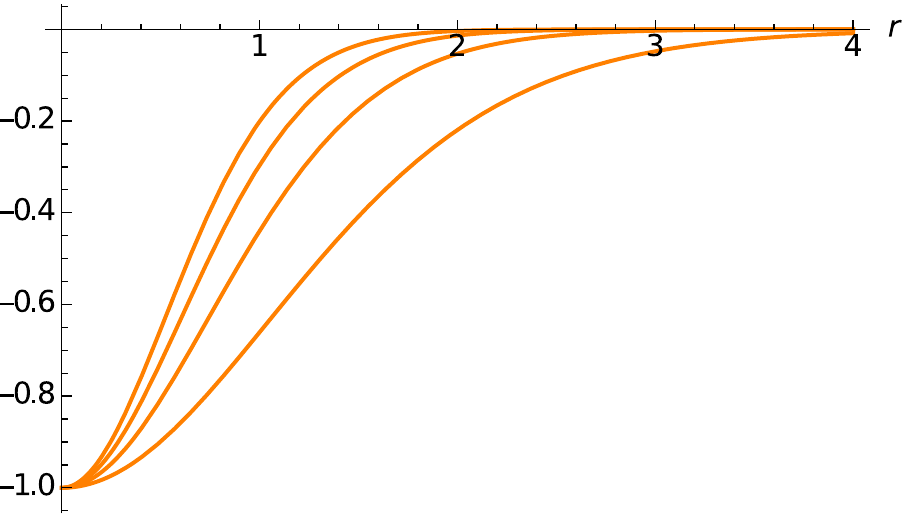}}
\subfloat[$\mathcal{E}$]{\includegraphics[width=0.49\linewidth]{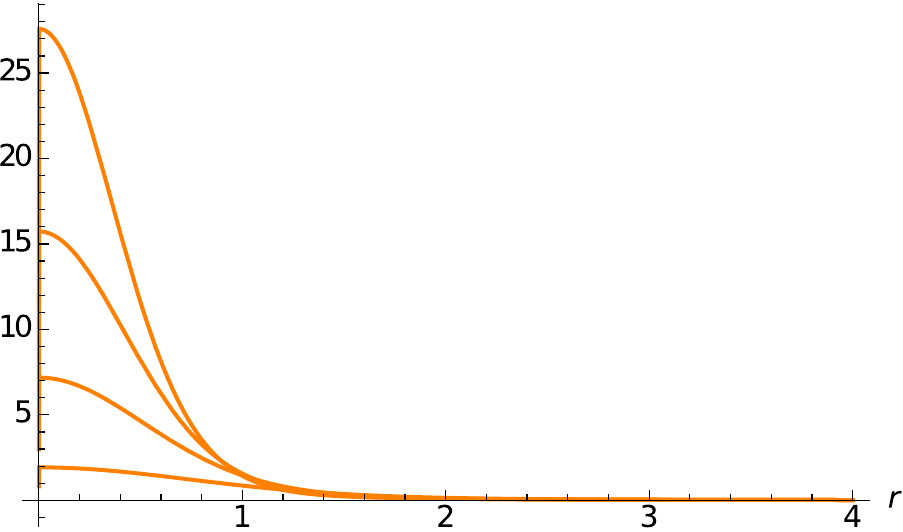}}}
\mbox{
\subfloat[$4\pi r^2\mathcal{E}$]{\includegraphics[width=0.49\linewidth]{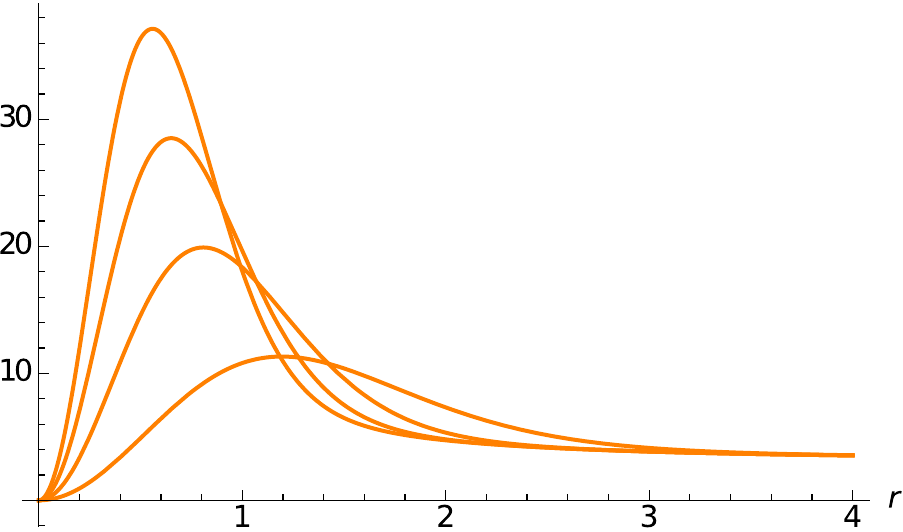}}
\subfloat[$\mathcal{B}$]{\includegraphics[width=0.49\linewidth]{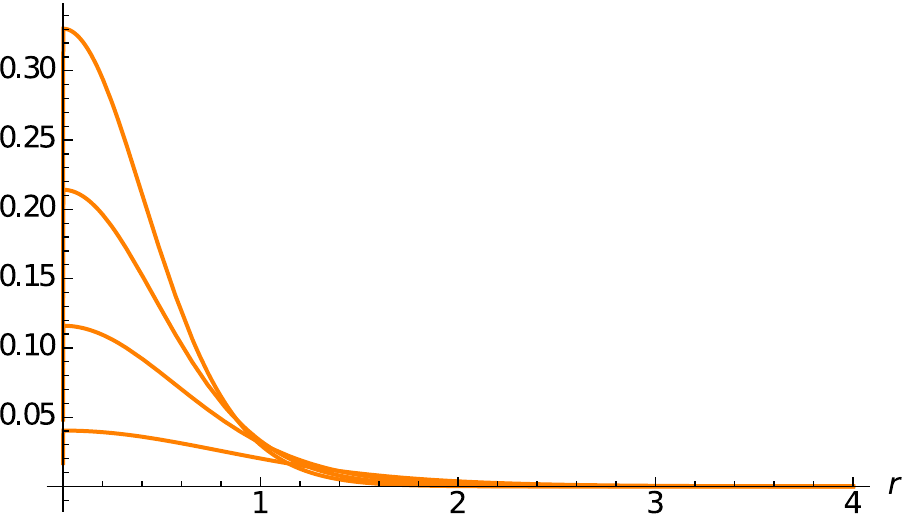}}}
\caption{An isolated half Skyrmion (a global monopole) in the 2+4
  model for various masses $m$: (a) the radial field profile $n_4=\cos
  f$. (b) and (c) the energy density $\mathcal{E}$ and $4\pi
  r^2\mathcal{E}$, respectively. The latter shows the divergence in
  the total energy. (d) is the baryon charge density. The chosen parameters
  are $c_2=\tfrac{1}{4}$, $c_4=1$ and $m=1,2,3,4$. }
\label{fig:c4_halfsk}
\end{center}
\end{figure}

\begin{figure}[!htb]
\begin{center}
\mbox{
\subfloat[$n_4=\cos f$]{\includegraphics[width=0.49\linewidth]{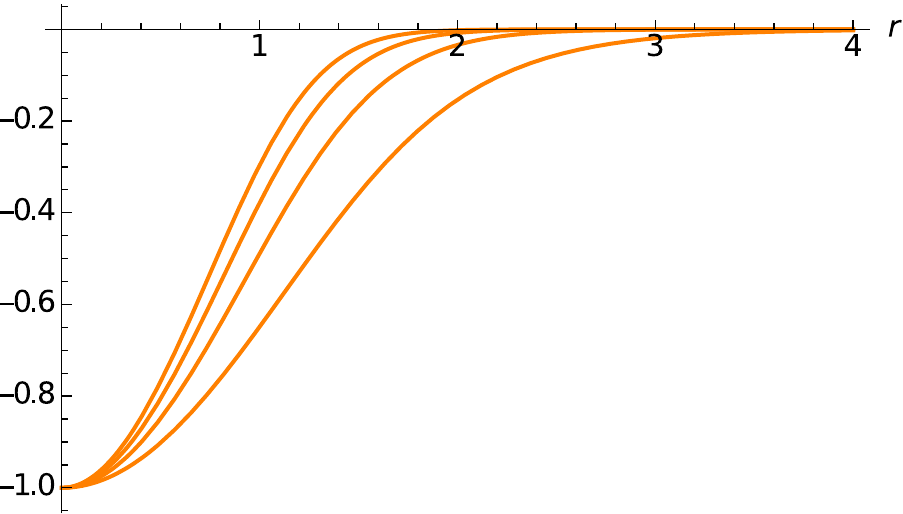}}
\subfloat[$\mathcal{E}$]{\includegraphics[width=0.49\linewidth]{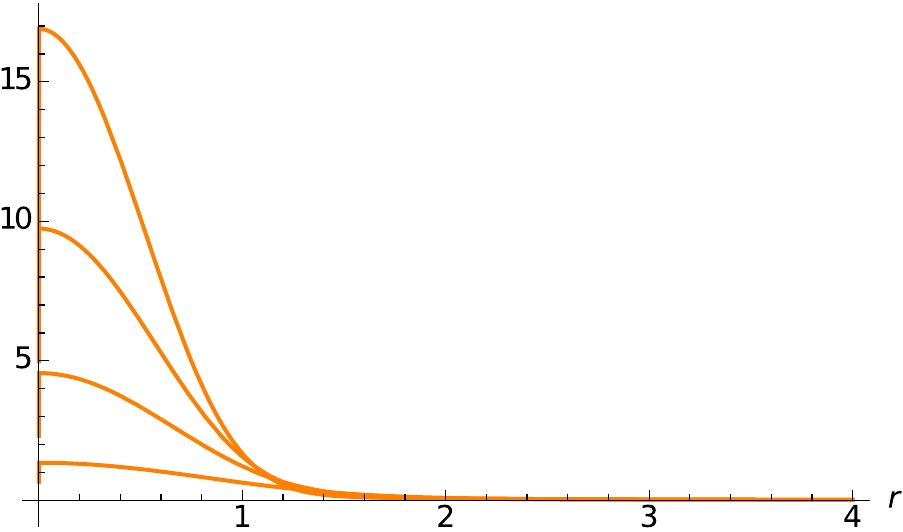}}}
\mbox{
\subfloat[$4\pi r^2\mathcal{E}$]{\includegraphics[width=0.49\linewidth]{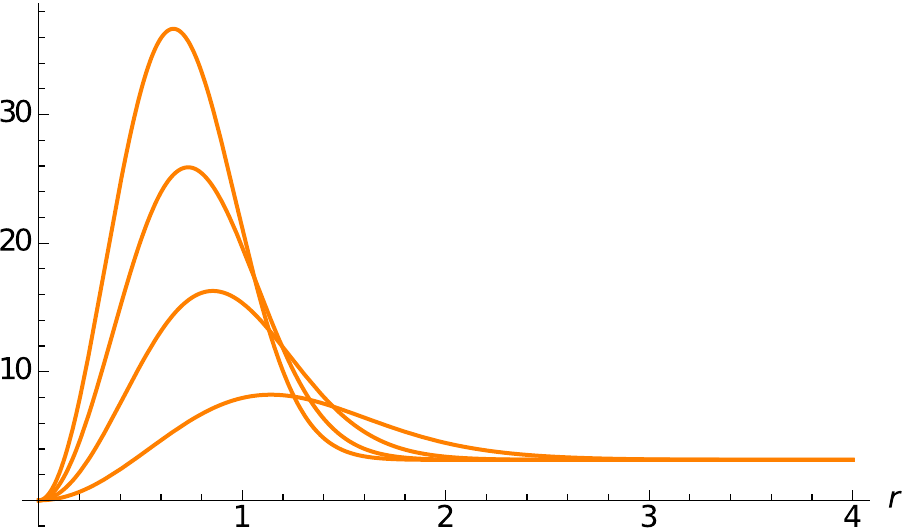}}
\subfloat[$\mathcal{B}$]{\includegraphics[width=0.49\linewidth]{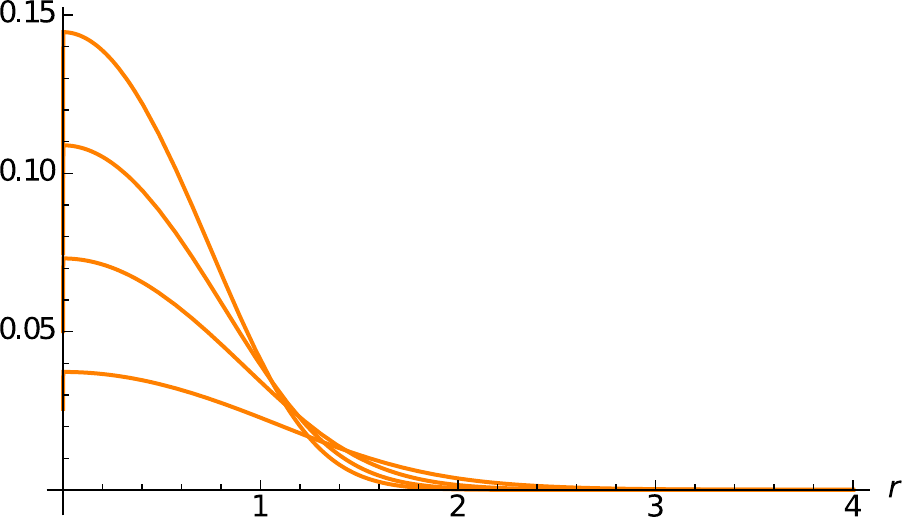}}}
\caption{An isolated half Skyrmion (a global monopole) in the 2+6
  model for various masses $m$: (a) the radial field profile $n_4=\cos
  f$. (b) and (c) the energy density $\mathcal{E}$ and $4\pi
  r^2\mathcal{E}$, respectively. The latter shows the divergence in
  the total energy. (d) is the baryon charge density. The chosen parameters
  are $c_2=\tfrac{1}{4}$, $c_6=1$ and $m=1,2,3,4$. }
\label{fig:c6_halfsk}
\end{center}
\end{figure}

In Figs.~\ref{fig:c4_halfsk} and \ref{fig:c6_halfsk} are shown
numerical solutions for the half Skyrmion winding on only the southern
hemisphere of the target space (i.e.~$n_4\in[-1,0]$) for the 2+4 and
2+6 model, respectively. 
The half Skyrmion can live in isolation only at the cost of an
infinite total energy. Figs.~\ref{fig:c4_halfsk}c and
\ref{fig:c6_halfsk}c show that the energy density multiplied by $r^2$,
i.e.~$4\pi r^2\mathcal{E}$ goes to a constant and hence the total
energy picks up a linear divergence in the radial integral 
$E\propto R_{\rm max}$, where $R_{\rm max}$ is the radial cut off of
the integral. All the baryon charge densities integrate numerically to
one half.  

%%%%%%%%%%%%%%%%%%%%%%%%%%
\section{Higher baryon numbers \label{sec:higher}}

In this section we make a first attempt to make solutions with higher
baryon numbers, i.e.~$B>1$. As we \emph{ab initio} do not know the shape of
the global minimizers, we use the relaxation method with different
initial guesses. The two guesses we choose here are the axially
symmetric Ansatz 
\beq
\mathbf{n} = \left(
  -\cos f(r),
  \sin\theta\sin B\phi\sin f(r),
  \sin\theta\cos B\phi\sin f(r),
  \cos\theta\sin f(r)
\right),
\label{eq:axial_symmetric_initial_guess}
\eeq
and the following Ansatz using a rational map $R$:
\beq
\mathbf{n} = \left(
-\cos f(r),
\frac{R + \bar{R}}{1+R\bar{R}}\sin f(r),
\frac{i(\bar{R}-R)}{1+R\bar{R}}\sin f(r),
\frac{1 - R\bar{R}}{1+R\bar{R}}\sin f(r)
\right), 
\label{eq:rational_map_initial_guess}
\eeq
with the symmetries found to minimize the Skyrmions without the
potential \eqref{eq:pot}. For the appropriate rational maps, $R$, see 
Refs.~\cite{Battye:1997qq,Houghton:1997kg}.
Note again that we have chosen a particular value on the vacuum
manifold, i.e.~$n_1=-1$ which by O(3) symmetry is equivalent to any
other choice. 

\begin{figure}[!tb]
\begin{center}
\captionsetup[subfloat]{labelformat=empty}
\mbox{
\subfloat{\includegraphics[width=0.44\linewidth]{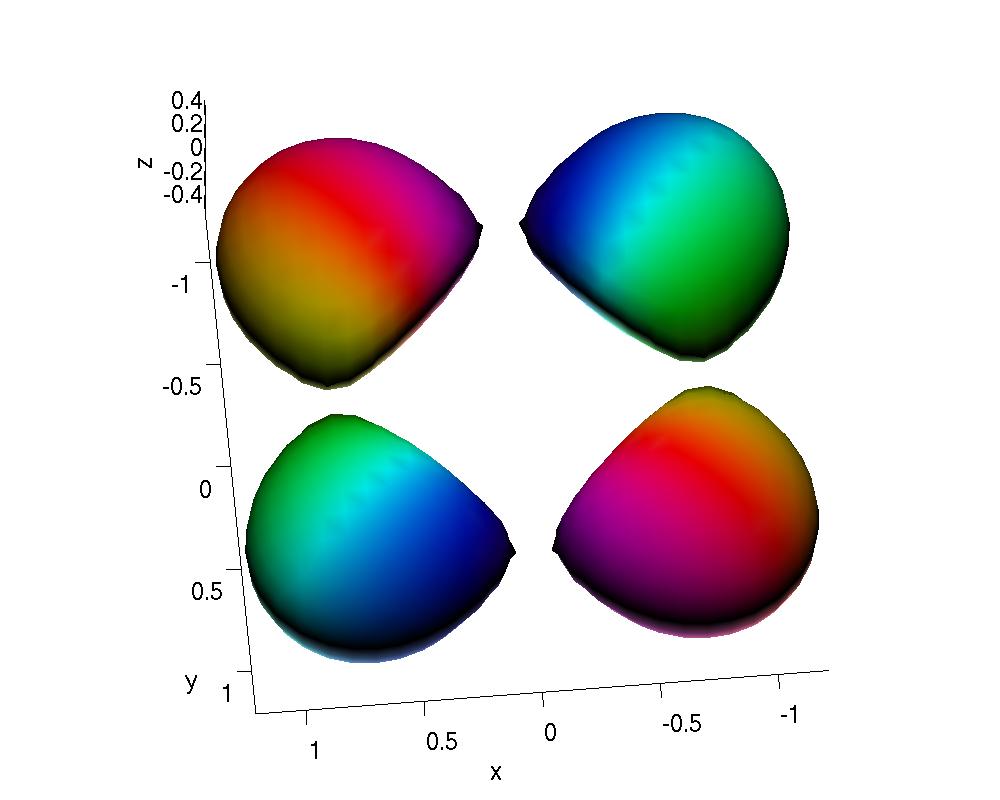}}
\subfloat{\includegraphics[width=0.44\linewidth]{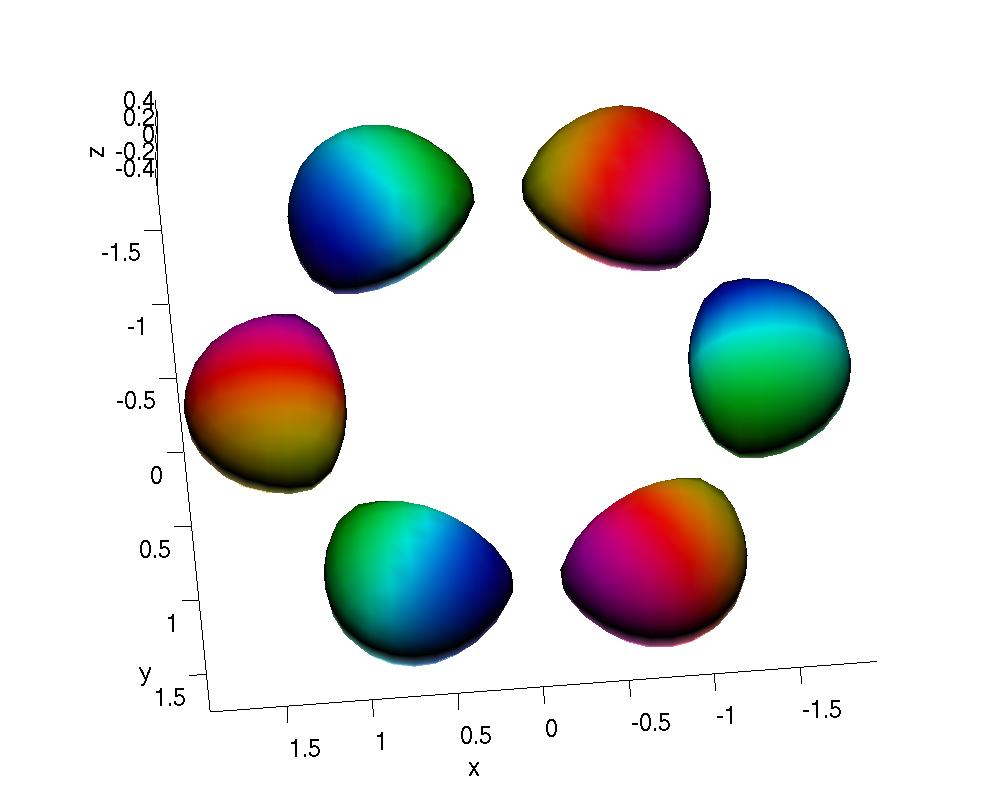}}}
\mbox{
\subfloat{\includegraphics[width=0.44\linewidth]{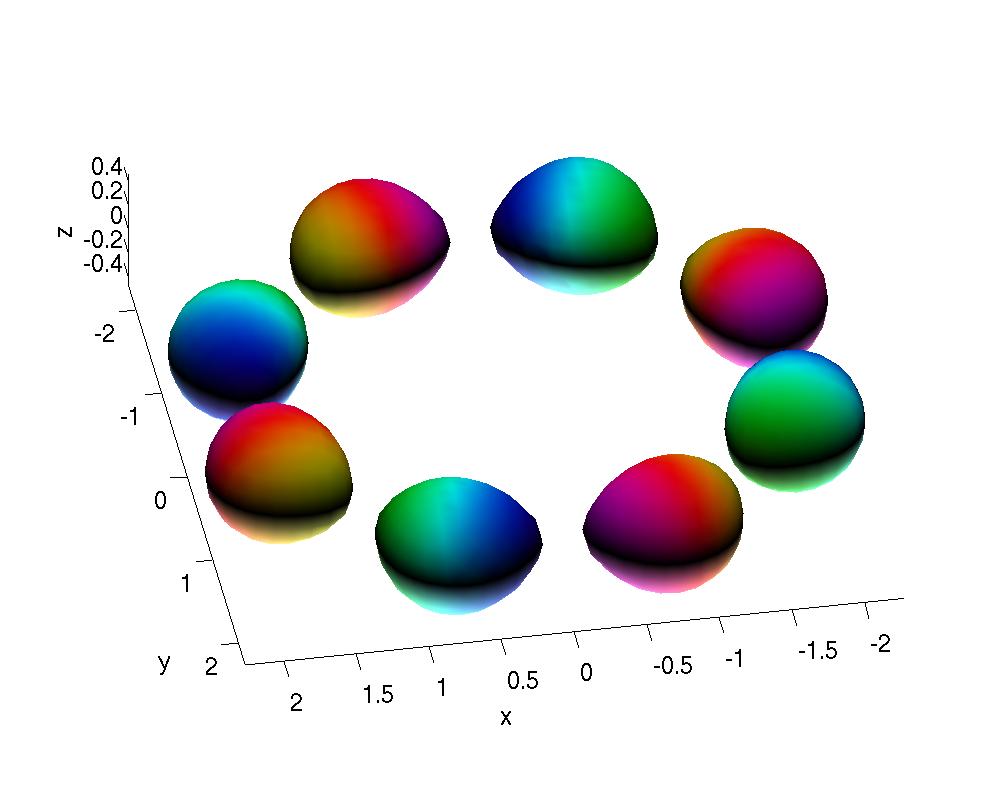}}
\subfloat{\includegraphics[width=0.44\linewidth]{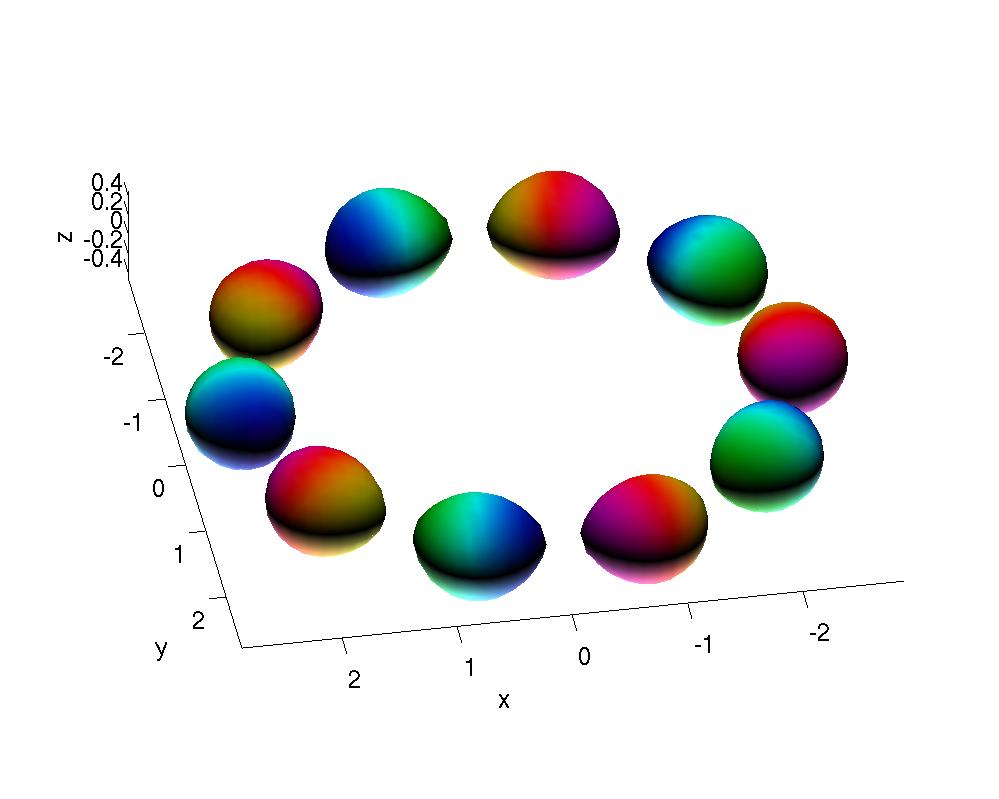}}}
\mbox{
\subfloat{\includegraphics[width=0.44\linewidth]{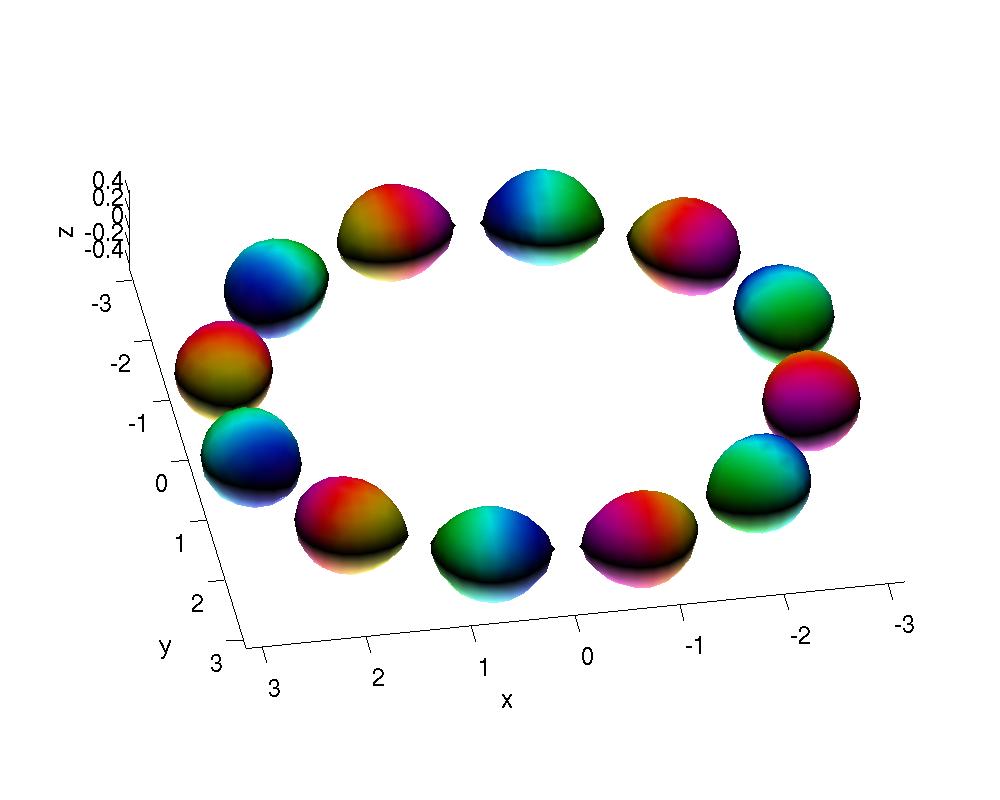}}}
\caption{Isosurfaces showing the half-maximum of the baryon charge
  density in the 2+4 model for baryon numbers $B=2,3,4,5,6$ with
  $c_2=\tfrac{1}{4}$, $c_4=1$ and fixed mass $m=4$. The color scheme
  is the same as that in Fig.~\ref{fig:M4B1}. 
} 
\label{fig:M4B23456}
\end{center}
\end{figure}

We begin with the 2+4 model and calculate the numerical solutions for
the first six baryon numbers, i.e.~$B=2,3,4,5,6$ ($B=1$ was already
made in Sec.~\ref{sec:molecule}). For concreteness, we
fix the parameters $c_2=\tfrac{1}{4}$, $c_4=1$ and $m=4$,
corresponding to a canonical mass $m^{\rm canonical}=16$.
We find that the lowest-energy states take the shapes of rings (beads
on rings) and show the numerical solutions as isosurfaces at the
half-maximum baryon charge densities in Fig.~\ref{fig:M4B23456}. 
These solutions are made with the axially-symmetric initial guess
\eqref{eq:axial_symmetric_initial_guess}. 
We again use the numerically integrated baryon charge density, 
$B^{\rm numerical}$ as a handle on the precision of the numerical
solution, see Tab.~\ref{tab:M4B23456}. In this table we also show the
total energy per unit baryon charge, $E^{\rm numerical}/B$, which
tells us whether the higher-charged solution is stable or only
metastable. We find that all the higher-charged solutions we have
calculated, namely $B=2,3,4,5,6$ are in fact stable 
(at least among this type of configurations).

\begin{table}[!htb]
\begin{center}
\caption{The numerically integrated baryon charge and energy per unit
  baryon charge for higher baryon numbers in the 2+4 model. }
\label{tab:M4B23456}
\begin{tabular}{clc}
$B$ & $B^{\rm numerical}$ & $E^{\rm numerical}/B$\\
\hline\hline
1 & $0.99967$ & $65.52(7)$\\
2 & $1.9994$ & $61.5(3)$\\
3 & $2.9979$ & $61.0(6)$\\
4 & $3.9971$ & $60.8(9)$\\
5 & $4.9964$ & $60.7(9)$\\
6 & $5.9957$ & $60.5(5)$
\end{tabular}
\end{center}
\end{table}

\begin{figure}[!ptb]
\begin{center}
\captionsetup[subfloat]{labelformat=empty}
\mbox{
\subfloat{\includegraphics[width=0.4\linewidth]{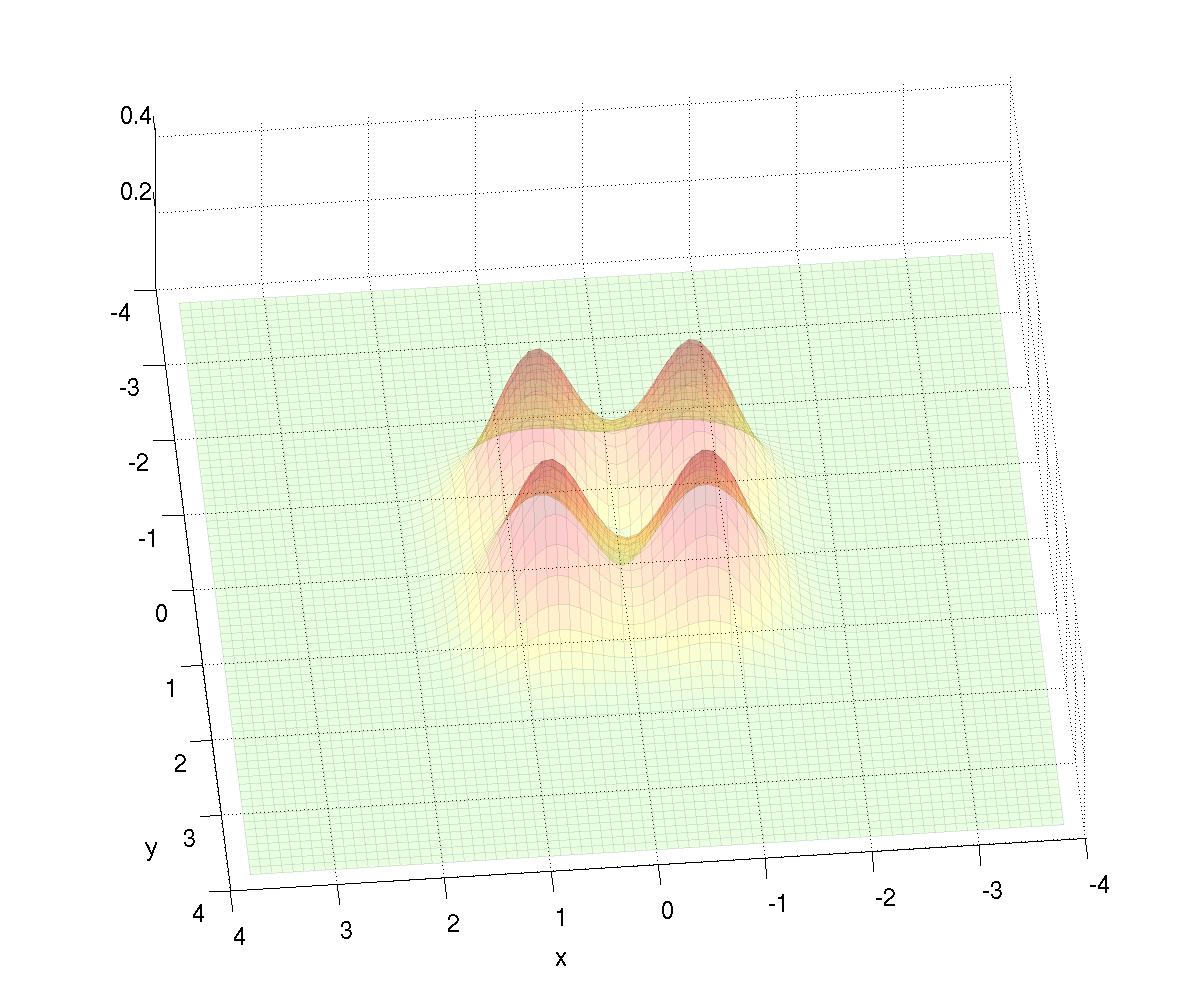}}
\subfloat{\includegraphics[width=0.4\linewidth]{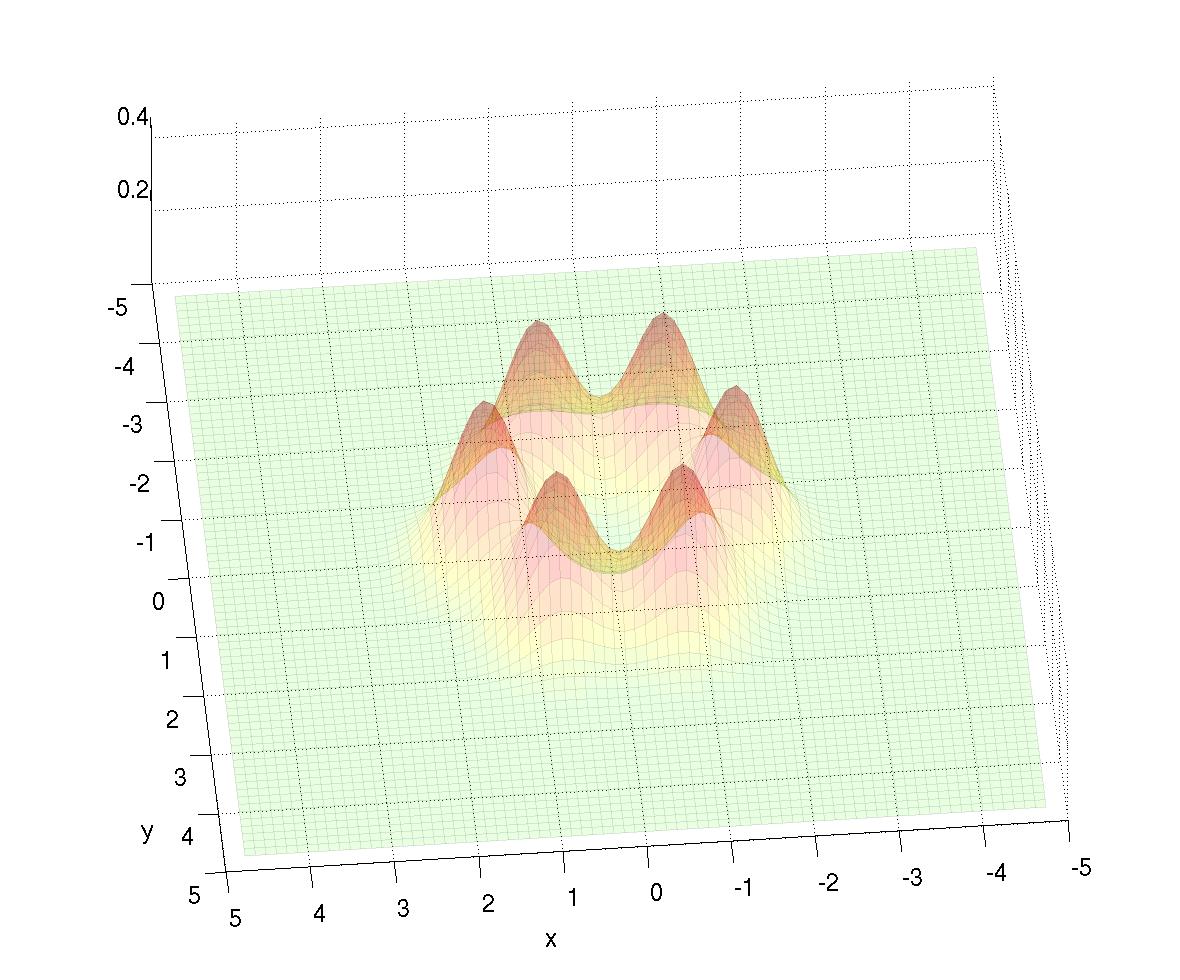}}}
\mbox{
\subfloat{\includegraphics[width=0.4\linewidth]{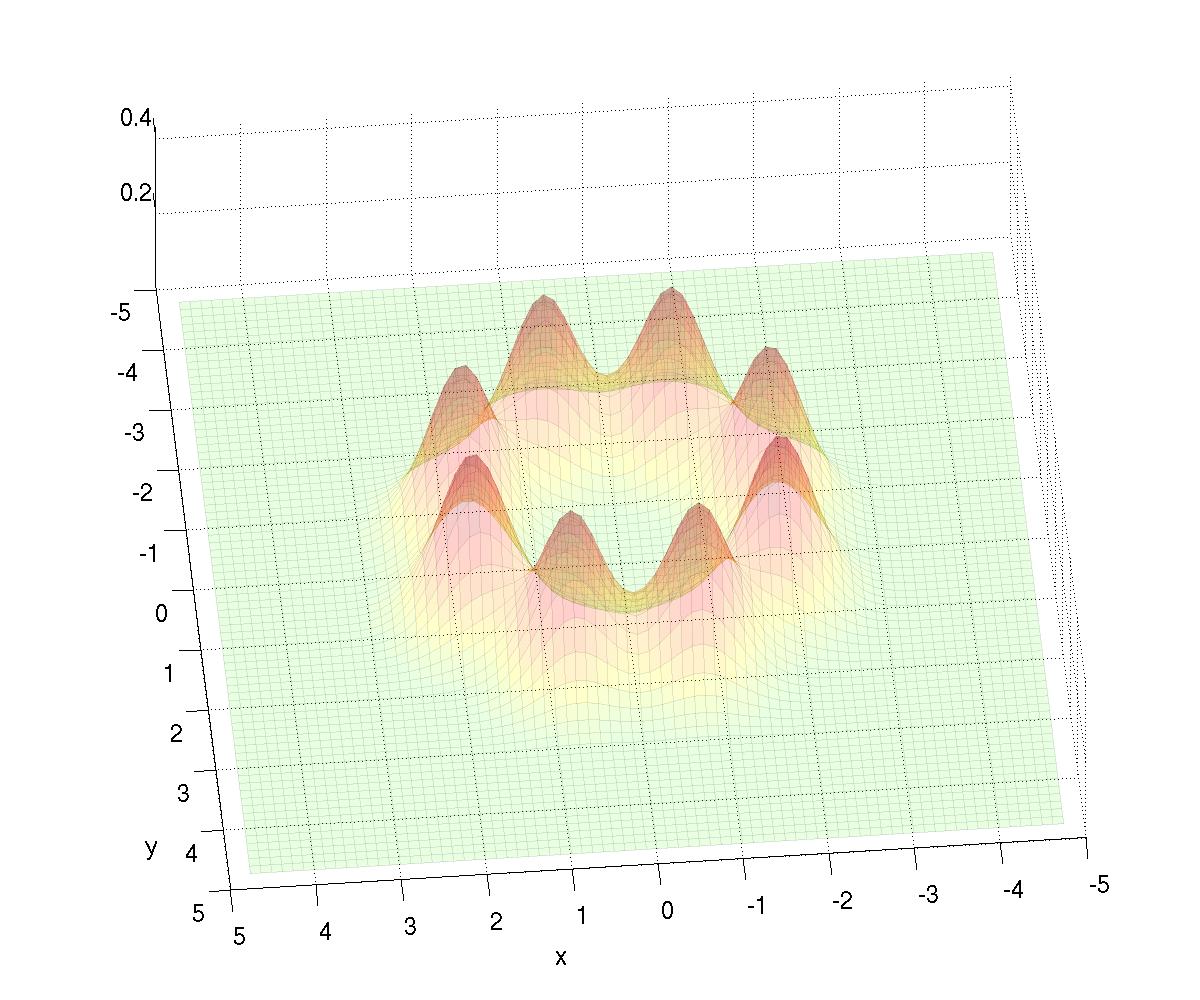}}
\subfloat{\includegraphics[width=0.4\linewidth]{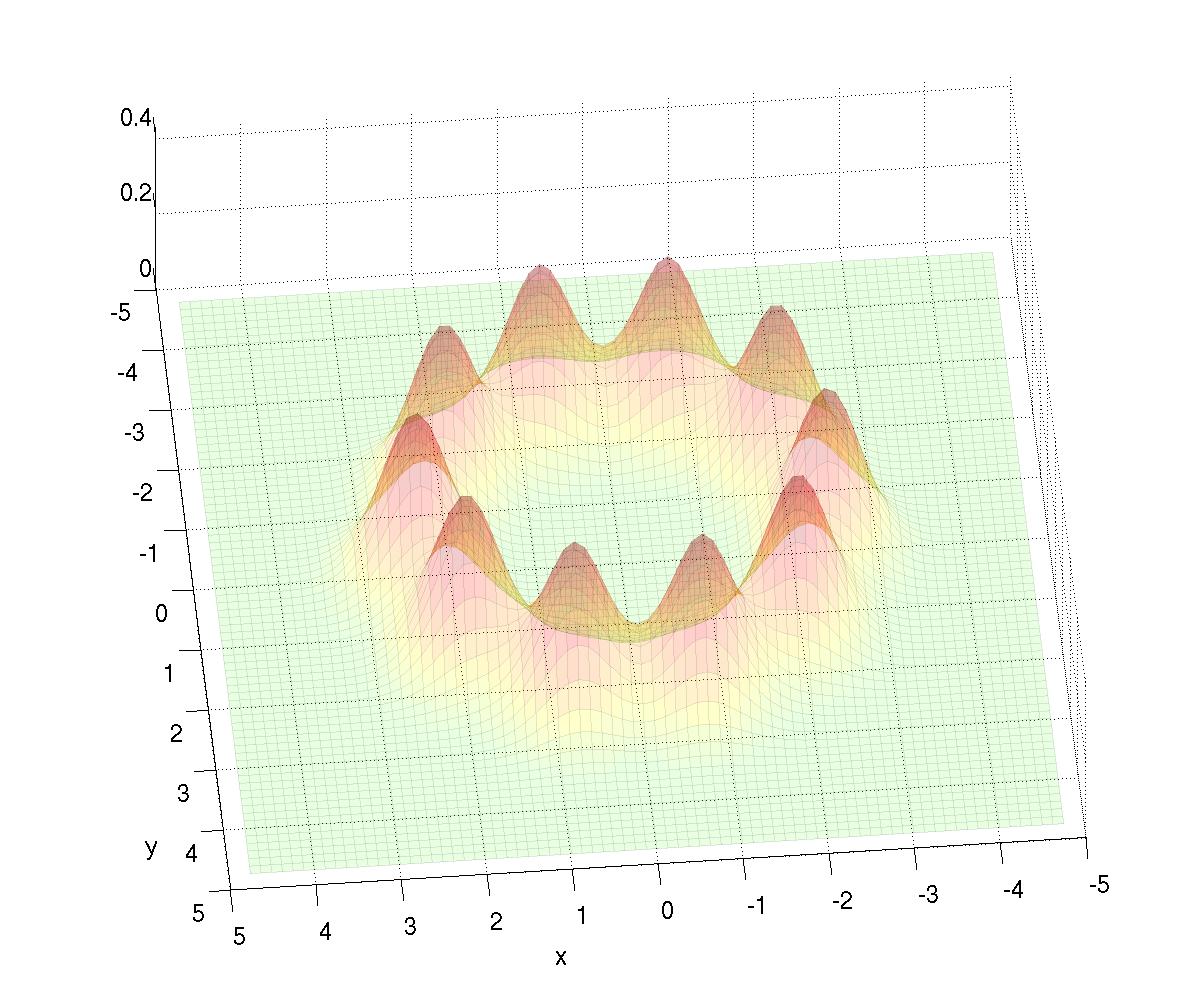}}}
\mbox{
\subfloat{\includegraphics[width=0.4\linewidth]{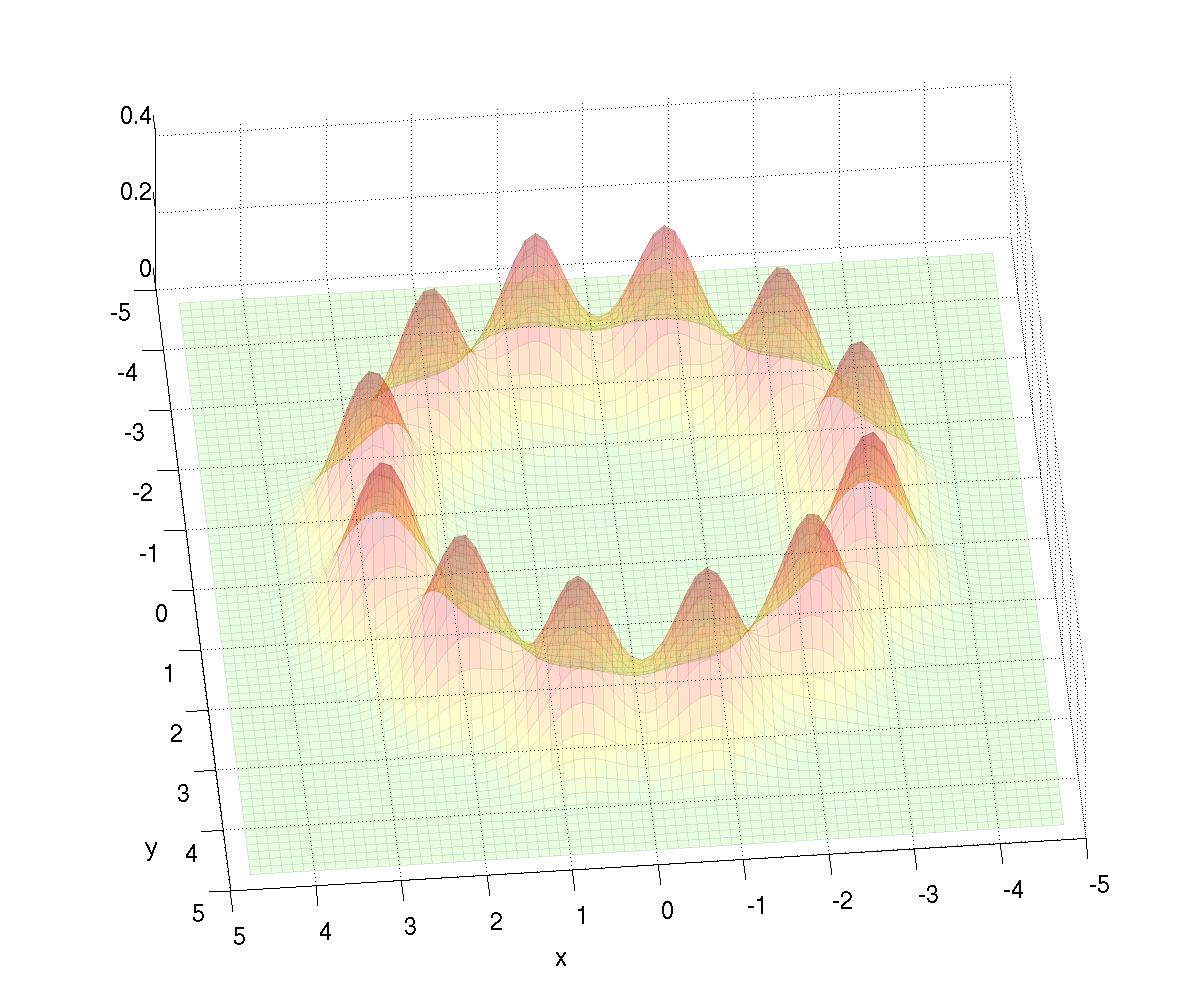}}}
\caption{Baryon charge density at a spatial slice through the
  $B=2,3,4,5,6$ molecules at $z=0$ in the 2+4 model for various
  choices of $(c_2,c_4)$ and for fixed mass $m=4$. 
} 
\label{fig:M4B23456_baryonslice}
\end{center}
\end{figure}

\begin{figure}[!ptb]
\begin{center}
\captionsetup[subfloat]{labelformat=empty}
\mbox{
\subfloat{\includegraphics[width=0.4\linewidth]{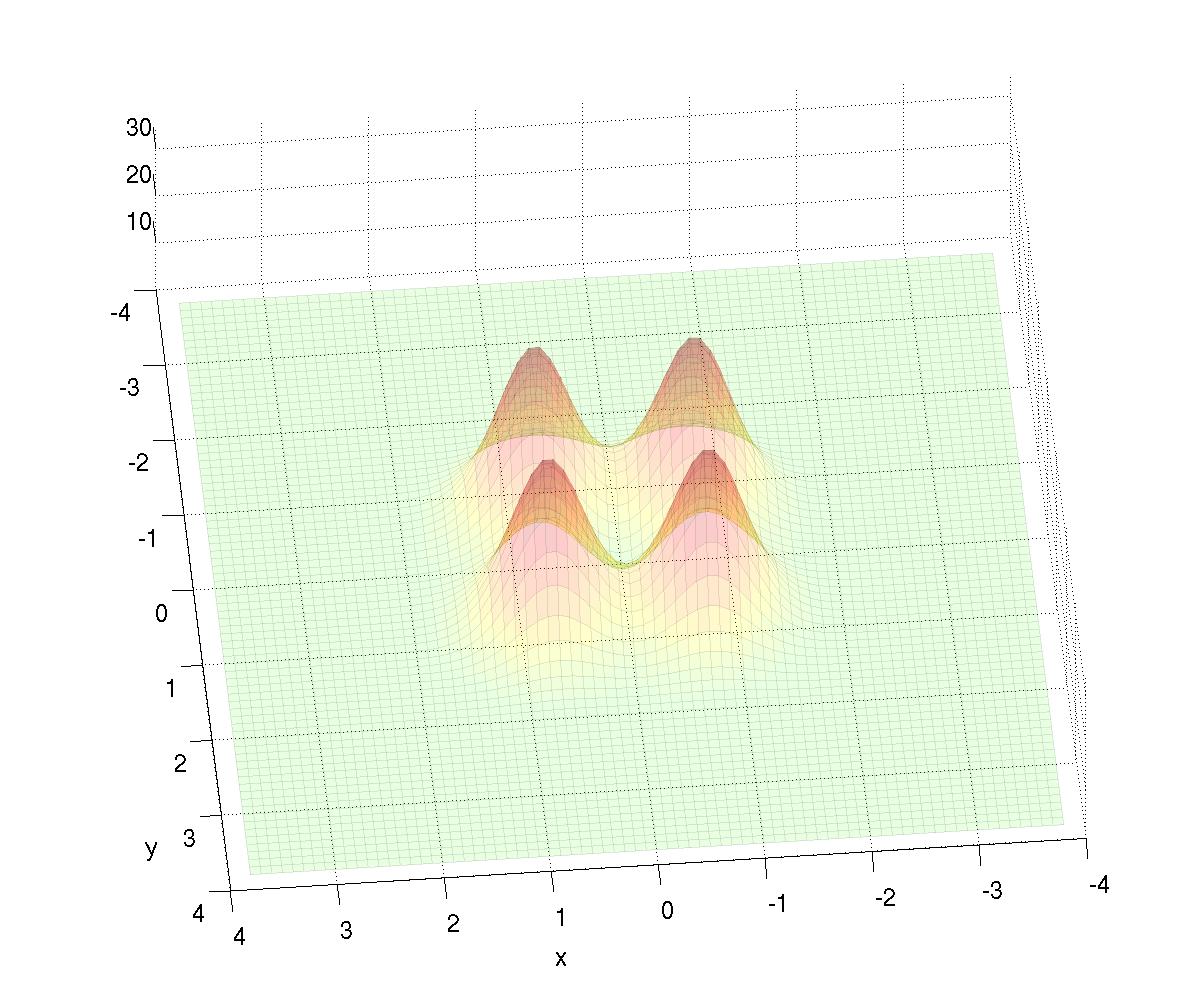}}
\subfloat{\includegraphics[width=0.4\linewidth]{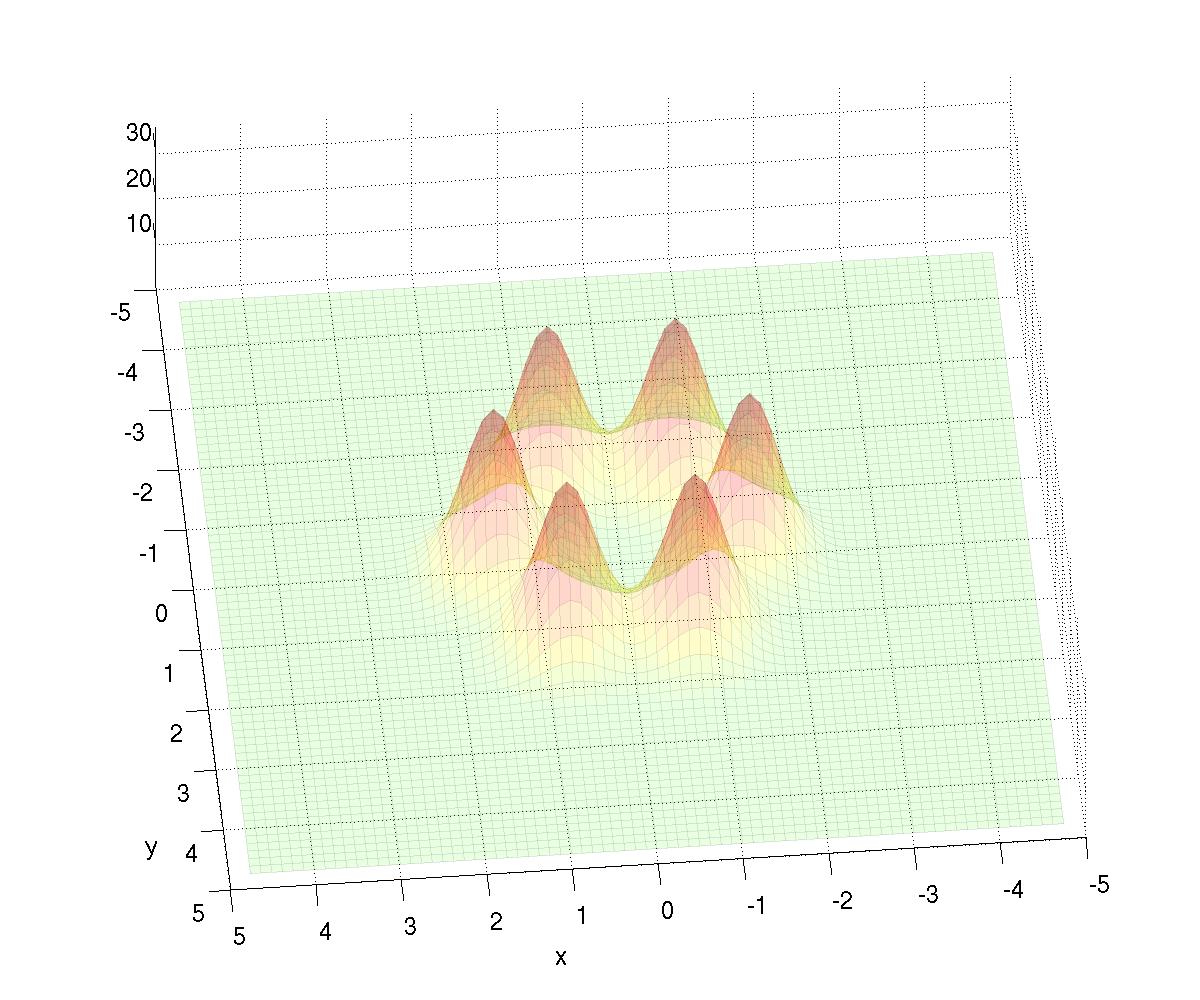}}}
\mbox{
\subfloat{\includegraphics[width=0.4\linewidth]{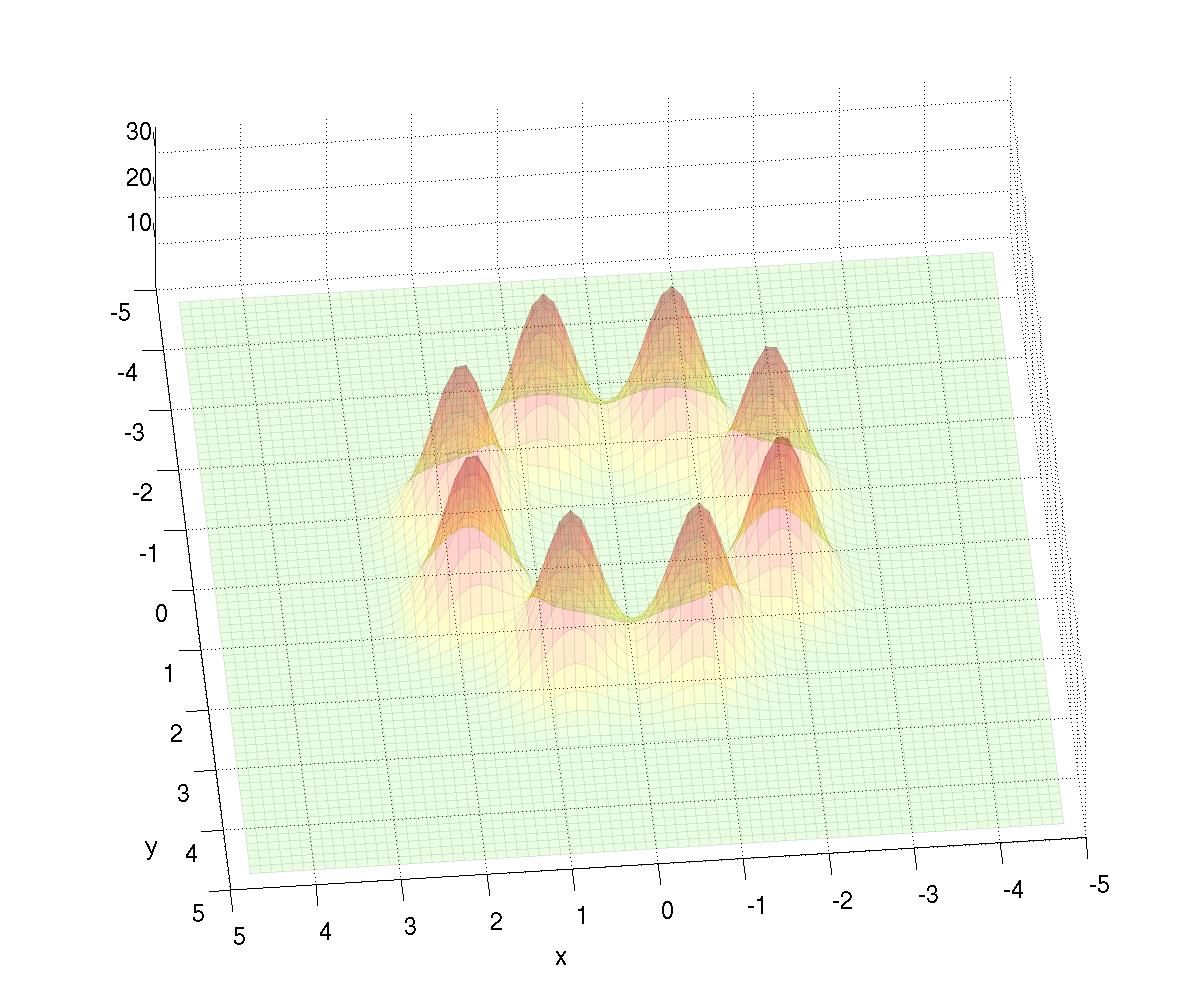}}
\subfloat{\includegraphics[width=0.4\linewidth]{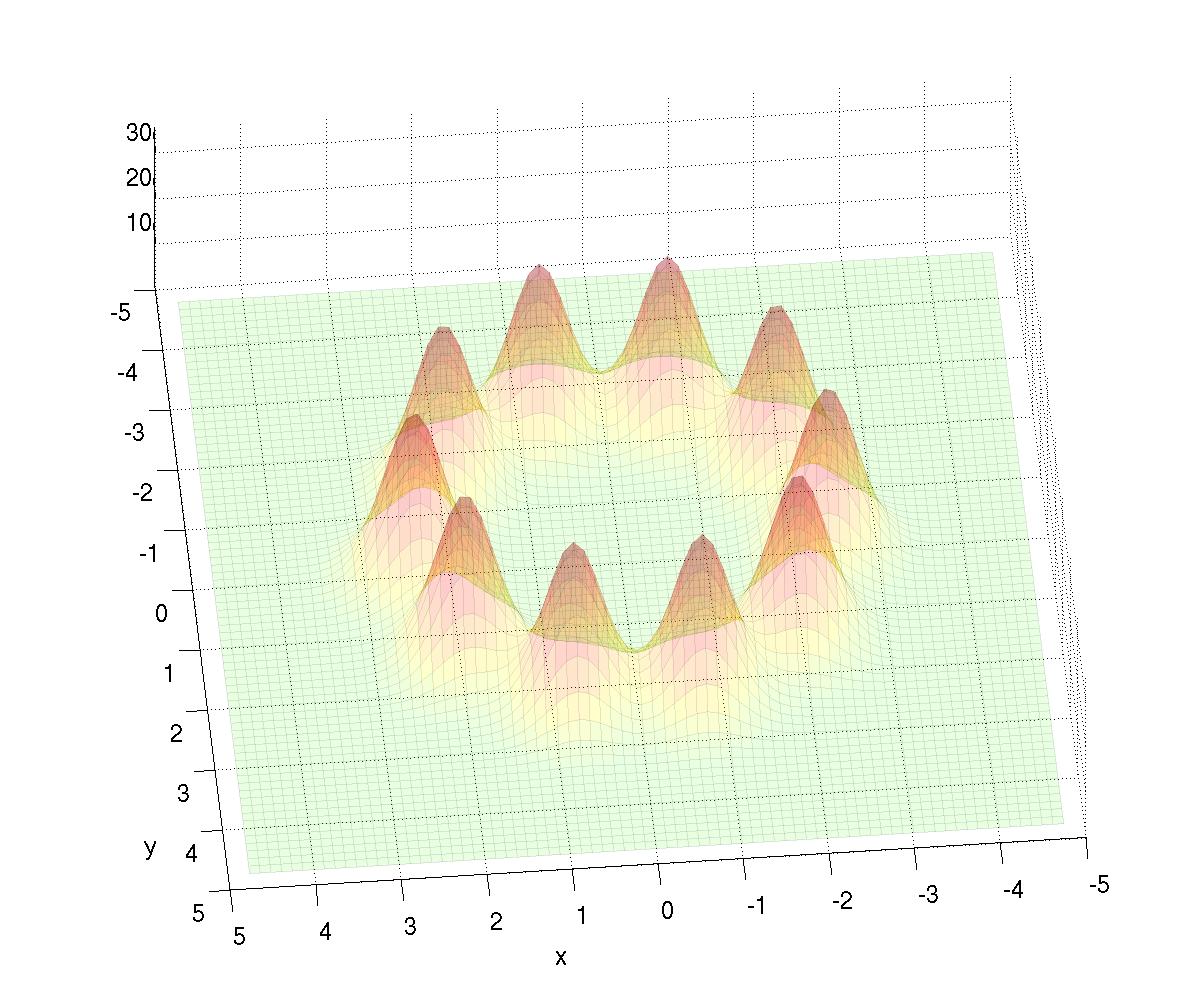}}}
\mbox{
\subfloat{\includegraphics[width=0.4\linewidth]{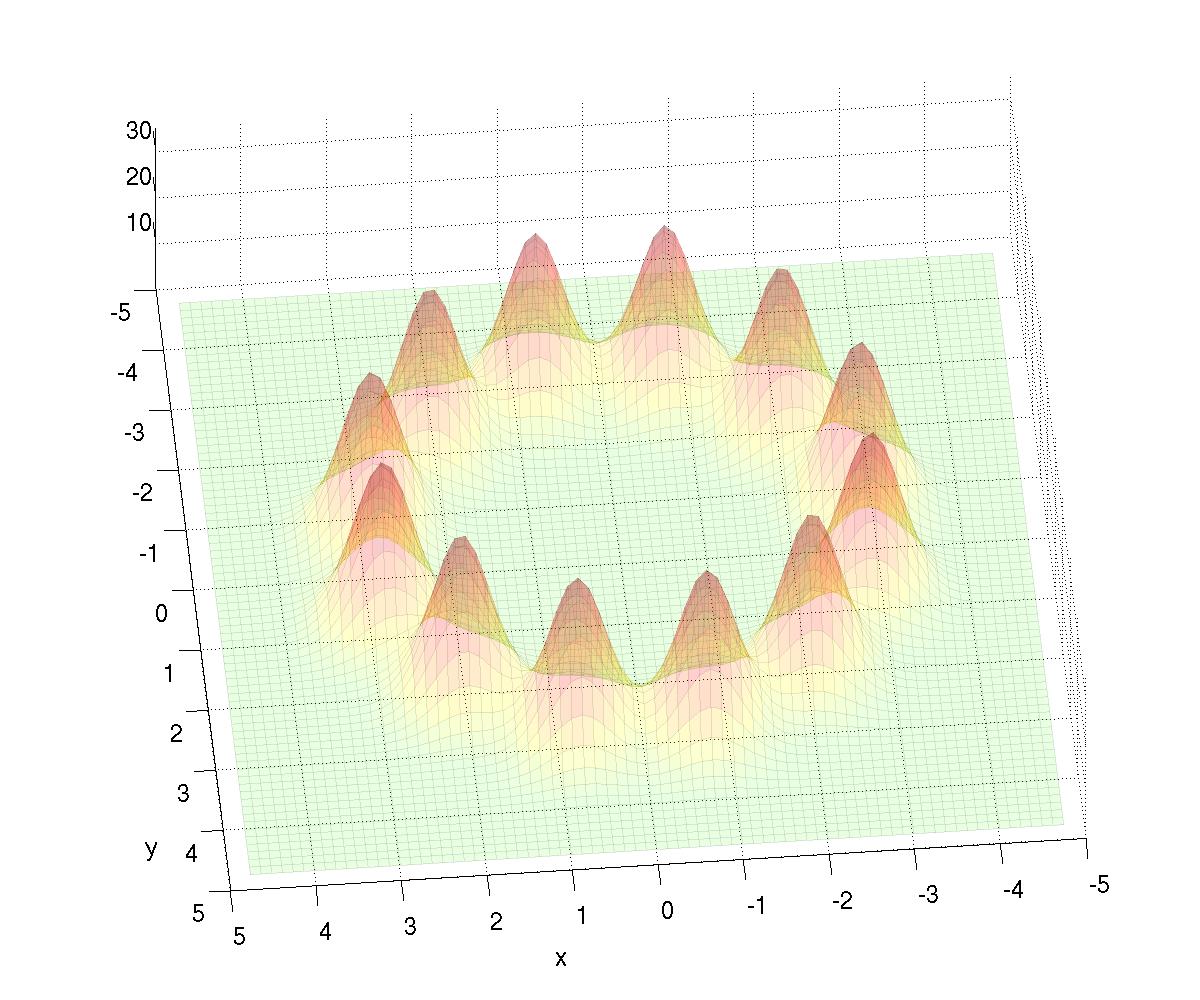}}}
\caption{Energy density at a spatial slice through the
  $B=2,3,4,5,6$ molecules at $z=0$ in the 2+4 model for various
  choices of $(c_2,c_4)$ and for fixed mass $m=4$. 
} 
\label{fig:M4B23456_energyslice}
\end{center}
\end{figure}

The isosurface shows the three-dimensional structure of the solution,
but not the profile shape of the baryon-charge density or energy
density. Therefore, as before, we show cross sections at $z=0$ of the
baryon charge density and energy density in
Fig.~\ref{fig:M4B23456_baryonslice} and
\ref{fig:M4B23456_energyslice}, respectively. 
Note that again the molecular shape is slightly more pronounced in the 
energy density than in the baryon charge density, viz.~the depth of
the valleys between the peaks are deeper. 
Finally, we notice that the molecular shape (by which we mean that
half a unit of baryon charge is spatially localized) is far more
pronounced for higher baryon charges $B>1$ than for $B=1$, for the
same coefficients, i.e.~$c_2=\tfrac{1}{4}$, $c_4=1$ and $m=4$.

As we mentioned already, we do not \emph{ab initio} know the
symmetries or spatial structure of the \emph{global} energy-minimizing
solutions. Therefore we have tried also different initial conditions
to check whether we can obtain lower-energy solutions with the same
baryon numbers compared to those obtained with the axially-symmetric
initial guess \eqref{eq:axial_symmetric_initial_guess}. Explicitly, we
have tried the rational-map Ansatz with the rational maps that
minimize the energy without the potential \eqref{eq:pot}. 
We found these solutions to have higher energies than those shown here
(or being spatially disconnected) and more details are given in
Appendix \ref{app:rational_map_2+4}.

\begin{figure}[!tb]
\begin{center}
\captionsetup[subfloat]{labelformat=empty}
\mbox{
\subfloat{\includegraphics[width=0.45\linewidth]{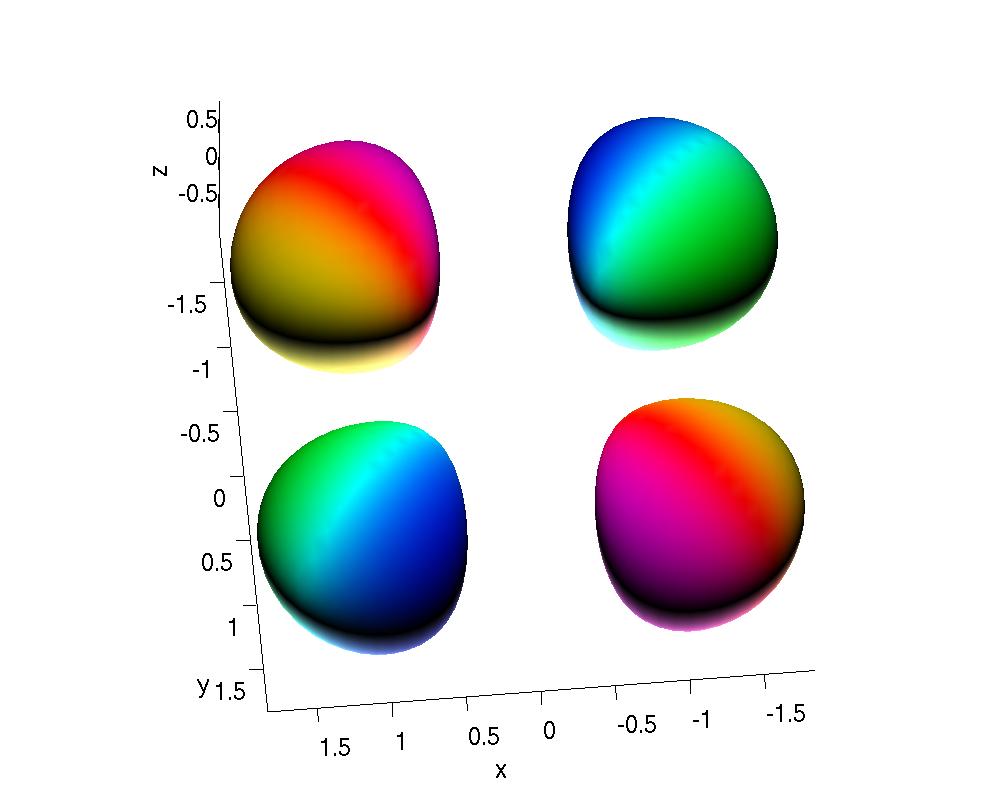}}
\subfloat{\includegraphics[width=0.45\linewidth]{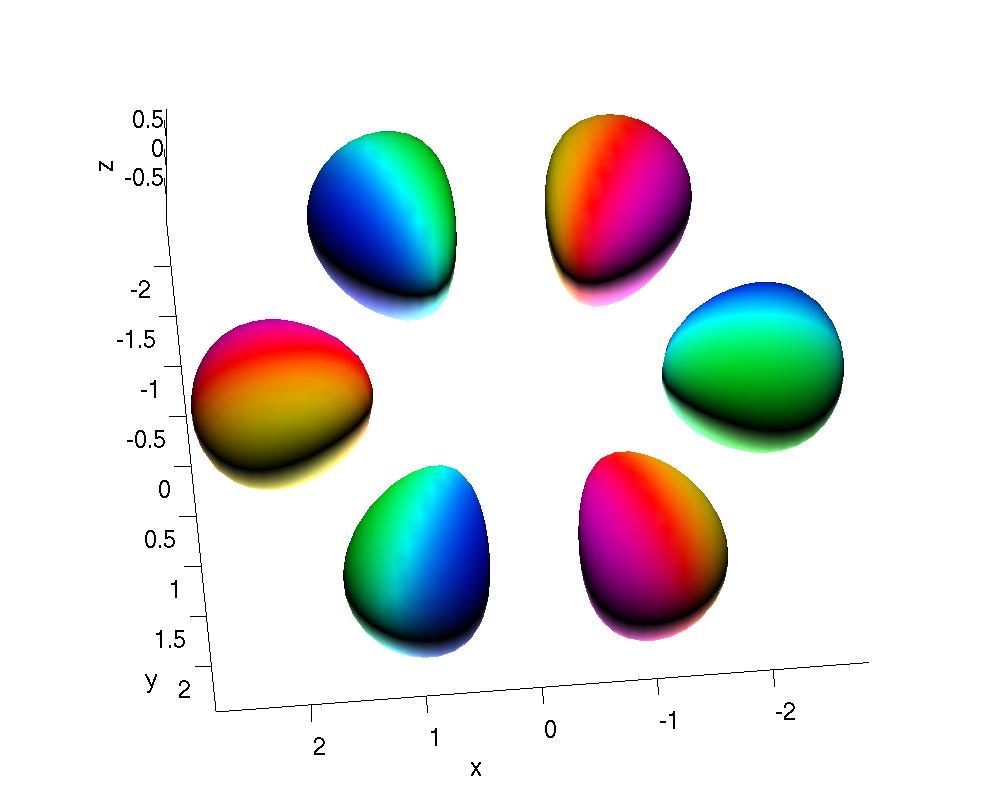}}}
\mbox{
\subfloat{\includegraphics[width=0.45\linewidth]{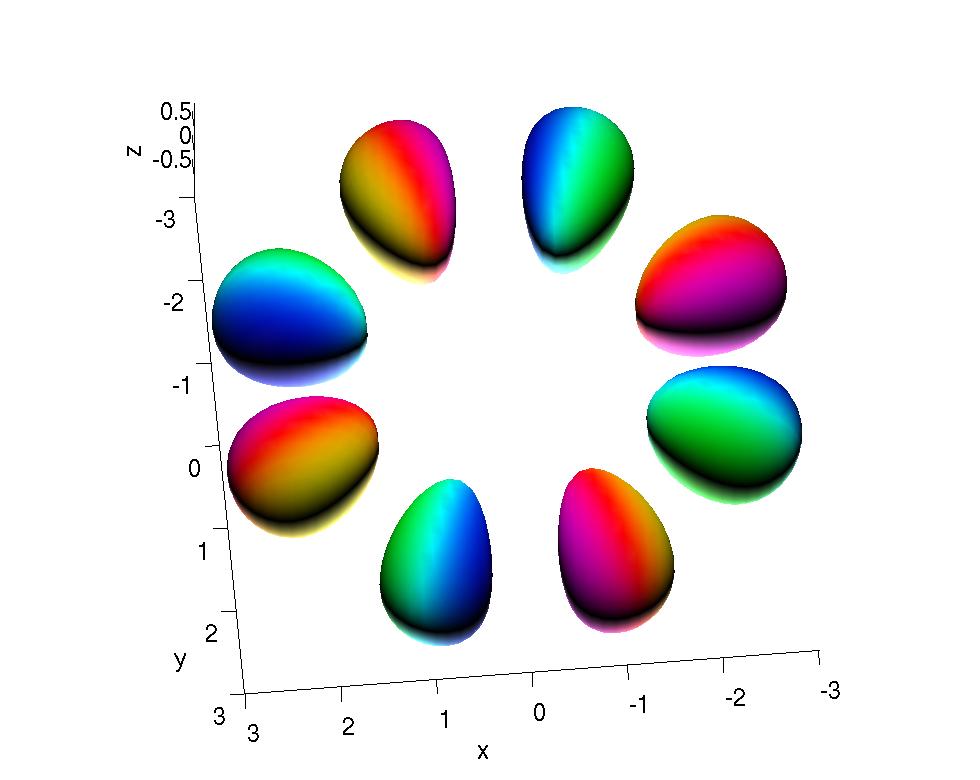}}
\subfloat{\includegraphics[width=0.45\linewidth]{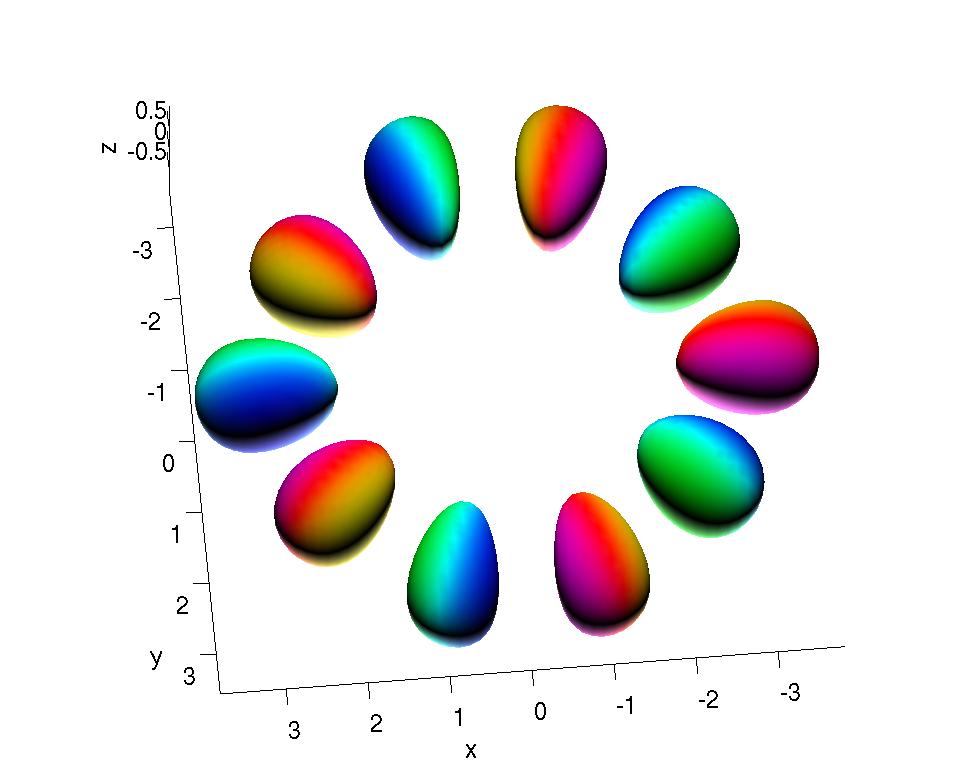}}}
\caption{Isosurfaces showing the half-maximum of the baryon charge
  density in the 2+6 model for baryon numbers $B=2,3,4,5$ with
  $c_2=\tfrac{1}{4}$, $c_6=1$ and fixed mass $m=4$. The color scheme
  is the same as that in Fig.~\ref{fig:M4B1}. 
}
\label{fig:M6B2345}
\end{center}
\end{figure}

We will now consider the 2+6 model and calculate numerical solutions
for the first five baryon numbers, i.e.~$B=2,3,4,5$ ($B=1$ was made in
Sec.~\ref{sec:molecule}). For concreteness, we fix here the parameters
$c_2=\tfrac{1}{4}$, $c_6=1$ and $m=4$, corresponding to a canonical
mass $m^{\rm canonical}=8\sqrt{2}$.
As in the case of the 2+4 model, we find again that the lowest-energy
states take the shapes of rings (beads on rings). 
These solutions are made with the axially-symmetric initial guess
\eqref{eq:axial_symmetric_initial_guess}. 
In Fig.~\ref{fig:M6B2345} are shown the isosurfaces of the baryon
charge densities at their respective half-maximum values of the
numerical solutions. 
The numerically integrated baryon charge and total energy per unit
baryon charge are displayed in Tab.~\ref{tab:M6B2345}. We find as in
the case of the 2+4 model, that all the calculated solutions with
higher-baryon numbers, $B>1$, are stable (as opposed to metastable) 
among this type of configurations.

\begin{table}[!htb]
\begin{center}
\caption{The numerically integrated baryon charge and energy per unit
  baryon charge for higher baryon numbers in the 2+6 model.}
\label{tab:M6B2345}
\begin{tabular}{clc}
$B$ & $B^{\rm numerical}$ & $E^{\rm numerical}/B$\\
\hline\hline
1 & $0.99986$ & $61.72(8)$\\
2 & $1.9998$ & $59.2(7)$\\
3 & $2.9990$ & $59.1(6)$\\
4 & $3.9985$ & $59.0(7)$\\
5 & $4.9981$ & $58.9(1)$
\end{tabular}
\end{center}
\end{table}

\begin{figure}[!tbp]
\begin{center}
\captionsetup[subfloat]{labelformat=empty}
\mbox{
\subfloat{\includegraphics[width=0.4\linewidth]{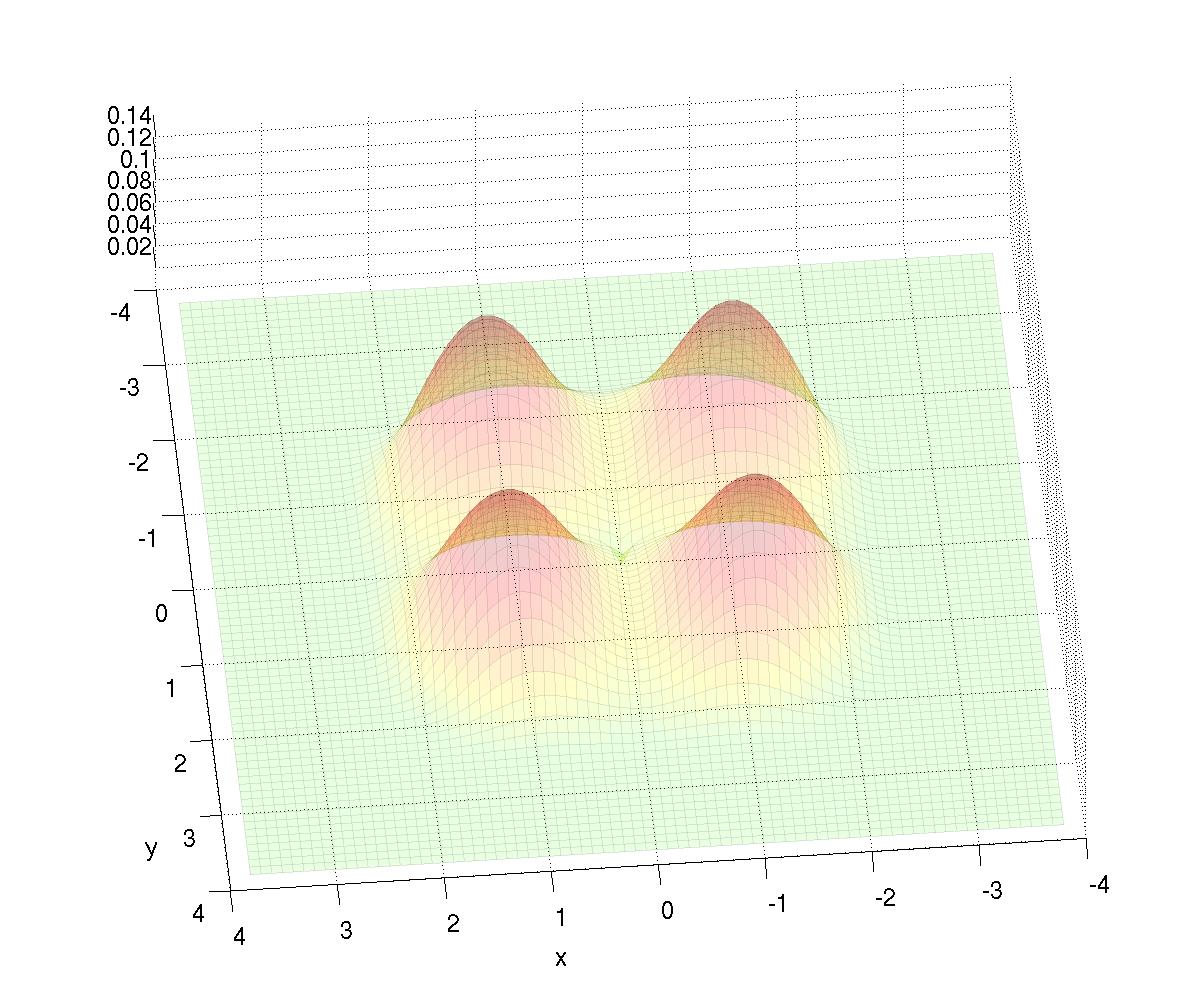}}
\subfloat{\includegraphics[width=0.4\linewidth]{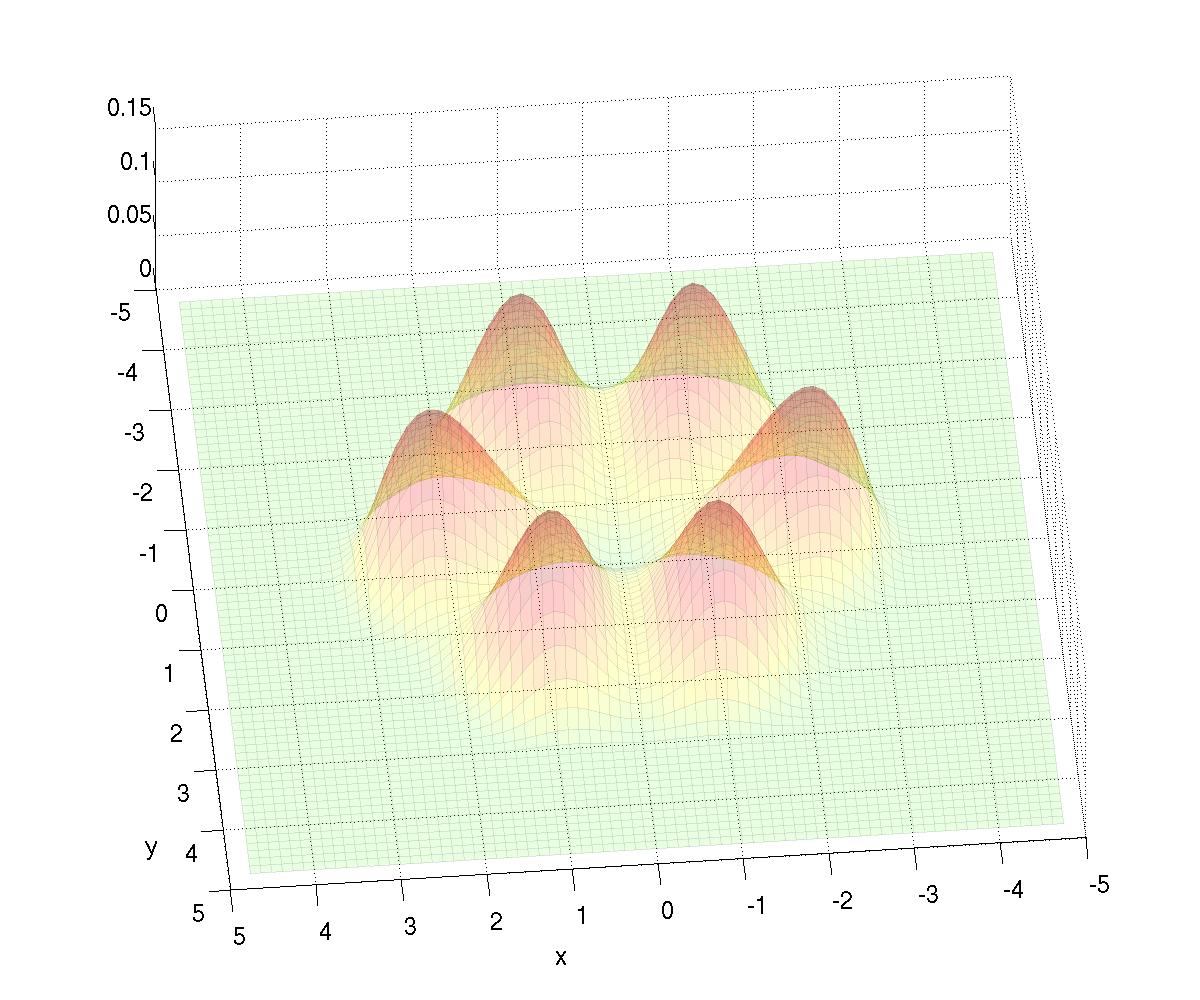}}}
\mbox{
\subfloat{\includegraphics[width=0.4\linewidth]{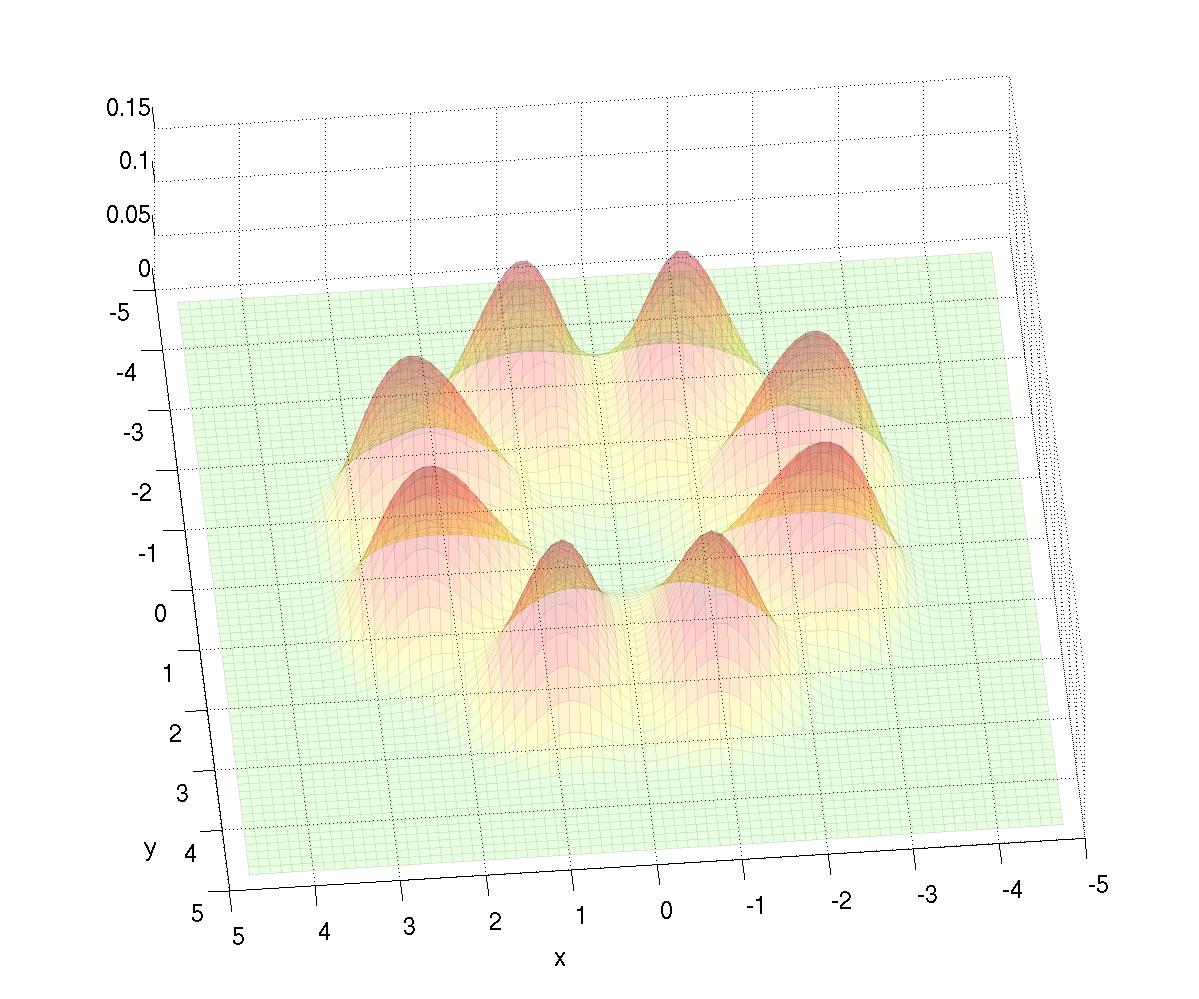}}
\subfloat{\includegraphics[width=0.4\linewidth]{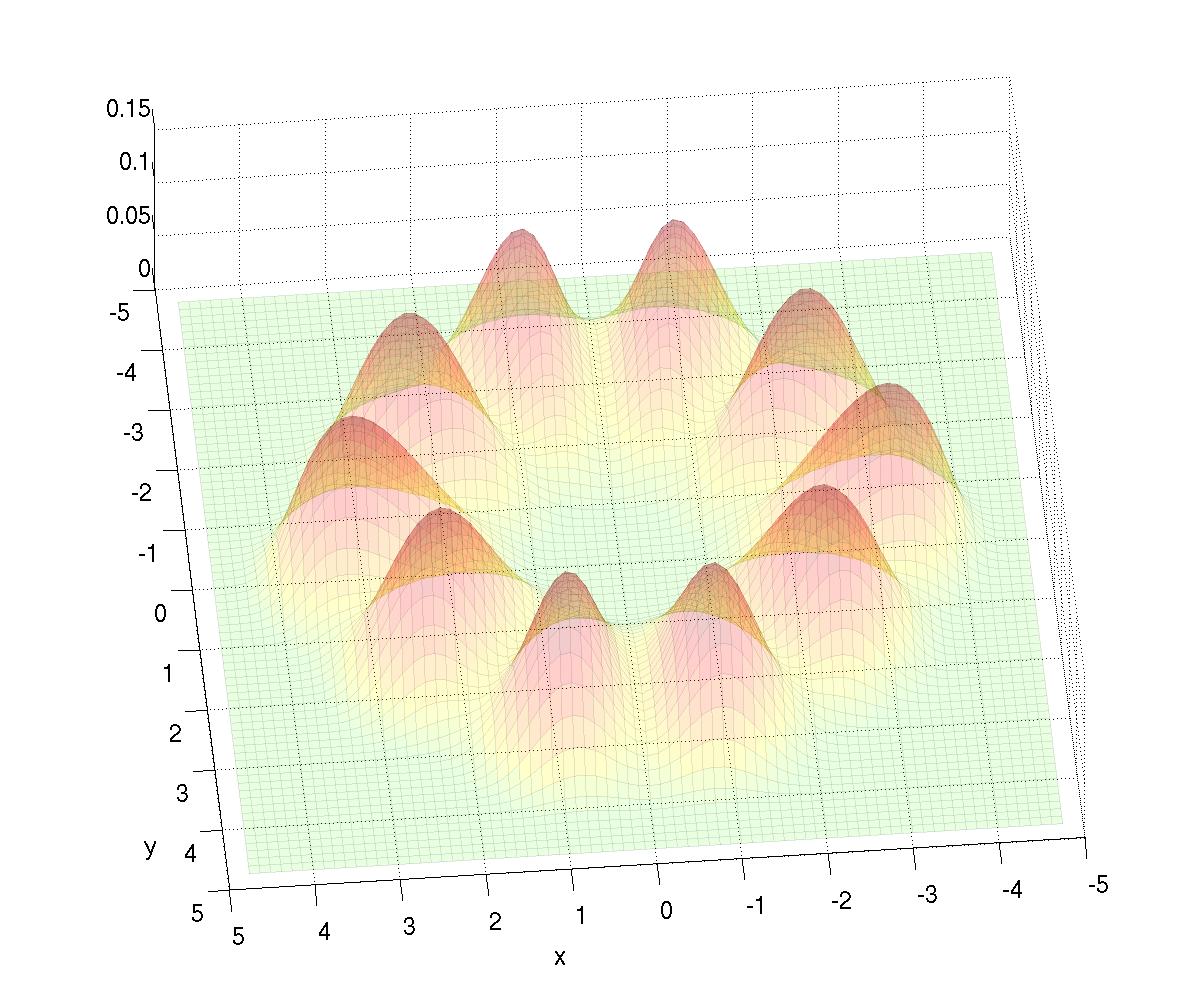}}}
\caption{Baryon charge density at a spatial slice through the
  $B=2,3,4,5$ molecules at $z=0$ in the 2+6 model for various
  choices of $(c_2,c_6)$ and for fixed mass $m=4$. 
} 
\label{fig:M6B2345_baryonslice}
\end{center}
\end{figure}

\begin{figure}[!tbp]
\begin{center}
\captionsetup[subfloat]{labelformat=empty}
\mbox{
\subfloat{\includegraphics[width=0.4\linewidth]{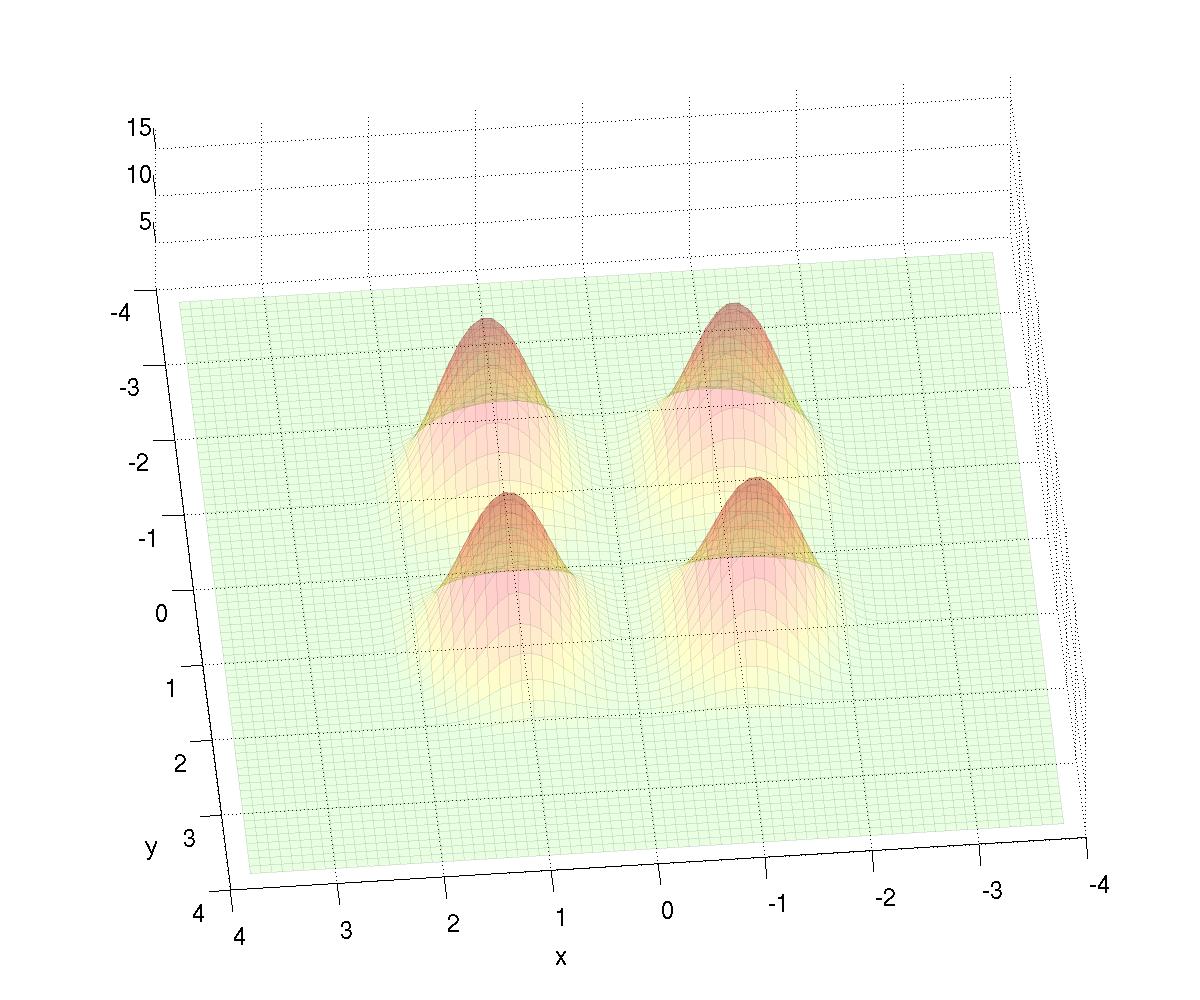}}
\subfloat{\includegraphics[width=0.4\linewidth]{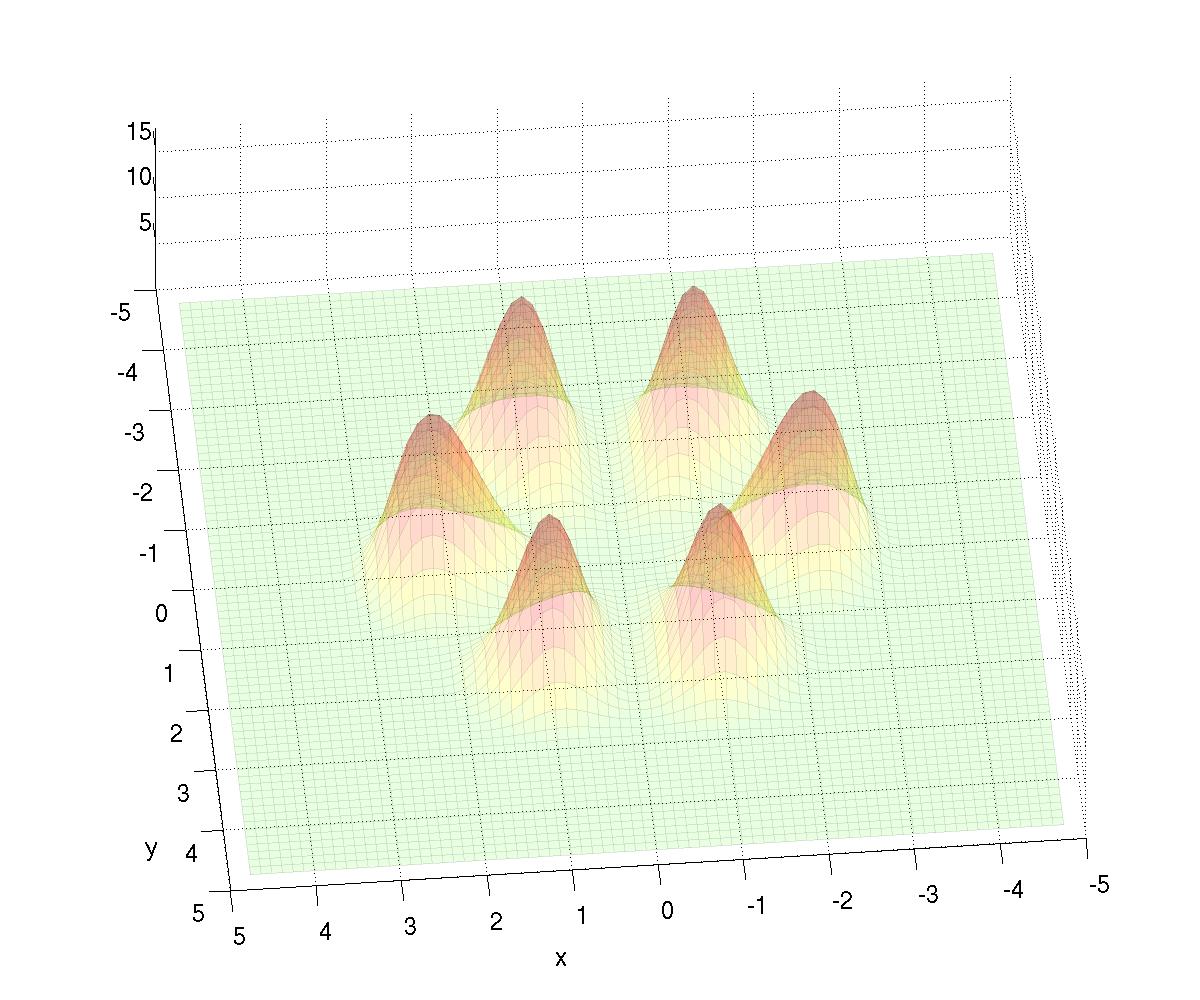}}}
\mbox{
\subfloat{\includegraphics[width=0.4\linewidth]{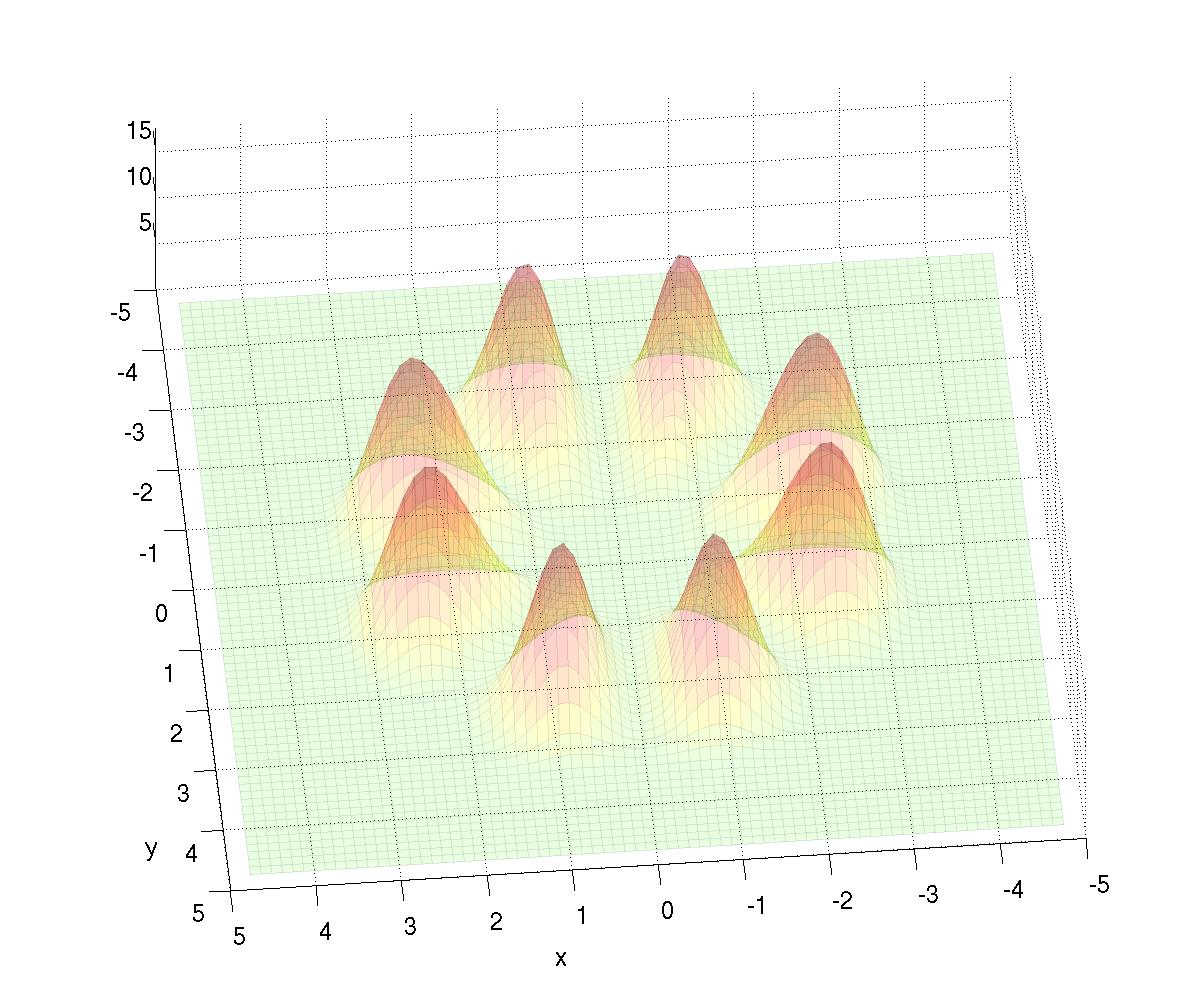}}
\subfloat{\includegraphics[width=0.4\linewidth]{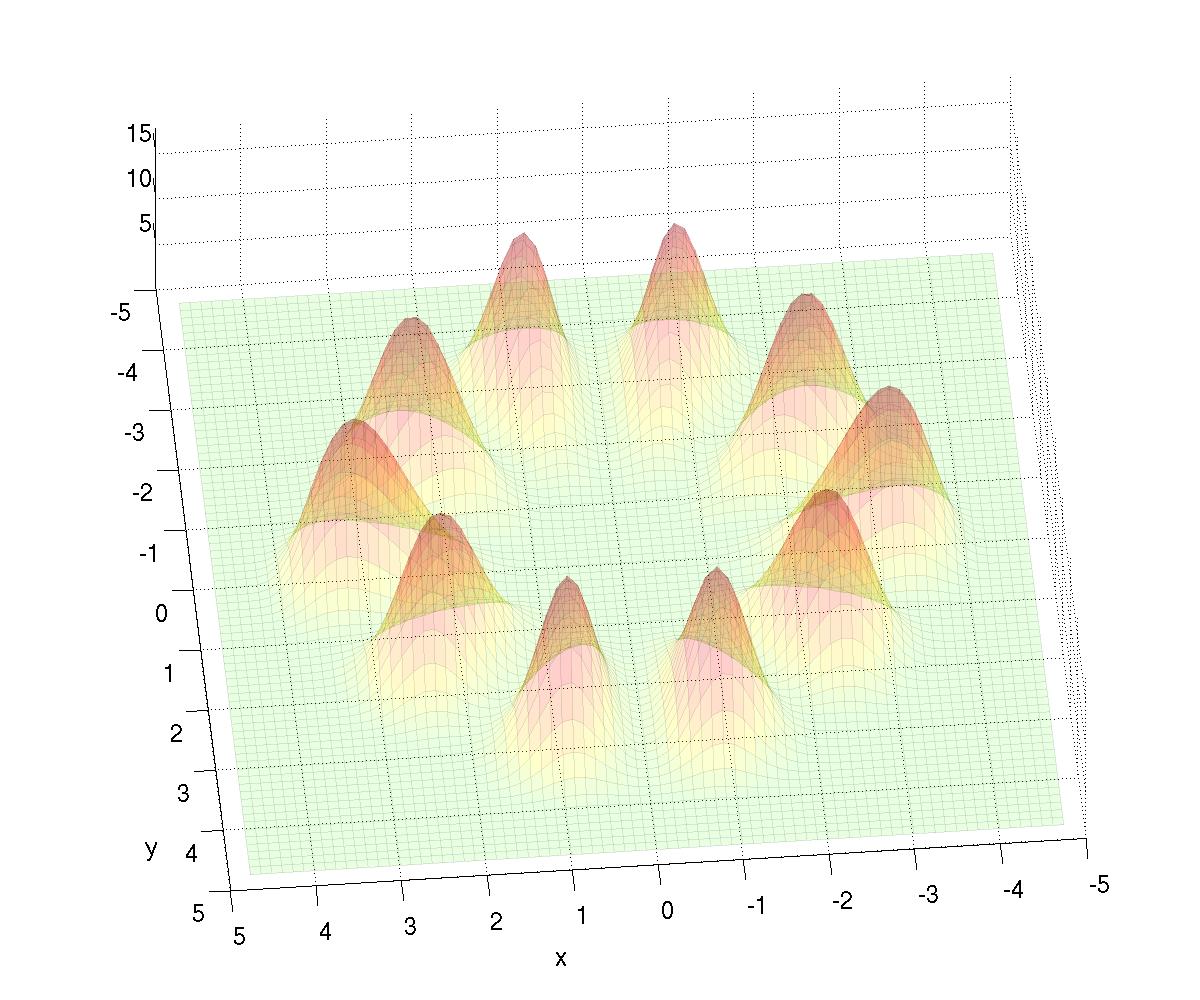}}}
\caption{Energy density at a spatial slice through the
  $B=2,3,4,5$ molecules at $z=0$ in the 2+6 model for various
  choices of $(c_2,c_6)$ and for fixed mass $m=4$. 
} 
\label{fig:M6B2345_energyslice}
\end{center}
\end{figure}

The cross sections at $z=0$ of the baryon charge densities and energy
densities are shown in Fig.~\ref{fig:M6B2345_baryonslice} and
\ref{fig:M6B2345_energyslice}, respectively. 
Note that again the molecular shape is slightly more pronounced in the 
energy density than in the baryon charge density, viz.~the depth of
the valleys between the peaks are deeper. In fact, the slices of the 
energy densities suggest that the beads are not spatially
connected. This is, however, not true. From the slices of baryon
charge densities, we see that the beads are connected and from
Tab.~\ref{tab:M6B2345}, we can see that there is in fact a small
binding energy. Asymptotically, however, for $B\to\infty$ the binding 
energy may go to zero (in the 2+6 model for the given parameters). 

%%%%%%%%%%%%%%%%%%%%%%%%%%
\section{Molecules with unequal fractions \label{sec:fractional}}

In this section, we show that by a modification of the potential
\eqref{eq:pot}, we can create a molecule, still with two components,
but with unevenly distributed baryon charge. The modified potential is 
\beq
V = \frac{1}{2} m^2 (n_4 - c)^2,
\eeq
giving rise to a molecule with two components with localized baryon
charge $(1+c)/2$ and $(1-c)/2$, respectively. 
This potential thus requires different asymptotic boundary conditions
and hence different initial guesses. Since we consider only the $B=1$
sector here, a modified hedgehog Ansatz is sufficient
\begin{equation}
\mathbf{n} = \left(
  c\, \hat{x}\sin f(r) - \sqrt{1-c^2}\cos f(r),
  \hat{y}\sin f(r),
  \hat{z}\sin f(r),
  c \cos f(r) + \sqrt{1-c^2}\hat{x}\sin f(r)
\right),
\end{equation}
which satisfies the vacuum equation at $r\to\infty$ provided
$f(\infty)=0$. 

\begin{figure}[!tbp]
\begin{center}
\mbox{
\subfloat[isosurface]{\includegraphics[width=0.33\linewidth]{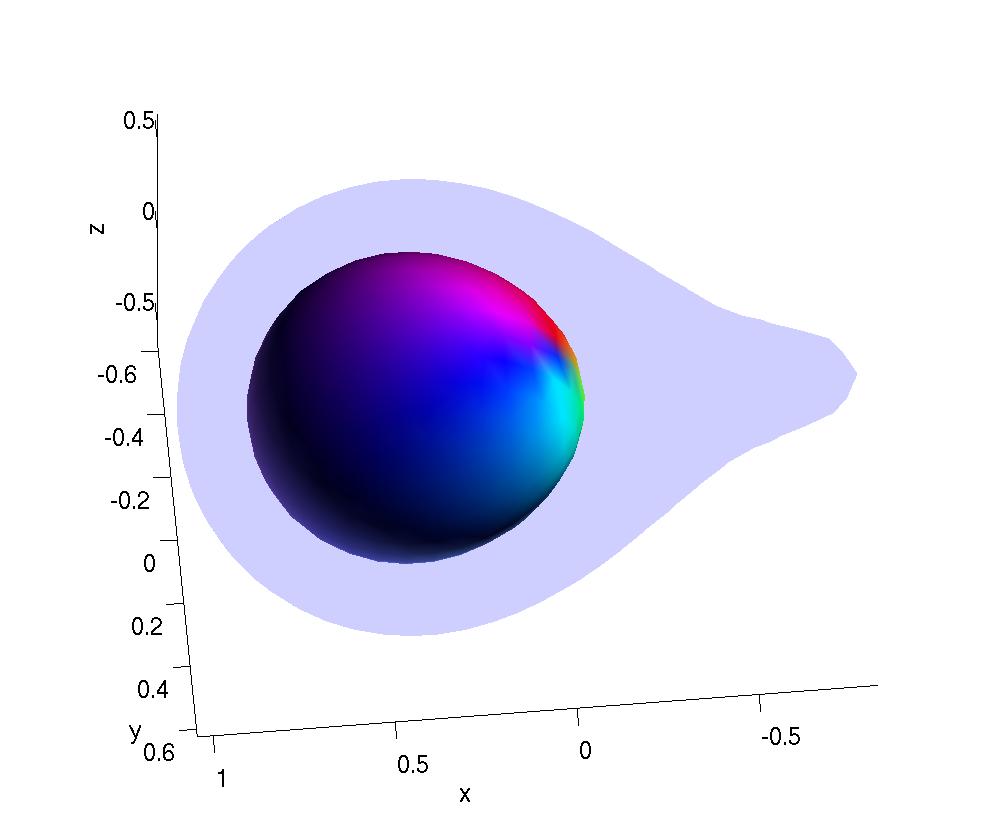}}
\subfloat[baryon charge density]{\includegraphics[width=0.33\linewidth]{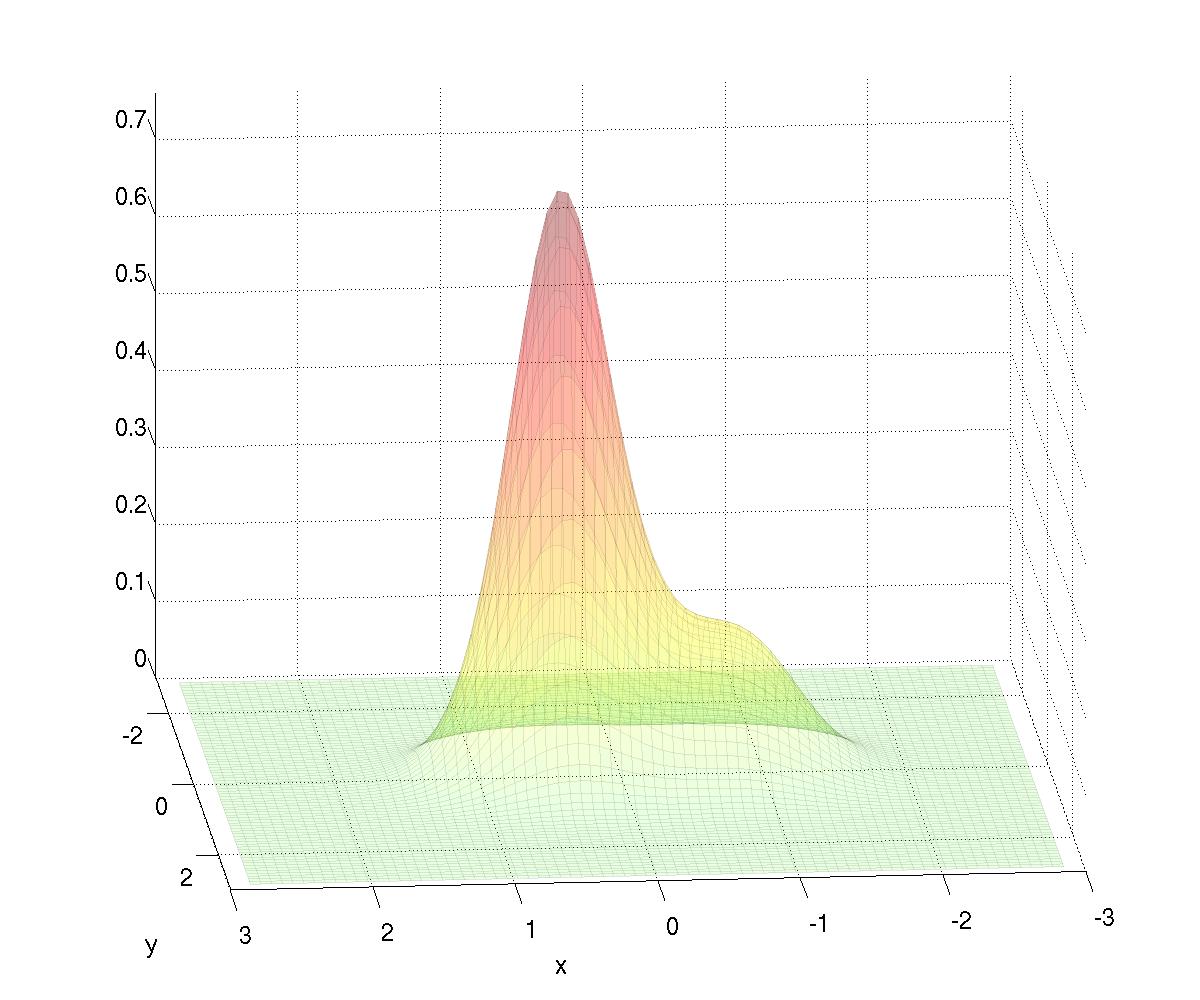}}
\subfloat[energy density]{\includegraphics[width=0.33\linewidth]{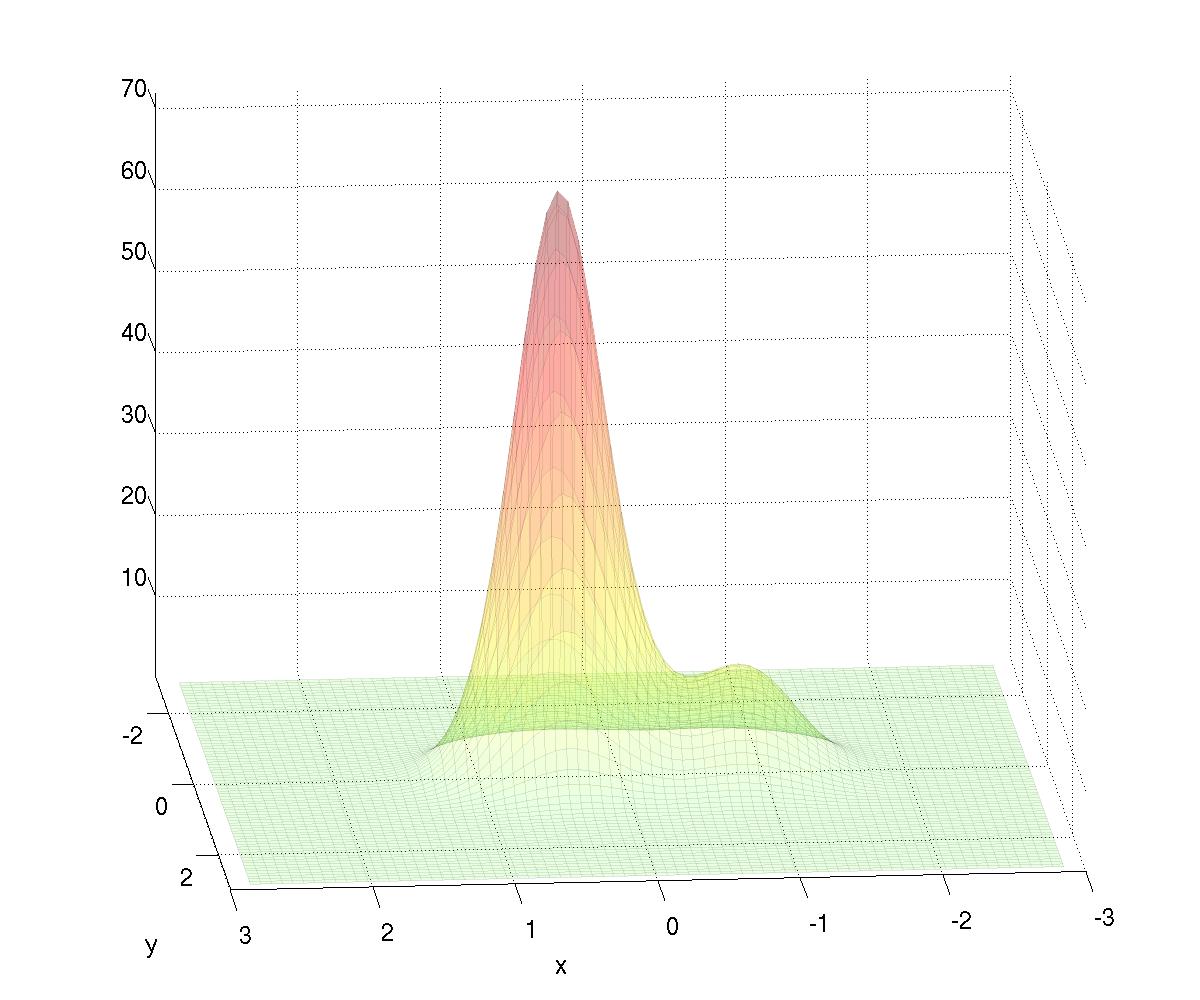}}}
\caption{Unequal fractional molecule with $c=\tfrac{1}{3}$ giving a
  molecule in the 2+4 model with two components of charge
  $\tfrac{2}{3}$ and $\tfrac{1}{3}$, respectively. 
  (a) shows the isosurface of the baryon charge density at
  half-maximum value and there is an added shadow which is an
  isosurface at one quarter of the maximum value. (b) and (c) show
  $xy$-slices (at $z=0$) of the baryon charge density and the energy
  density, respectively. The parameters are $c_2=\tfrac{1}{4}$,
  $c_4=1$ and $m=4$.
  The color scheme is the same as that in Fig.~\ref{fig:M4B1}.
  The numerically integrated
  baryon charge is $B^{\rm numerical}=0.99974$. } 
\label{fig:M4B1c13}
\end{center}
\end{figure}

\begin{figure}[!tbp]
\begin{center}
\mbox{
\subfloat[isosurface]{\includegraphics[width=0.33\linewidth]{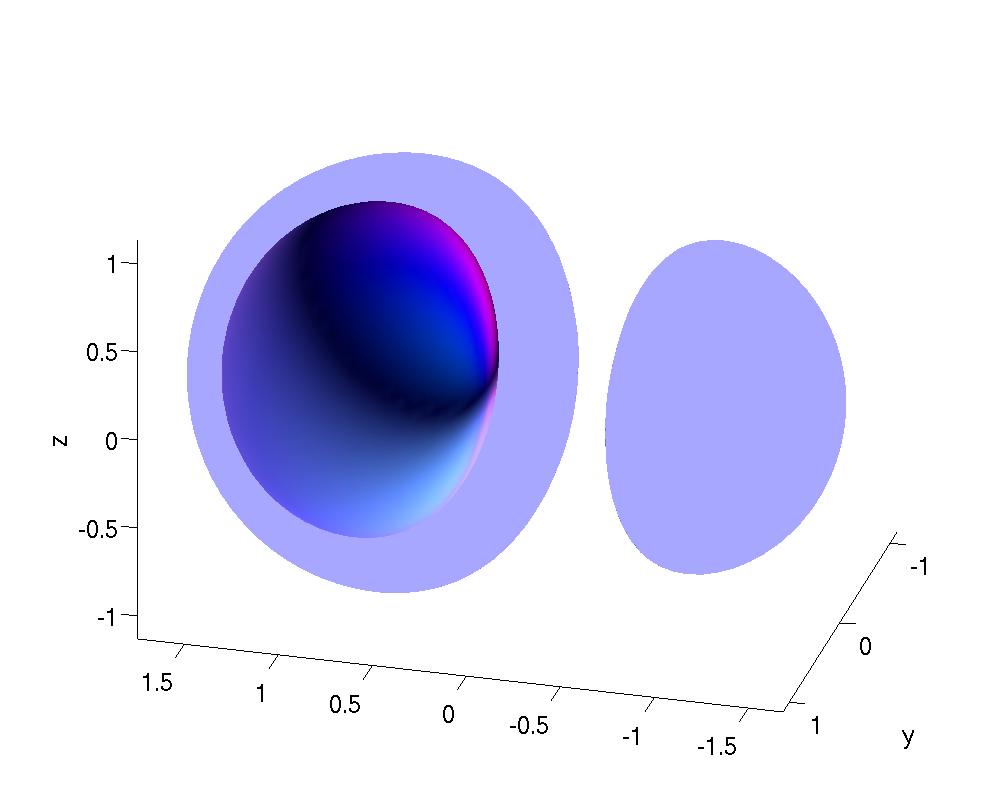}}
\subfloat[baryon charge density]{\includegraphics[width=0.33\linewidth]{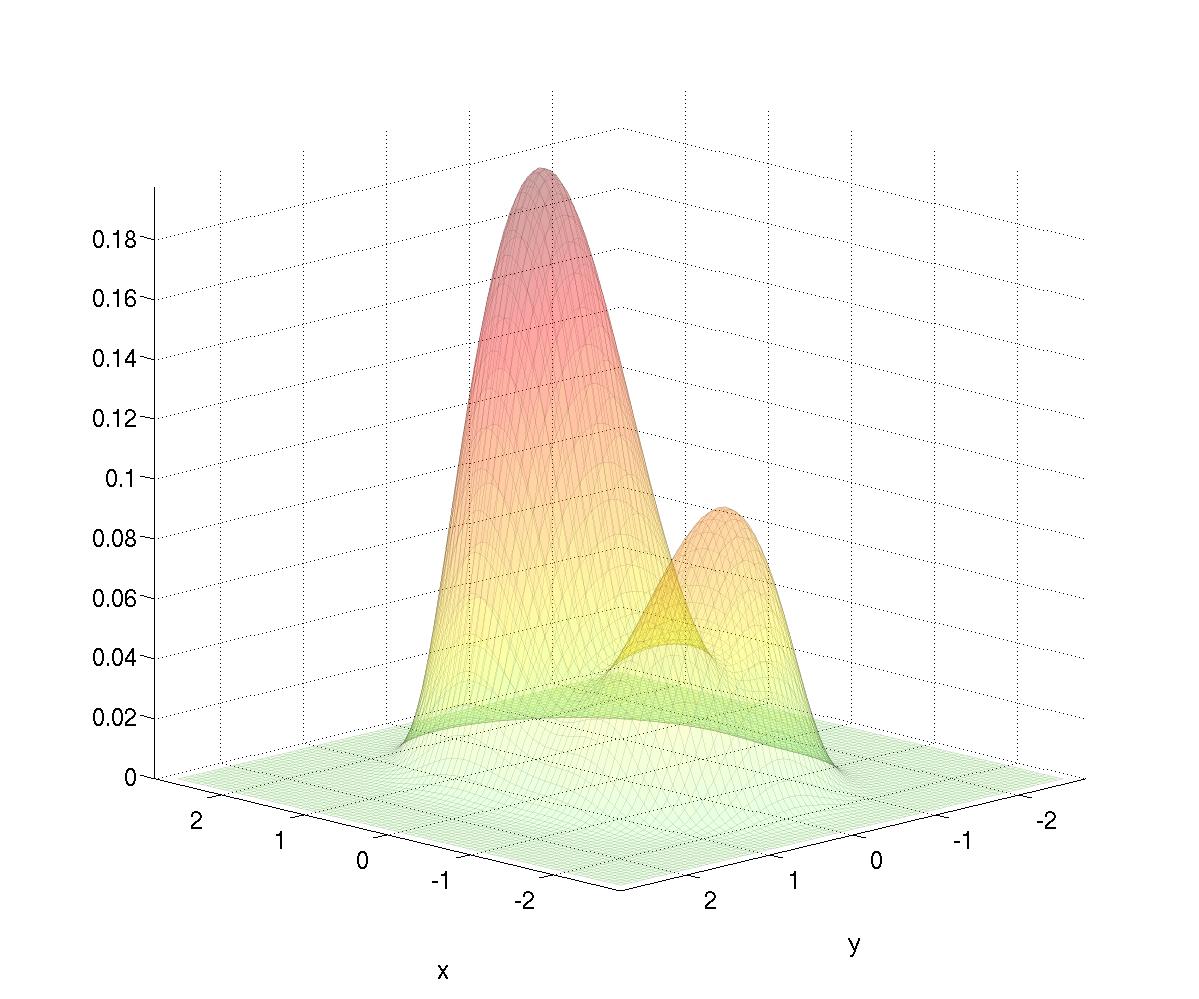}}
\subfloat[energy density]{\includegraphics[width=0.33\linewidth]{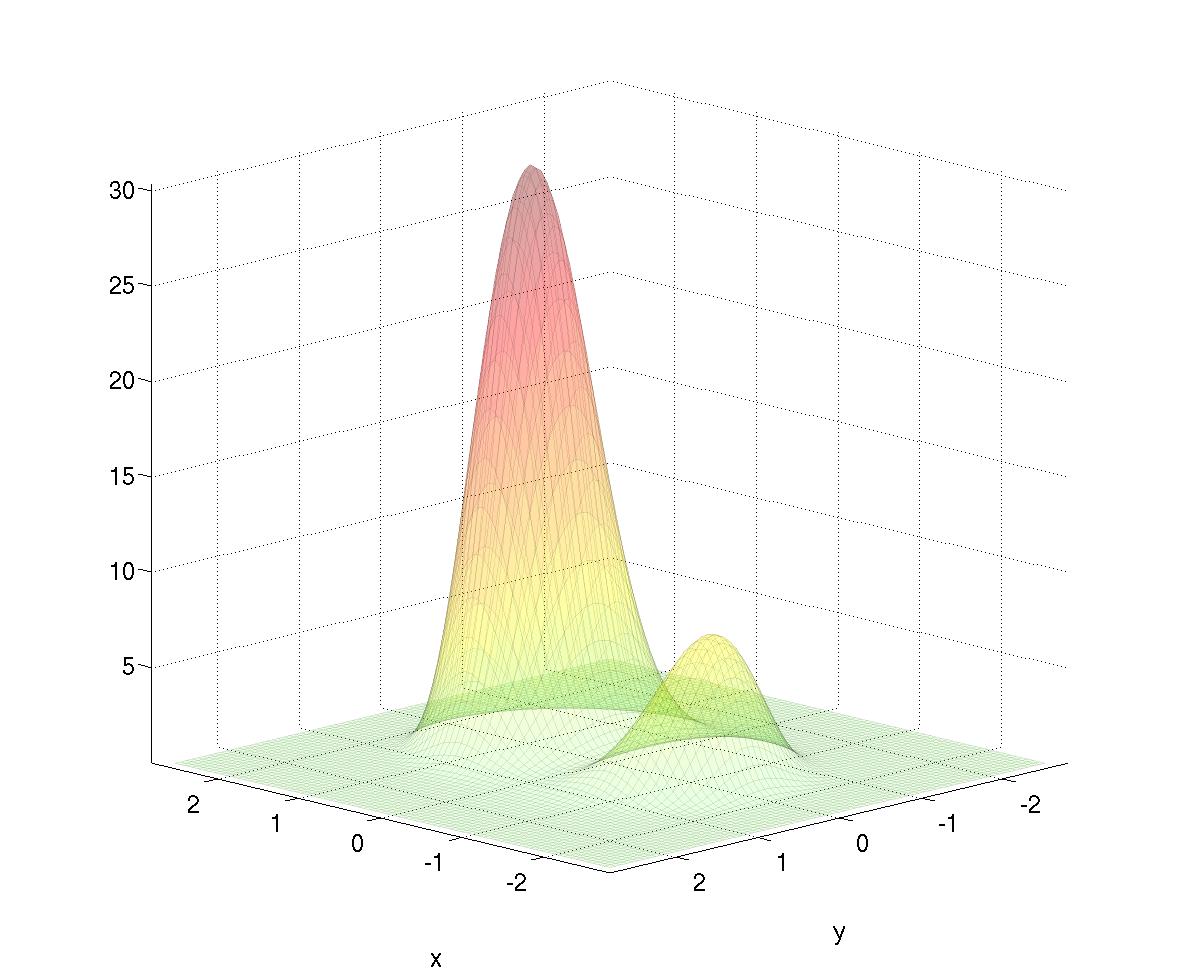}}}
\caption{Unequal fractional molecule with $c=\tfrac{1}{3}$ giving a
  molecule in the 2+6 model with two components of charge
  $\tfrac{2}{3}$ and $\tfrac{1}{3}$, respectively. 
  (a) shows the isosurface of the baryon charge density at
  half-maximum value and there is an added shadow which is an
  isosurface at one quarter of the maximum value. (b) and (c) show
  $xy$-slices (at $z=0$) of the baryon charge density and the energy
  density, respectively. The parameters are $c_2=\tfrac{1}{4}$,
  $c_6=1$ and $m=4$.
  The color scheme is the same as that in Fig.~\ref{fig:M4B1}.
  The numerically integrated
  baryon charge is $B^{\rm numerical}=0.99995$. } 
\label{fig:M6B1c13}
\end{center}
\end{figure}

In the following we will choose a concrete example, setting
$c=\tfrac{1}{3}$, which is interesting, because the two components of
the baryon should contain $\tfrac{2}{3}$ and $\tfrac{1}{3}$ of the
unit baryon charge, respectively. 
For concreteness, we will keep the parameters that we have used in
Sec.~\ref{sec:higher}, namely: $c_2=\tfrac{1}{4}$, $c_4=1$ ($c_6=1$)
and $m=4$ in the case of the 2+4 (2+6) model. The numerical solutions
are shown in Figs.~\ref{fig:M4B1c13} and \ref{fig:M6B1c13}.

%%%%%%%%%%%%%%%%%%%%%
\section{Summary and Discussion \label{sec:summary} }

We have constructed fractional Skyrmions and their 
molecules in the Skyrme model 
with the potential term, $V=m^2n_4^2$. 
As for higher-derivative terms, we have considered 
the conventional fourth-order derivative term, viz.~the Skyrme term 
or the sixth-order derivative term being the baryon number current
squared.
One molecule 
consists of a pair of a global monopole and anti-monopole, each 
with half a baryon number.
Since an isolated global monopole 
has divergent energy in an infinite space,
the constituents are confined. 
We have also constructed an isolated fractional Skyrmion 
as a global monopole (which thus has a divergent total energy). 
We have then constructed Skyrmion solutions with 
higher baryon numbers up to $B=6$, 
and have found that configurations in the form of 
beads on rings are energetically stable.
We have also found other metastable configurations
for $B=3$ and $B=5$, but exhausting all possible metastable
configurations remains as a future problem.
Finally by considering the potential term 
$V\sim m^2 (n_4 -c)^2$, 
we have found that 
fractional Skyrmions have 
baryon numbers that are not equal to one half.
As an example, 
we have constructed a molecule with 
fractional Skyrmions with the baryon numbers $1/3 + 2/3$.

We have shown that our choice of the potential \eqref{eq:pot} as an
effective low-energy potential is able to describe the half-Skyrmion
phase which should be present in QCD at high density
\cite{Ma:2013ooa}. 
In Ref.~\cite{Ma:2013ooa}, which studies a holographic model, quite
different from our low-energy effective field theory model, the
$\omega$ and $\rho$ mesons were needed to consistently describe the
high density phase where the half Skyrmions exist.
Whether our simple model with our choice of effective potential is
able to capture the relevant phenomenological features of QCD at high
density is a very interesting and important question that is left as a 
future work.

As a lower dimensional analog, 
there exists a fractional baby-Skyrmion molecule
consisting of a pair of 
a global vortex and anti-vortex with half $\pi_2$ charges 
in the O(3) model with
the XY (or easy-plane) potential term $V =m^2 n_3^2$ in $d=2+1$ dimensions.
In this model too,  
baby Skyrmions as beads on 
a ring were found as (meta)stable configurations
\cite{Kobayashi:2013aja,Kobayashi:2013wra}.  
However, the lowest-energy configurations 
of higher topological numbers are of the form of  
square lattices of fractional lumps 
\cite{Jaykka:2010bq}.
We expect a cubic lattice of fractional Skyrmions 
for higher baryon numbers 
in our 3+1 dimensional case.

If we add a potential 
$V_2 = m_2^2 n_1$ with $m_2 < m$ 
in the O(3) model with 
the XY-potential term $V =m^2 n_3^2$ in $d=2+1$ dimensions, 
vortices are connected by a sine-Gordon soliton.
Therefore, fractional lumps are linearly confined by 
the sine-Gordon soliton. 
This happens in fact in two-gap (or two-component) superconductors  
described by a Landau-Ginzburg Lagrangian 
in the form of 
a U(1) gauge theory with two charged scalar fields 
$\phi_1$ and $\phi_2$.
The two fields are coupled through 
the Josephson term 
$L_{\rm J} = \gamma \phi_1^* \phi_2$ 
in the case of two-gap superconductors, 
but not for two-component superconductors.
In either case, it admits a  molecule of 
half-quantized vortices \cite{Babaev:2002}.
In the presence of the Josephson term,  
fractional vortices are connected \cite{Goryo:2007} 
by a sine-Gordon kink \cite{Tanaka:2001}.
In the strong gauge-coupling limit, 
the model reduces to an O(3) sigma model 
with the potential term 
$V =m^2 n_3^2$ complemented by 
$V_2 = \gamma n_1$ induced by 
the Josephson term \cite{Nitta:2012xq,Kobayashi:2013ju}, 
and the molecule reduces to 
a fractional baby-Skyrmion molecule
mentioned above \cite{footnote:BEC}.
In the same way,
if we add a potential
that breaks the SO(3) symmetry, possessed by the vacuum, 
such as  $V_2 = m_2^2 n_1$ $(m_2 < m)$,
fractional Skyrmions constituting a molecule 
will be connected by 
a baby-Skyrmion string, 
realizing linear confinement 
of fractional Skyrmions.

The Bogomol'nyi-Prasad-Sommerfield (BPS) Skyrme model, proposed
recently \cite{Adam:2010fg}, consists of only the sixth-order
derivative term as well as appropriate potentials. This model admits
exact solutions with compact support. By choosing the potential in
this paper, we may be able to construct exact solutions of a
fractional Skyrmion molecule.

Fractional Skyrmions  
in the O(4) model (or the Skyrme model) 
on ${\mathbb R}^2\times S^1$
were discussed in Ref.~\cite{Nitta:2014vpa},
in which our potential term is related to 
the boundary condition where the field $n_4$ changes 
the sign along $S^1$.
A rather different origin of fractional topological charge was also
found for vortices and lumps \cite{Eto:2009bz}. 
A unified understanding of fractional topological charges 
will be an important future work.

\section*{Acknowledgments}

The work of M.~N.~is supported in part by Grant-in-Aid for Scientific Research 
No.~25400268
and by the ``Topological Quantum Phenomena'' 
Grant-in-Aid for Scientific Research 
on Innovative Areas (No.~25103720)  
from the Ministry of Education, Culture, Sports, Science and Technology 
(MEXT) of Japan. 
S.~B.~G.~thanks Keio University for hospitality during which this
project took shape. 
S.~B.~G.~thanks the Recruitment Program of High-end Foreign Experts for 
support.

\begin{appendix}

\section{Rational map Ansatz in the 2+4 model\label{app:rational_map_2+4}}

\begin{figure}[!tb]
\begin{center}
\captionsetup[subfloat]{labelformat=empty}
\mbox{
\subfloat{\includegraphics[width=0.45\linewidth]{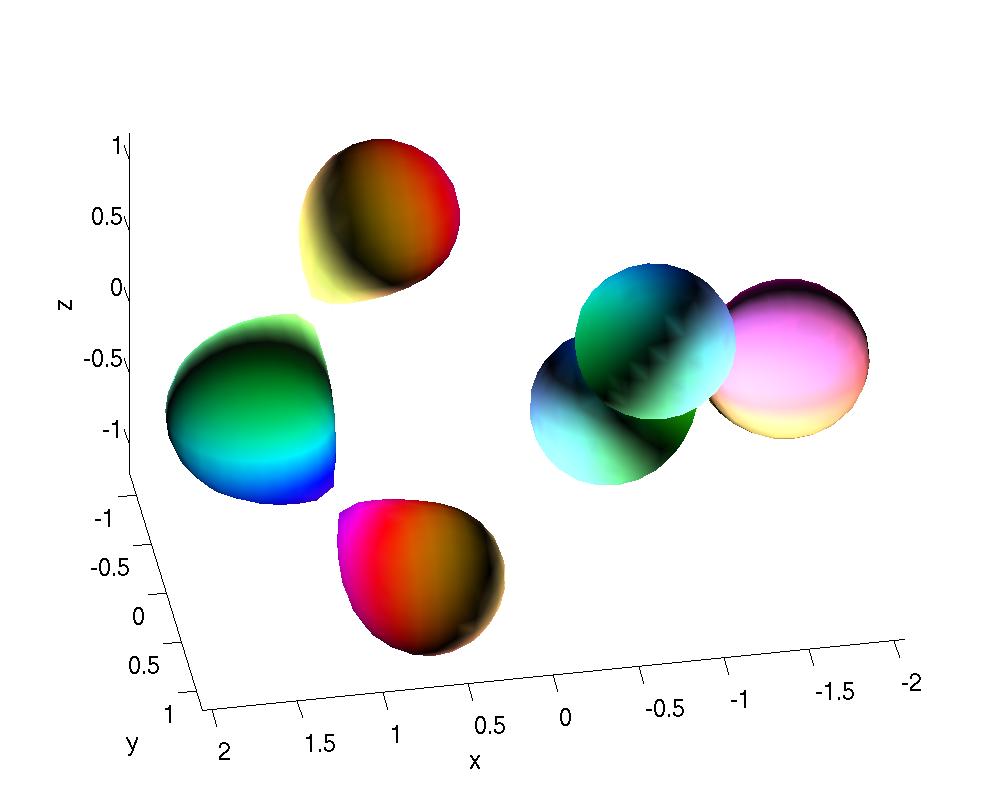}}
\subfloat{\includegraphics[width=0.45\linewidth]{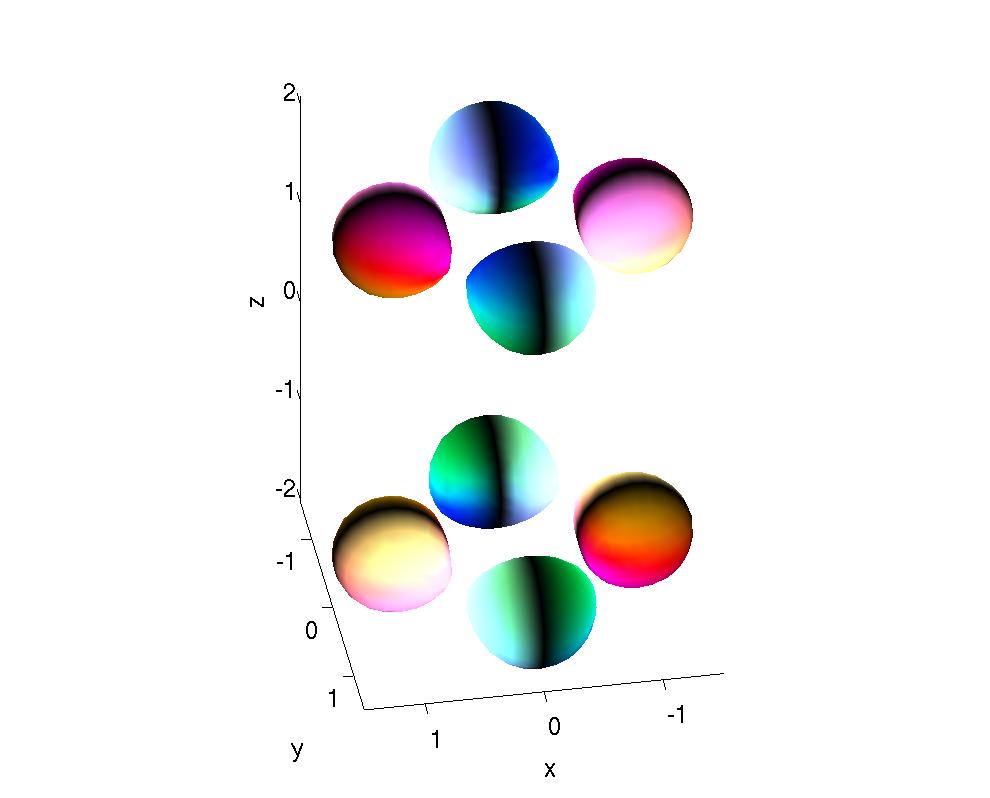}}}
\mbox{
\subfloat{\includegraphics[width=0.45\linewidth]{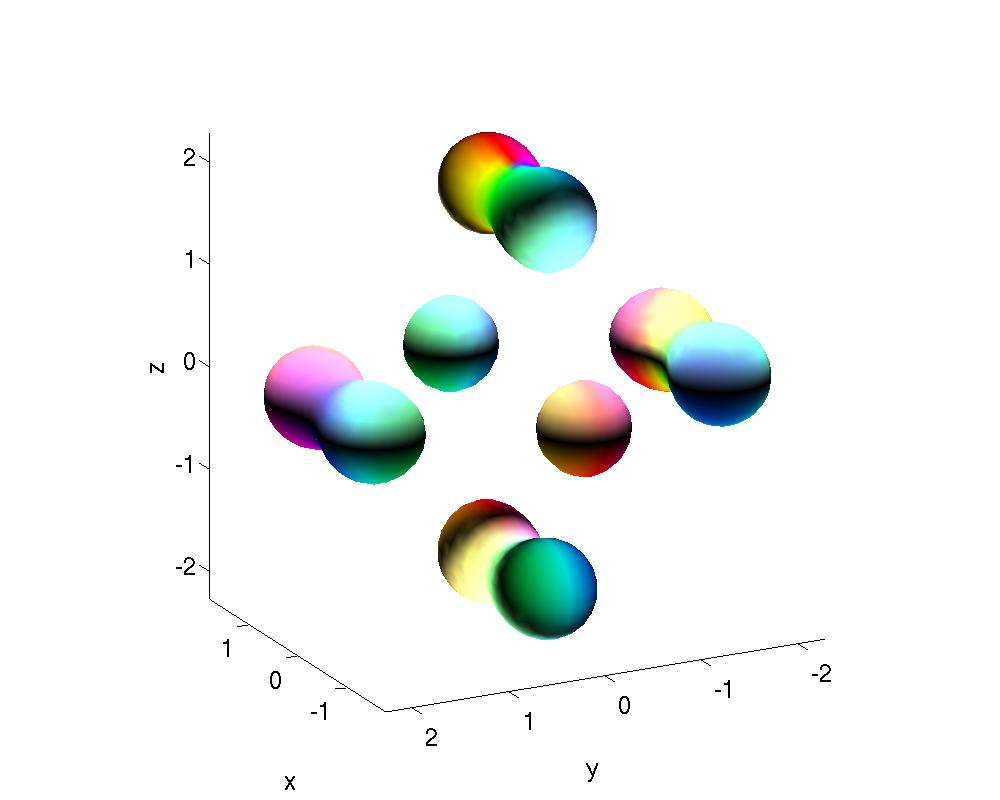}}}
\caption{Isosurfaces showing the half-maximum of the baryon charge
  density in the 2+4 model for baryon numbers $B=3,4,5$ with
  $c_2=\tfrac{1}{4}$, $c_4=1$ and fixed mass $m=4$. These numerical
  solutions have been obtained using the rational-map Ansatz
  \eqref{eq:rational_map_initial_guess} as initial guesses. The color
  scheme is the same as that in Fig.~\ref{fig:M4B1}. 
} 
\label{fig:M4B345_rational_map}
\end{center}
\end{figure}

In this section we investigate numerical solutions with different
initial guesses to be used by the relaxation method, in particular, we
consider the rational-map Ansatz \eqref{eq:rational_map_initial_guess}
with an appropriate rational map $R$. For $B=1$ and $B=2$, the axially
symmetric Ansatz \eqref{eq:axial_symmetric_initial_guess} is a very
educated guess due to the high level of symmetry. For $B=3$ and higher
$B$, without the potential \eqref{eq:pot}, the lowest-energy state
turns out to have a discrete symmetry instead of axial symmetry
\cite{Battye:1997qq,Houghton:1997kg}. As already mentioned, we do not
\emph{a priori} know what symmetry or shapes the minimizer of the
energy may have, and therefore we try the minimizers which are found
to be the lowest-energy states without the potential \eqref{eq:pot} as 
initial guess also here, i.e.~with the potential \eqref{eq:pot}. 
For concreteness, we use the 2+4 model with $c_2=\tfrac{1}{4}$,
$c_4=1$ and $m=4$ and calculate numerical solutions for $B=3,4,5$. The
isosurfaces of the baryon charge density at half-maximum values are
shown in Fig.~\ref{fig:M4B345_rational_map}.

\begin{figure}[!tb]
\begin{center}
\captionsetup[subfloat]{labelformat=empty}
\mbox{
\subfloat[baryon charge density]{\includegraphics[width=0.45\linewidth]{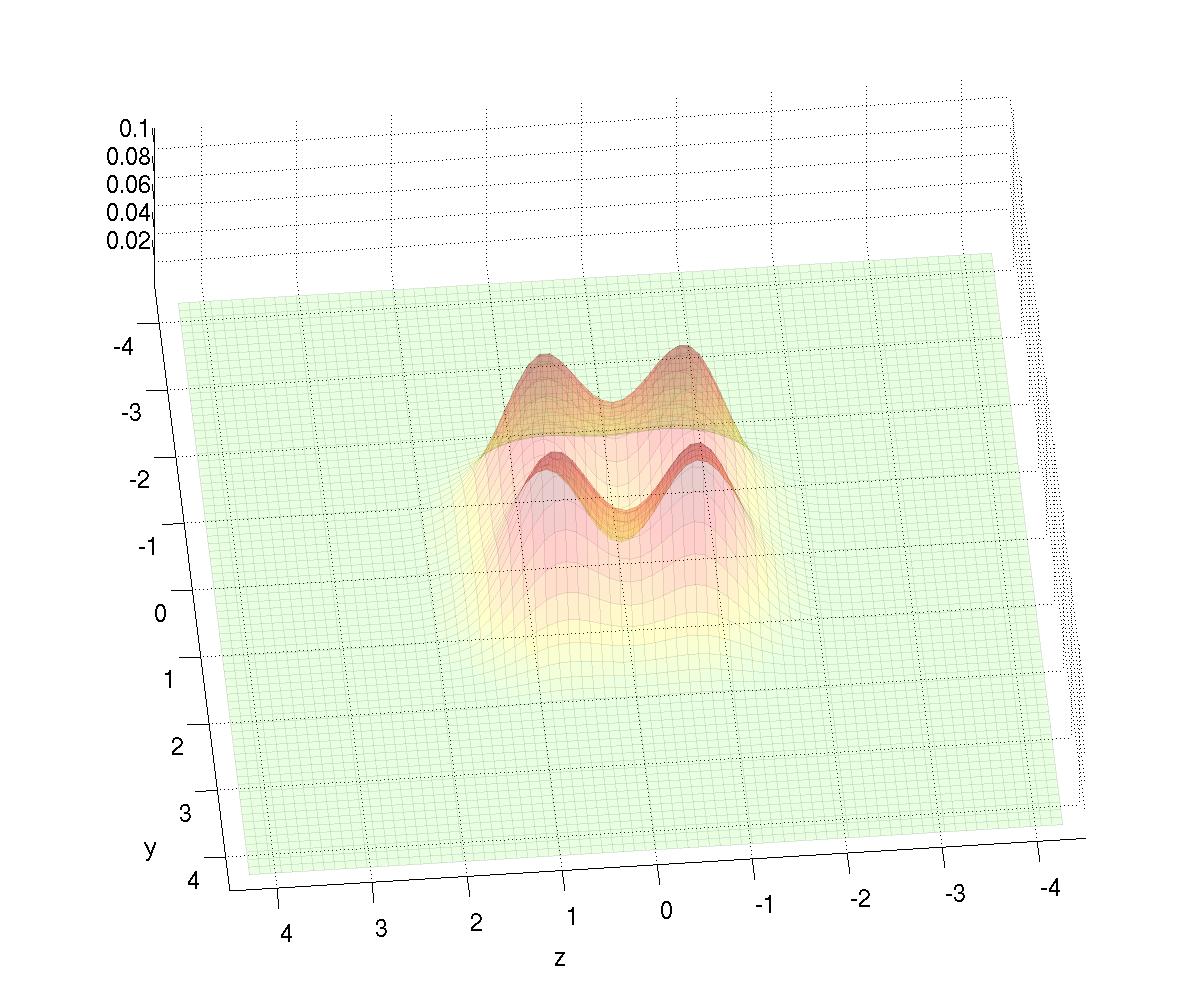}}
\subfloat[energy density]{\includegraphics[width=0.45\linewidth]{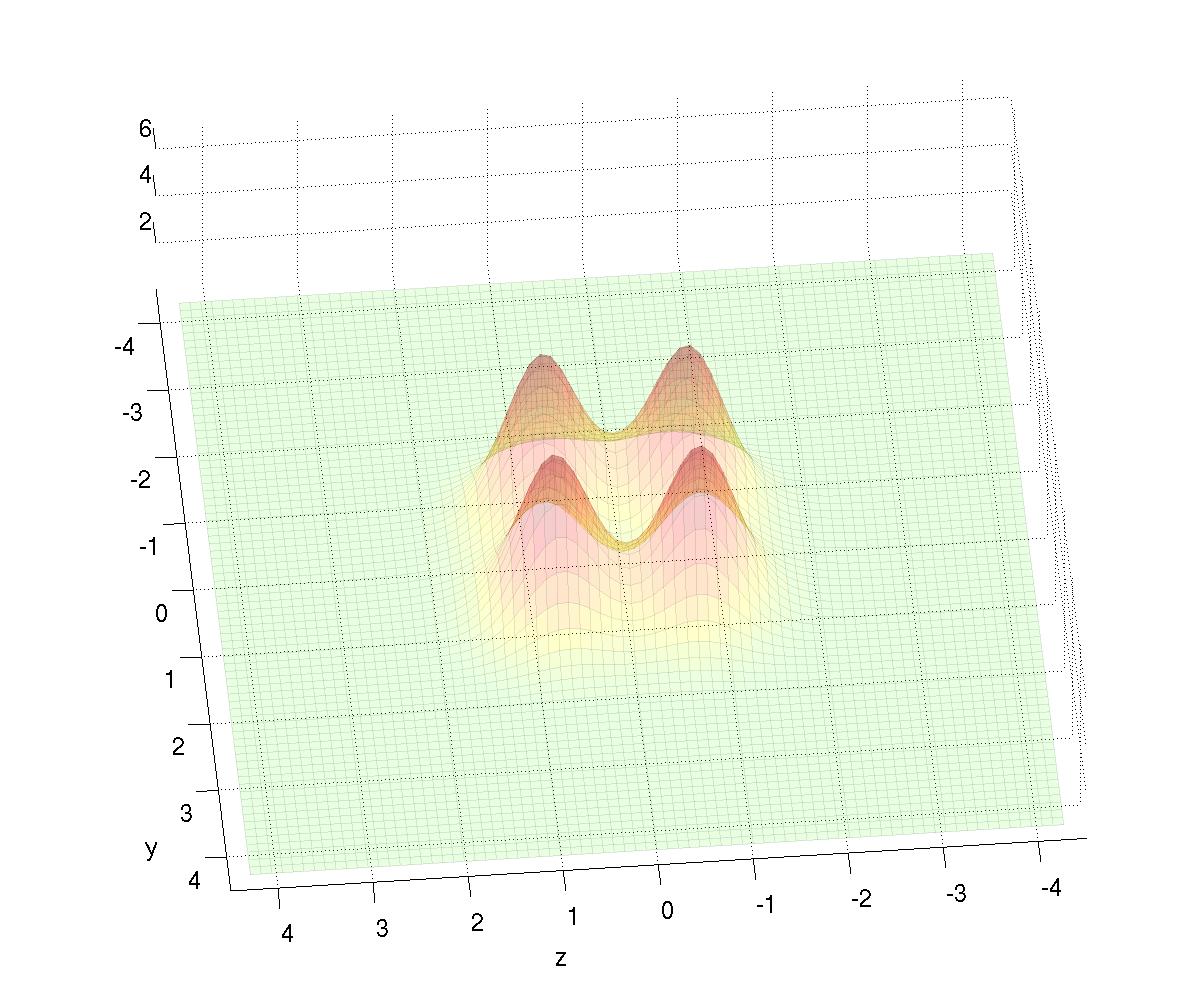}}}
\caption{Cross sections at $x=0$ showing the baryon charge density and
  energy density in the 2+4 model for baryon numbers $B=3$ with
  $c_2=\tfrac{1}{4}$, $c_4=1$ and fixed mass $m=4$. This numerical
  solution has been obtained using the rational-map Ansatz
  \eqref{eq:rational_map_initial_guess} as an initial guess. 
} 
\label{fig:M4B345_rational_map_slices}
\end{center}
\end{figure}

The first case, $B=3$, found using the rational map with a tetrahedral
symmetry turns out to give a numerical solution which is only
metastable, see Tab.~\ref{tab:M4B345_rational_map} for a comparison of
the energies of the numerical solutions. 
In order to demonstrate that the two clusters of three half units of
Skyrme charge are really connected, we display a cross section of the
baryon charge density and energy density at $x=0$ in
Fig.~\ref{fig:M4B345_rational_map_slices}. 
Of course, no half unit of Skyrme charge can be spatially localized
(without a tail) and thus neither can three half units. 
Both the baryon charge density and the energy density on the
$yz$ slices in Fig.~\ref{fig:M4B345_rational_map_slices} have maximum
values around one quarter of their respective global maximum values
(i.e.~in the three-dimensional space). 

\begin{table}[!htb]
\begin{center}
\caption{A comparison of the numerically integrated baryon charge and
  energy per unit baryon charge for higher baryon numbers in the 2+4
  model for different initial guesses. The ${}^\star$ denotes a
  solution that is spatially disconnected.  }
\label{tab:M4B345_rational_map}
\begin{tabular}{llc}
$B$ (initial guess) & $B^{\rm numerical}$ & $E^{\rm numerical}/B$\\
\hline\hline
3 (axially symmetric) & $2.9979$ & $61.0(6)$\\
3 (tetrahedral) & $2.9986$ & $61.1(7)$\\
4 (axially symmetric) & $3.9971$ & $60.8(9)$\\
$4^\star$ (cubic) & $3.9971$ & $61.5(2)$\\
5 (axial) & $4.9964$ & $60.7(9)$\\
5 (octahedral) & $4.9964$ & $61.5(4)$
\end{tabular}
\end{center}
\end{table}

The case, $B=4$, found using the rational map with a cubic symmetry
gave rise to a solution that split up into two $B=2$ axially symmetric
solutions; i.e.~they are both chargewise and energetically
separated. Due to the $B=2$ solution with axial symmetry having a
larger energy per unit baryon charge than the $B=4$ solution, see
Tab.~\ref{tab:M4B23456}, two $B=2$ solutions thus have a higher energy
than the $B=4$ solution being a ring (with beads). 

The last case, i.e.~$B=5$ is like the $B=3$ case spatially connected,
but is made with a rational map having an octahedral symmetry. It is
however energetically only metastable, viz.~it has a higher energy
than the solution made using the axially symmetric Ansatz. 

The bottom line is that the lowest-energy solutions for the molecules
are all found using the axially symmetric Ansatz
\eqref{eq:axial_symmetric_initial_guess}. Using the rational-map
Ansatz \eqref{eq:rational_map_initial_guess}, we have found metastable
solutions for the $B=3,5$ cases, which however are energetically prone
to decay.

\section{Low-mass limit of the 2+4 model\label{app:low_mass_limit}}

\begin{figure}[!tb]
\begin{center}
\captionsetup[subfloat]{labelformat=empty}
\mbox{
\subfloat[$m=0$]{\includegraphics[width=0.24\linewidth]{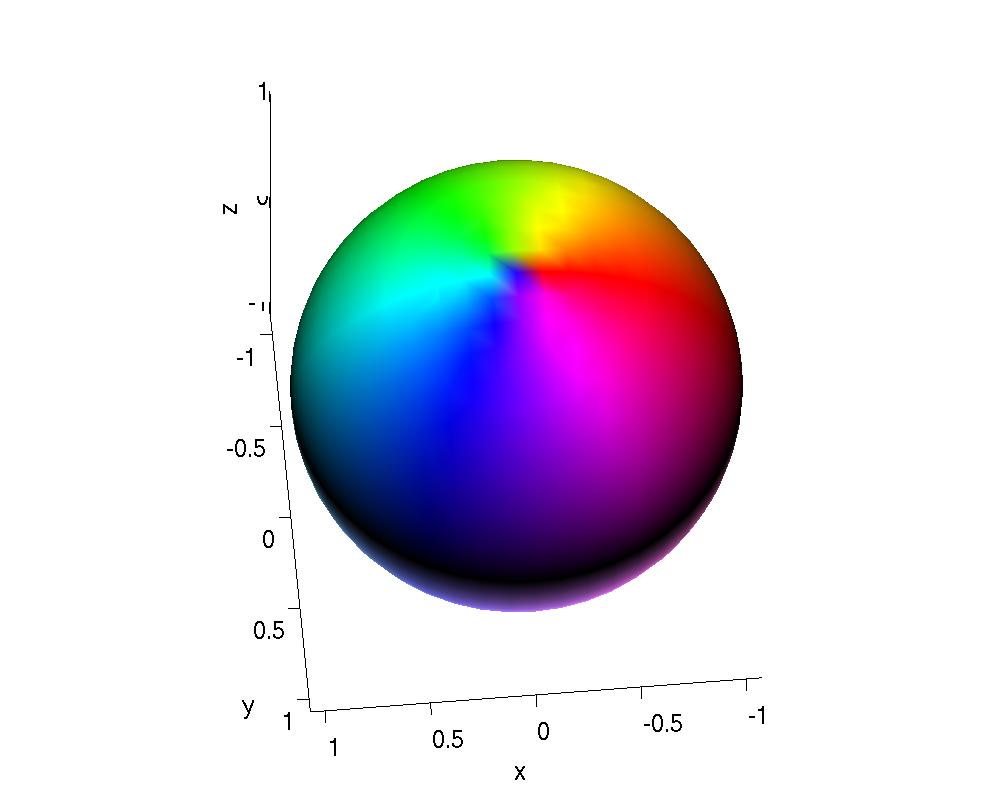}}
\subfloat[$m=\tfrac{1}{2}$]{\includegraphics[width=0.24\linewidth]{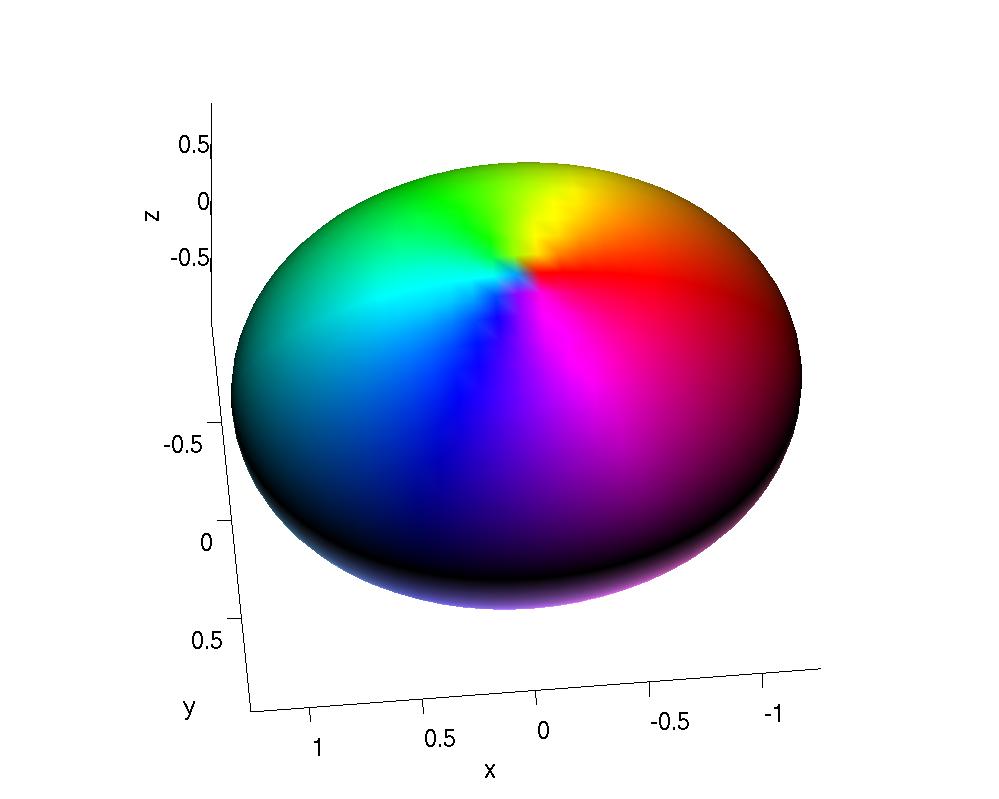}}
\subfloat[$m=1$]{\includegraphics[width=0.24\linewidth]{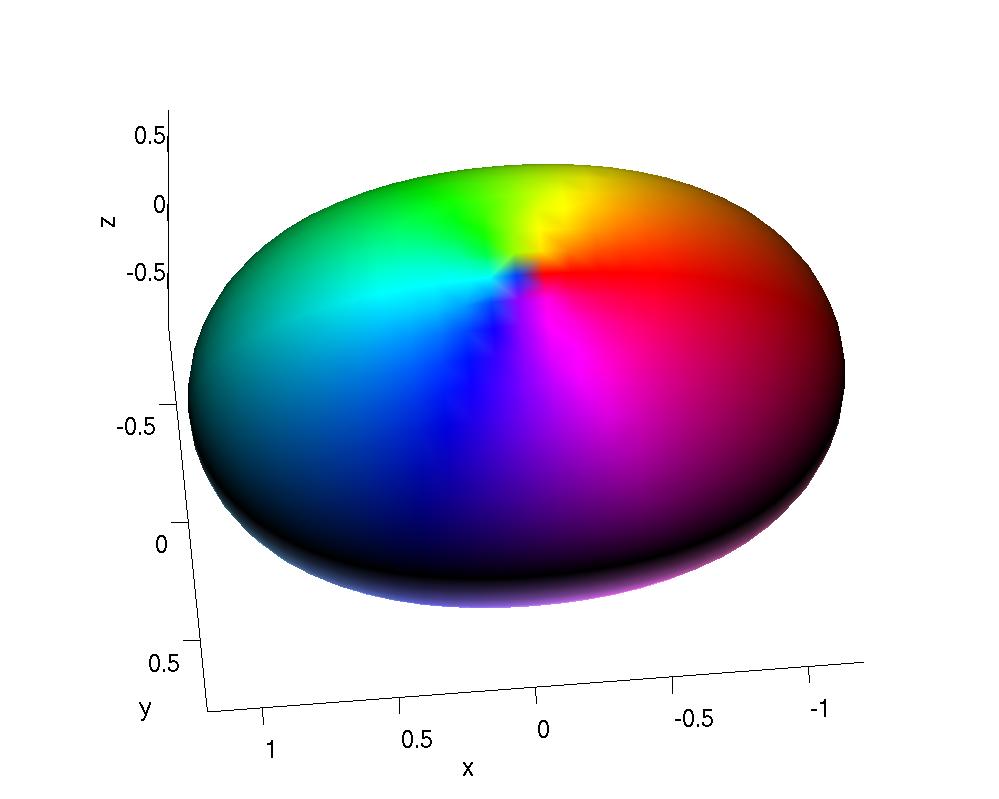}}
\subfloat[$m=\tfrac{3}{2}$]{\includegraphics[width=0.24\linewidth]{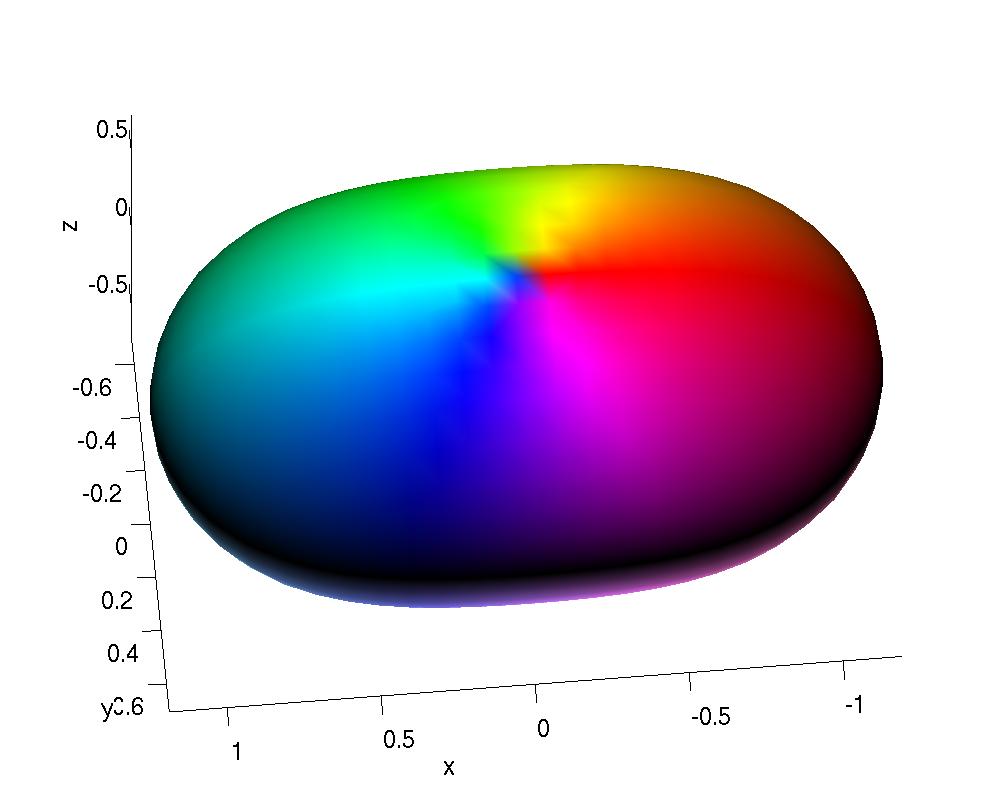}}}
\mbox{
\subfloat[$m=2$]{\includegraphics[width=0.24\linewidth]{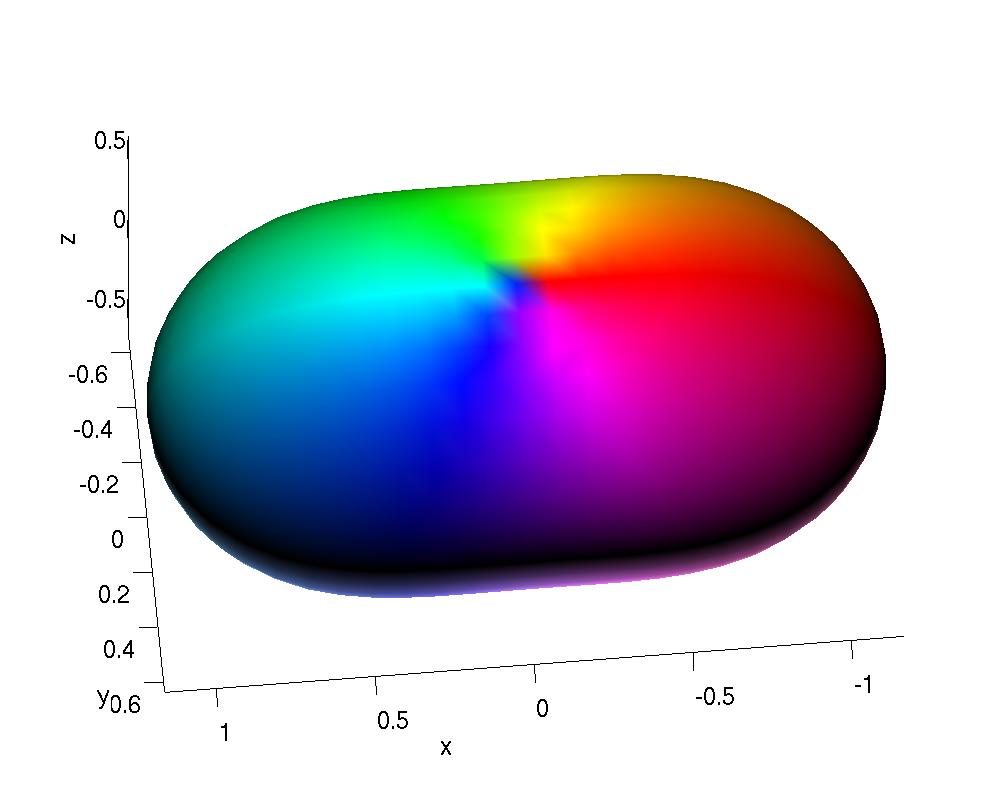}}
\subfloat[$m=\tfrac{5}{2}$]{\includegraphics[width=0.24\linewidth]{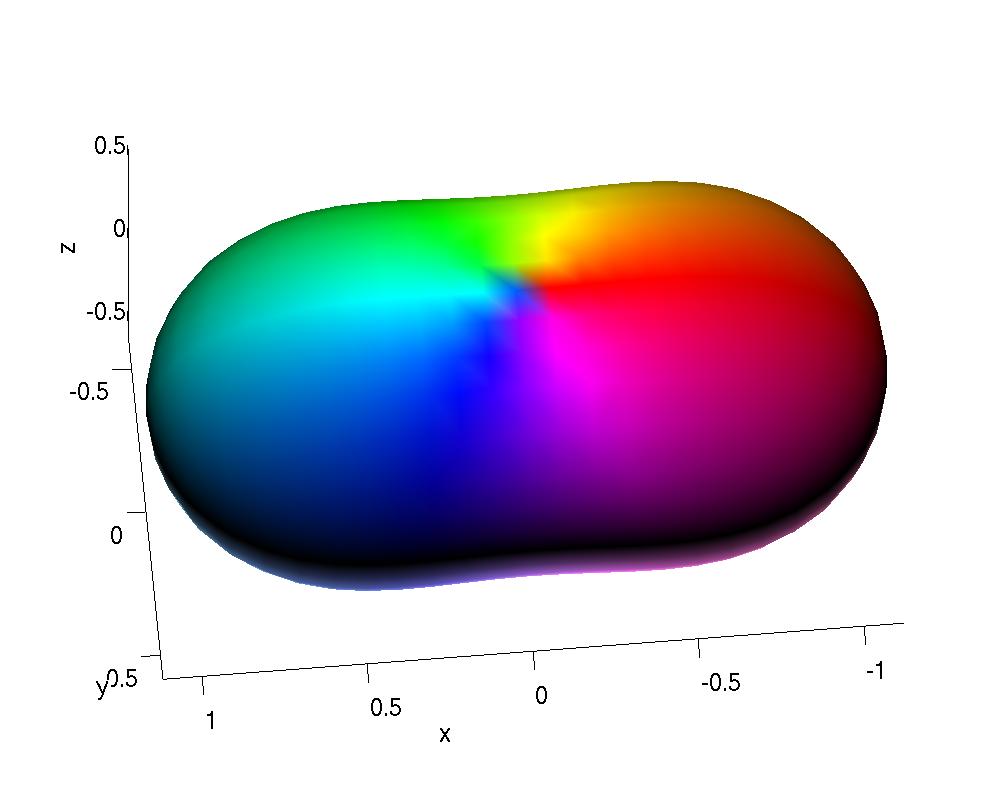}}
\subfloat[$m=3$]{\includegraphics[width=0.24\linewidth]{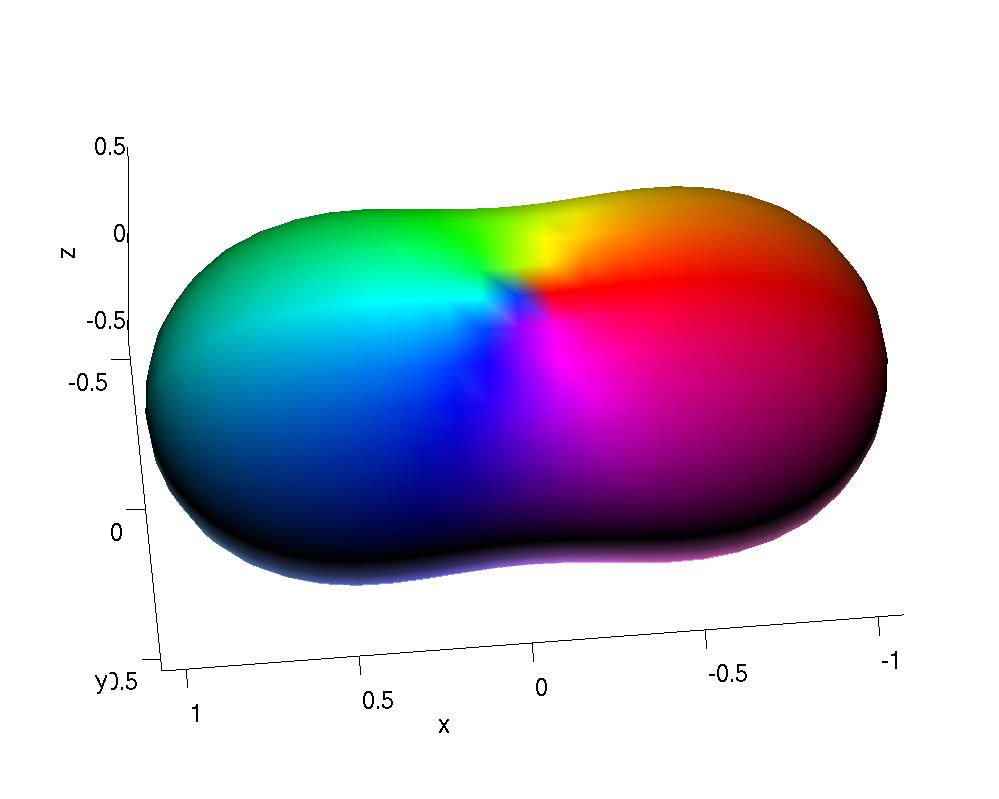}}
\subfloat[$m=\tfrac{7}{2}$]{\includegraphics[width=0.24\linewidth]{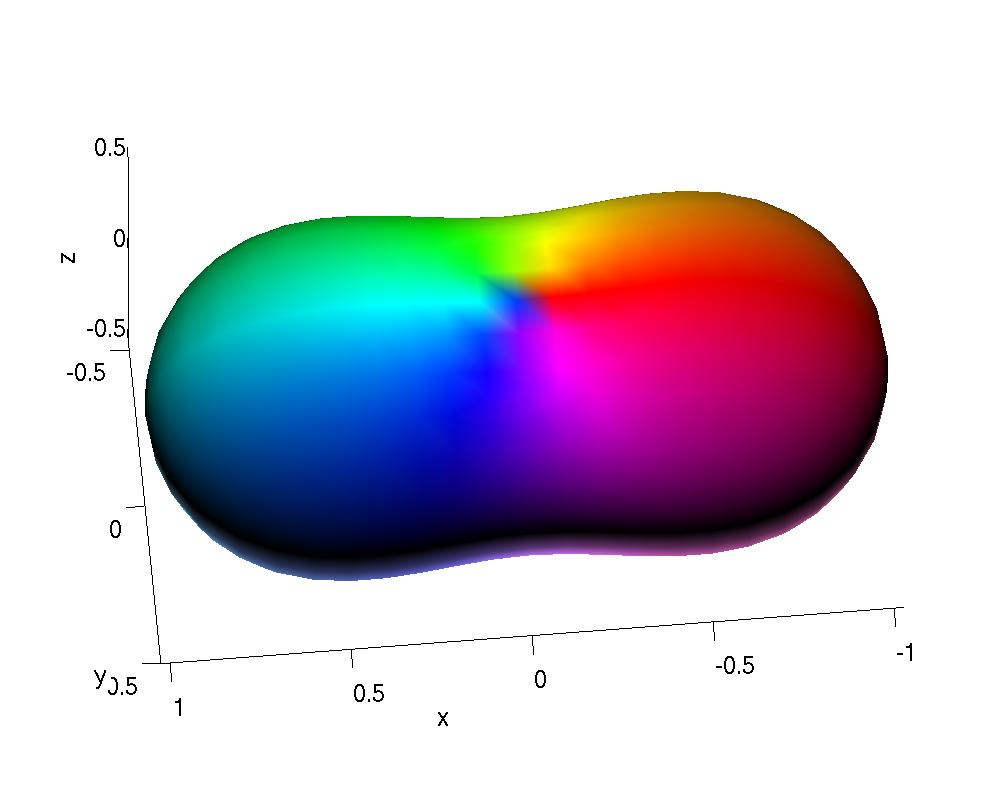}}}
\caption{Isosurfaces showing the half-maximum of the baryon charge
  density in the 2+4 model for a single baryon with masses
  $m=0,\tfrac{1}{2},1,\tfrac{3}{2},2,\tfrac{5}{2},3,\tfrac{7}{2}$ and
  the usual fixed coefficients $c_2=\tfrac{1}{4}$, $c_4=1$. The color
  scheme is the same as that in Fig.~\ref{fig:M4B1}. 
} 
\label{fig:M4B1low_mass}
\end{center}
\end{figure}

For completeness, we provide a series of numerical solutions for the
2+4 model, for the, in this paper, most studied case;
i.e.~$c_2=\tfrac{1}{4}$, $c_4=1$ with various masses
$m=0,\tfrac{1}{2},1,\tfrac{3}{2},2,\tfrac{5}{2},3,\tfrac{7}{2}$
whereas the case of $m=4$ is used throughout the paper. 
The isosurfaces of the baryon charge densities of the solutions are
shown in Fig.~\ref{fig:M4B1low_mass}. The baryon charges, masses,
dipole moments and sizes are given in Tab.~\ref{tab:M4B1low_mass}. 

\begin{table}[!htb]
\caption{The numerically integrated baryon charge, the numerically
  integrated energy and the numerically integrated baryonic dipole
  moment for the various masses shown in
  Fig.~\ref{fig:M4B1low_mass}. }
\label{tab:M4B1low_mass}
\begin{tabular}{ccccc}
$m$ & $B^{\rm numerical}$ & $E^{\rm numerical}$ & $\mathfrak{p}^B$ &
  $\mathfrak{s}^B$\\ 
\hline
\hline
$0$ & $0.9963$ & $35.731$ & $0$ & $1.876$\\
$\tfrac{1}{2}$ & $0.9895$ & $38.044$ & $0.334$ & $1.763$\\
$1$ & $0.9997$ & $43.930$ & $0.364$ & $1.441$\\
$\tfrac{3}{2}$ & $0.9999$ & $48.457$ & $0.398$ & $1.271$\\
$2$ & $0.9999$ & $52.446$ & $0.419$ & $1.157$\\
$\tfrac{5}{2}$ & $0.9998$ & $56.069$ & $0.432$ & $1.072$\\
$3$ & $0.9998$ & $59.421$ & $0.439$ & $1.005$\\
$\tfrac{7}{2}$ & $0.9997$ & $62.556$ & $0.443$ & $0.950$\\
$4$ & $0.9997$ & $65.527$ & $0.478$ & $0.901$
\end{tabular}
\end{table}

\end{appendix}

%%%%%%%%%% References %%%%%%%%%%%%%%%%%%%%%%%%%
\newcommand{\J}[4]{{\sl #1} {\bf #2} (#3) #4}
\newcommand{\andJ}[3]{{\bf #1} (#2) #3}
\newcommand{\AP}{Ann.\ Phys.\ (N.Y.)}
\newcommand{\MPL}{Mod.\ Phys.\ Lett.}
\newcommand{\NP}{Nucl.\ Phys.}
\newcommand{\PL}{Phys.\ Lett.}
\newcommand{\PR}{ Phys.\ Rev.}
\newcommand{\PRL}{Phys.\ Rev.\ Lett.}
\newcommand{\PTP}{Prog.\ Theor.\ Phys.}
\newcommand{\hep}[1]{{\tt hep-th/{#1}}}
%%%%%%%%%%%%%%%%%%%%%%%%%%%%%%%%%%%%%%%%%%%%%%%


\begin{thebibliography}{100}


\bibitem{Skyrme:1962vh} 
  T.~H.~R.~Skyrme,
  ``A Unified Field Theory of Mesons and Baryons,''
  Nucl.\ Phys.\  {\bf 31}, 556 (1962);  
  %%CITATION = NUPHA,31,556;%%
  %\bibitem{Skyrme:1961vq} 
  %T.~H.~R.~Skyrme,
  ``A Nonlinear field theory,''  
  Proc.\ Roy.\ Soc.\ Lond.\ A {\bf 260}, 127 (1961).  
  %%CITATION = PRSLA,A260,127;%%

\bibitem{Adkins:1983ya} 
  G.~S.~Adkins, C.~R.~Nappi and E.~Witten,
  ``Static Properties of Nucleons in the Skyrme Model,''  
  Nucl.\ Phys.\ B {\bf 228}, 552 (1983).  
  %%CITATION = NUPHA,B228,552;%%

\bibitem{Witten:1983tw} 
  E.~Witten,
  ``Global Aspects of Current Algebra,''
  Nucl.\ Phys.\ B {\bf 223}, 422 (1983);
  %%CITATION = NUPHA,B223,422;%%
  %2107 citations counted in INSPIRE as of 17 Jan 2015
  %\bibitem{Witten:1983tx} 
  %E.~Witten,
  ``Current Algebra, Baryons, and Quark Confinement,''
  Nucl.\ Phys.\ B {\bf 223}, 433 (1983).
  %%CITATION = NUPHA,B223,433;%%
  %1113 citations counted in INSPIRE as of 17 Jan 2015

\bibitem{Sakai:2004cn} 
  T.~Sakai and S.~Sugimoto,
  ``Low energy hadron physics in holographic QCD,''  
  Prog.\ Theor.\ Phys.\  {\bf 113}, 843 (2005)  [hep-th/0412141]; 
  %%CITATION = HEP-TH/0412141;%%
  %\bibitem{Sakai:2005yt} 
  %T.~Sakai and S.~Sugimoto,
  ``More on a holographic dual of QCD,''  
  Prog.\ Theor.\ Phys.\  {\bf 114}, 1083 (2005)  [hep-th/0507073].  
  %%CITATION = HEP-TH/0507073;%%

\bibitem{Hata:2007mb} 
  H.~Hata, T.~Sakai, S.~Sugimoto and S.~Yamato,
  ``Baryons from instantons in holographic QCD,''  
  Prog.\ Theor.\ Phys.\  {\bf 117}, 1157 (2007)  [hep-th/0701280 [hep-th]]. 
  %%CITATION = HEP-TH/0701280;%%

\bibitem{Ma:2013ooa} 
  Y.~L.~Ma, M.~Harada, H.~K.~Lee, Y.~Oh, B.~Y.~Park and M.~Rho,
  ``Dense baryonic matter in the hidden local symmetry approach: Half-skyrmions and nucleon mass,''
  Phys.\ Rev.\ D {\bf 88}, no. 1, 014016 (2013)
  [arXiv:1304.5638 [hep-ph]];
  %%CITATION = ARXIV:1304.5638;%%
  %10 citations counted in INSPIRE as of 06 Nov 2014
  %\bibitem{Ma:2013ela} 
  %Y.~L.~Ma, M.~Harada, H.~K.~Lee, Y.~Oh, B.~Y.~Park and M.~Rho,
  ``Dense baryonic matter in conformally-compensated hidden local symmetry: Vector manifestation and chiral symmetry restoration,''
  Phys.\ Rev.\ D {\bf 90}, 034015 (2014)
  [arXiv:1308.6476 [hep-ph]];
  %%CITATION = ARXIV:1308.6476;%%
  %2 citations counted in INSPIRE as of 06 Nov 2014
  %\bibitem{Ma:2013vga} 
  Y.~L.~Ma, M.~Harada, H.~K.~Lee, Y.~Oh and M.~Rho,
  ``Skyrmions, half-skyrmions and nucleon mass in dense baryonic matter,''
  Int.\ J.\ Mod.\ Phys.\ Conf.\ Ser.\  {\bf 29}, 1460238 (2014)
  [arXiv:1312.2290 [hep-ph]].
  %%CITATION = ARXIV:1312.2290;%%
  %2 citations counted in INSPIRE as of 06 Nov 2014

%%%%%%%%%%%%% baby 

\bibitem{Piette:1994ug}
  B.~M.~A.~Piette, B.~J.~Schroers and W.~J.~Zakrzewski,
  ``Multi - Solitons In A Two-Dimensional Skyrme Model,''
  Z.\ Phys.\  C {\bf 65}, 165 (1995)
  [arXiv:hep-th/9406160];
  %%CITATION = ZEPYA,C65,165;%%
  %\bibitem{Piette:1994mh}
  %B.~M.~A.~Piette, B.~J.~Schroers and W.~J.~Zakrzewski,
  ``Dynamics of baby skyrmions,''
  Nucl.\ Phys.\  B {\bf 439}, 205 (1995)
  [arXiv:hep-ph/9410256].
  %%CITATION = NUPHA,B439,205;%%

\bibitem{Weidig:1998ii}
  T.~Weidig,
  ``The baby Skyrme models and their multi-skyrmions,''
  Nonlinearity {\bf 12}, 1489-1503 (1999)
  [arXiv:hep-th/9811238].
  %%CITATION = HEP-TH/9811238;%%

\bibitem{Jaykka:2010bq} 
  J.~Jaykka and M.~Speight,
  ``Easy plane baby skyrmions,''
  Phys.\ Rev.\ D {\bf 82}, 125030 (2010)
  [arXiv:1010.2217 [hep-th]].
  %%CITATION = ARXIV:1010.2217;%%
  %17 citations counted in INSPIRE as of 06 Nov 2014

\bibitem{Kobayashi:2013aja} 
  M.~Kobayashi and M.~Nitta,
  ``Fractional vortex molecules and vortex polygons in a baby Skyrme model,''
  Phys.\ Rev.\ D {\bf 87}, no. 12, 125013 (2013)
  [arXiv:1307.0242 [hep-th]].
  %%CITATION = ARXIV:1307.0242;%%
  %6 citations counted in INSPIRE as of 06 Nov 2014

\bibitem{Kobayashi:2013wra} 
  M.~Kobayashi and M.~Nitta,
  ``Vortex polygons and their stabilities in Bose-Einstein condensates and field theory,''
  J.\ Low.\ Temp.\ Phys.\  {\bf 175}, 208 (2014)
  [arXiv:1307.1345 [cond-mat.quant-gas]].
  %%CITATION = ARXIV:1307.1345;%%
  %2 citations counted in INSPIRE as of 06 Nov 2014

\bibitem{Schroers:1995he} 
  B.~J.~Schroers,
  ``Bogomolny solitons in a gauged O(3) sigma model,''
  Phys.\ Lett.\ B {\bf 356}, 291 (1995)
  [hep-th/9506004];
  %%CITATION = HEP-TH/9506004;%%
  %43 citations counted in INSPIRE as of 06 Nov 2014
%\bibitem{Schroers:1996zy} 
 % B.~J.~Schroers,
  ``The Spectrum of Bogomol'nyi solitons in gauged linear sigma models,''
  Nucl.\ Phys.\ B {\bf 475}, 440 (1996)
  [hep-th/9603101].
  %%CITATION = HEP-TH/9603101;%%
  %23 citations counted in INSPIRE as of 06 Nov 2014

%\cite{Baptista:2004rk}
\bibitem{Baptista:2004rk} 
  J.~M.~Baptista,
  ``Vortex equations in Abelian gauged sigma-models,''
  Commun.\ Math.\ Phys.\  {\bf 261}, 161 (2006)
  [math/0411517 [math-dg]].
  %%CITATION = MATH/0411517;%%
  %20 citations counted in INSPIRE as of 17 Feb 2015


\bibitem{Nitta:2011um} 
  M.~Nitta and W.~Vinci,
  ``Decomposing Instantons in Two Dimensions,''
  J.\ Phys.\ A {\bf 45}, 175401 (2012)
  [arXiv:1108.5742 [hep-th]].
  %%CITATION = ARXIV:1108.5742;%%
  %6 citations counted in INSPIRE as of 06 Nov 2014

\bibitem{Alonso-Izquierdo:2014cza} 
  A.~Alonso-Izquierdo, W.~G.~Fuertes and J.~M.~Guilarte,
  ``Two Species of Vortices in a massive Gauged Non-linear Sigma Model,''
  arXiv:1409.8419 [hep-th].
  %%CITATION = ARXIV:1409.8419;%%

\bibitem{Adam:2010fg} 
  C.~Adam, J.~Sanchez-Guillen and A.~Wereszczynski,
  ``A Skyrme-type proposal for baryonic matter,''
  Phys.\ Lett.\ B {\bf 691}, 105 (2010)
  [arXiv:1001.4544 [hep-th]];
  %%CITATION = ARXIV:1001.4544;%%
  %\bibitem{Adam:2010ds} 
  %C.~Adam, J.~Sanchez-Guillen and A.~Wereszczynski,
  ``A BPS Skyrme model and baryons at large $N_c$,''
  Phys.\ Rev.\ D {\bf 82}, 085015 (2010)
  [arXiv:1007.1567 [hep-th]].
  %%CITATION = ARXIV:1007.1567;%%

\bibitem{Gudnason:2013qba} 
  S.~B.~Gudnason and M.~Nitta,
  ``Baryonic sphere: a spherical domain wall carrying baryon number,''
  Phys.\ Rev.\ D {\bf 89}, 025012 (2014)
  [arXiv:1311.4454 [hep-th]].
  %%CITATION = ARXIV:1311.4454;%%
  %6 citations counted in INSPIRE as of 23 Oct 2014

\bibitem{Gudnason:2014gla} 
  S.~B.~Gudnason and M.~Nitta,
  ``Effective field theories on solitons of generic shapes,''
  arXiv:1407.2822 [hep-th].
  %%CITATION = ARXIV:1407.2822;%%
  %1 citations counted in INSPIRE as of 23 Oct 2014

\bibitem{Gudnason:2014jga} 
  S.~B.~Gudnason and M.~Nitta,
  ``Baryonic Torii: Toroidal baryons in a generalized Skyrme model,''
  Phys.\ Rev.\ D {\bf 91}, 045027 (2015)
  [arXiv:1410.8407 [hep-th]].
  %%CITATION = ARXIV:1410.8407;%%

\bibitem{Battye:1997qq} 
  R.~A.~Battye and P.~M.~Sutcliffe,
  ``Symmetric skyrmions,''
  Phys.\ Rev.\ Lett.\  {\bf 79}, 363 (1997)
  [hep-th/9702089].
  %%CITATION = HEP-TH/9702089;%%
  %134 citations counted in INSPIRE as of 02 Jan 2015

\bibitem{Houghton:1997kg} 
  C.~J.~Houghton, N.~S.~Manton and P.~M.~Sutcliffe,
  ``Rational maps, monopoles and Skyrmions,''
  Nucl.\ Phys.\ B {\bf 510}, 507 (1998)
  [hep-th/9705151].
  %%CITATION = HEP-TH/9705151;%%
  %221 citations counted in INSPIRE as of 02 Jan 2015

\bibitem{Brihaye:1998} 
  Y.~Brihaye and D.~H.~Tchrakian,
  ``Solitons/instantons in  d-dimensional  gauged  Skyrme models,''
  Nonlinearity {\bf 11}, 891 (1998)
  [hep-th/9805059].

\bibitem{Brihaye:1998vr} 
  Y.~Brihaye, B.~Kleihaus and D.~H.~Tchrakian,
  ``Dyon - Skyrmion lumps,''
  J.\ Math.\ Phys.\  {\bf 40}, 1136 (1999)
  [hep-th/9805059].
  %%CITATION = HEP-TH/9805059;%%
  %16 citations counted in INSPIRE as of 06 Nov 2014

\bibitem{Kleihaus:1999ea} 
  B.~Kleihaus, D.~H.~Tchrakian and F.~Zimmerschied,
  ``Monopole skyrmions,''
  J.\ Math.\ Phys.\  {\bf 41}, 816 (2000)
  [hep-th/9907035].
  %%CITATION = HEP-TH/9907035;%%
  %4 citations counted in INSPIRE as of 06 Nov 2014

\bibitem{Brihaye:2000ku} 
  Y.~Brihaye, B.~Hartmann and D.~H.~Tchrakian,
  ``Monopoles and dyons in SO(3) gauged Skyrme models,''
  J.\ Math.\ Phys.\  {\bf 42}, 3270 (2001)
  [hep-th/0010152].
  %%CITATION = HEP-TH/0010152;%%
  %5 citations counted in INSPIRE as of 06 Nov 2014

\bibitem{Brihaye:2001je} 
  Y.~Brihaye, J.~Burzlaff, V.~Paturyan and D.~H.~Tchrakian,
  ``Comment on the soliton of the SO(3) gauged Skyrme model,''
  Nonlinearity {\bf 15}, 385 (2002)
  [hep-th/0109034].
  %%CITATION = HEP-TH/0109034;%%
  %3 citations counted in INSPIRE as of 06 Nov 2014

\bibitem{Grigoriev:2002qc} 
  D.~Y.~Grigoriev, P.~M.~Sutcliffe and D.~H.~Tchrakian,
  ``Skyrmed monopoles,''
  Phys.\ Lett.\ B {\bf 540}, 146 (2002)
  [hep-th/0206160].
  %%CITATION = HEP-TH/0206160;%%
  %11 citations counted in INSPIRE as of 25 gen 2015

\bibitem{Brihaye:2004pz} 
  Y.~Brihaye, C.~T.~Hill and C.~K.~Zachos,
  ``Bounding gauged skyrmion masses,''
  Phys.\ Rev.\ D {\bf 70}, 111502 (2004)
  [hep-th/0409222].
  %%CITATION = HEP-TH/0409222;%%

\bibitem{Brihaye:2007gk} 
  Y.~Brihaye, J.~Burzlaff and D.~H.~Tchrakian,
  ``Asymptotic analysis of the Skyrmed monopole,''
  Phys.\ Rev.\ D {\bf 77}, 107701 (2008)
  [arXiv:0712.0549 [hep-th]].
  %%CITATION = ARXIV:0712.0549;%%

\bibitem{Nitta:2012wi} 
  M.~Nitta,
  ``Correspondence between Skyrmions in 2+1 and 3+1 Dimensions,''  
  Phys.\ Rev.\ D {\bf 87}, 025013 (2013)  [arXiv:1210.2233 [hep-th]];  
  %%CITATION = ARXIV:1210.2233;%%
  %\bibitem{Nitta:2012rq} 
  %M.~Nitta,
  ``Matryoshka Skyrmions,''  
  Nucl.\  Phys.\ B {\bf 872}, 62 (2013)  [arXiv:1211.4916 [hep-th]].  
  %%CITATION = ARXIV:1211.4916;%%  
  %1 citations counted in INSPIRE as of 12 Apr 2013

\bibitem{Gudnason:2014nba} 
  S.~B.~Gudnason and M.~Nitta,
  ``Domain wall Skyrmions,''
  Phys.\ Rev.\ D {\bf 89}, 085022 (2014)
  [arXiv:1403.1245 [hep-th]].
  %%CITATION = ARXIV:1403.1245;%%

\bibitem{Gudnason:2014hsa} 
  S.~B.~Gudnason and M.~Nitta,
  ``Incarnations of Skyrmions,''
  Phys.\ Rev.\ D {\bf 90}, 085007 (2014)
  [arXiv:1407.7210 [hep-th]].
  %%CITATION = ARXIV:1407.7210;%%

\bibitem{Harland:2013rxa} 
  D.~Harland,
  ``Topological energy bounds for the Skyrme and Faddeev models with massive pions,''
  Phys.\ Lett.\ B {\bf 728}, 518 (2014)
  [arXiv:1311.2403 [hep-th]].
  %%CITATION = ARXIV:1311.2403;%%
  %8 citations counted in INSPIRE as of 23 Jan 2015


%%%%%%%%% discussion

\bibitem{Babaev:2002}
  E.~Babaev, 
  ``Vortices with Fractional Flux in Two-Gap Superconductors and in
  Extended Faddeev Model,''
  Phys.\ Rev.\ Lett.\ {\bf 89} (2002) 067001; 
  E.~Babaev, A.~Sudbo and N.~W.~Ashcroft,
  ``A superconductor to superfluid phase transition in liquid metallic
  hydrogen,''
  Nature {\bf 431}, 666 (2004); 
  %\bibitem{Smiseth:2004na}
  J.~Smiseth, E.~Smorgrav, E.~Babaev and A.~Sudbo,
  ``Field- and temperature induced topological phase transitions in
  the three-dimensional $N$-component London superconductor,''
  Phys.\ Rev.\  B {\bf 71}, 214509 (2005) 
  [arXiv:cond-mat/0411761];
  %%CITATION = PHRVA,B71,214509;%%
  E.~Babaev and N.~W.~Ashcroft,
  ``Violation of the London law and Onsager-Feynman quantization in
  multicomponent superconductors,''
  Nature Phys. {\bf 3}, 530 (2007).

\bibitem{Goryo:2007}
  J.~Goryo, S.~Soma and H.~Matsukawa, 
  ``Deconfinement of vortices with continuously variable fractions of
  the unit flux quanta in two-gap superconductors,'' 
  Euro Phys.\ Lett.\ {\bf 80}, 17002 (2007)
  [arXiv:cond-mat/0608015].

\bibitem{Tanaka:2001}
  Y.~Tanaka, 
  ``Phase instability in multi-band superconductors,''
  J.\ Phys.\ Soc.\ Jp.\ {\bf 70}, 2844 (2001);  
  %Y.~Tanaka,
  ``Soliton in Two-Band Superconductor,''
  Phys.\ Rev.\ Lett.\ {\bf 88}, 017002 (2001).

\bibitem{Nitta:2012xq} 
  M.~Nitta,
  ``Josephson vortices and the Atiyah-Manton construction,''  
  Phys.\ Rev.\ D {\bf 86}, 125004 (2012)  [arXiv:1207.6958 [hep-th]].  
  %%CITATION = ARXIV:1207.6958;%%  
  %4 citations counted in INSPIRE as of 20 Apr 2013

\bibitem{Kobayashi:2013ju} 
  M.~Kobayashi and M.~Nitta,
  ``Sine-Gordon kinks on a domain wall ring,''  
  Phys.\  Rev.\ D {\bf 87}, 085003 (2013)  [arXiv:1302.0989 [hep-th]].  
  %%CITATION = ARXIV:1302.0989;%%

\bibitem{footnote:BEC}
  K.~Kasamatsu, M.~Tsubota and M.~Ueda, 
  ``Vortex Molecules in Coherently Coupled Two-Component Bose-Einstein
  Condensates,''
  Phys.\ Rev.\ Lett\ {\bf 93}, 250406 (2004);
  %K.~Kasamatsu, M.~Tsubota and M.~Ueda, 
  ``Vortices in multicomponent Bose-Einstein condensates,''
  Int.\ J.\ Mod.\ Phys.\ {\bf B} 19, 1835 (2005);
  %\bibitem{Cipriani:2013nya} 
  M.~Cipriani and M.~Nitta,
  ``Crossover between integer and fractional vortex lattices in
  coherently coupled two-component Bose-Einstein condensates,'' 
  Phys.\ Rev.\ Lett.\  {\bf 111}, 170401 (2013)
  [arXiv:1303.2592 [cond-mat.quant-gas]];
  %%CITATION = ARXIV:1303.2592;%%
  %7 citations counted in INSPIRE as of 25 Jan 2015
  %\bibitem{Nitta:2013eaa} 
  M.~Nitta, M.~Eto and M.~Cipriani,
  ``Vortex molecules in Bose-Einstein condensates,''
  J.\ Low.\ Temp.\ Phys.\  {\bf 175}, 177 (2013)
  [arXiv:1307.4312 [cond-mat.quant-gas]].
  %%CITATION = ARXIV:1307.4312;%%
  %2 citations counted in INSPIRE as of 25 Jan 2015

\bibitem{Nitta:2014vpa} 
  M.~Nitta,
  ``Fractional instantons and bions in the O $(N)$ model with twisted boundary conditions,''
  JHEP {\bf 1503}, 108 (2015)
  [arXiv:1412.7681 [hep-th]].
  %%CITATION = ARXIV:1412.7681;%%
  %4 citations counted in INSPIRE as of 18 Apr 2015

\bibitem{Eto:2009bz} 
  M.~Eto, T.~Fujimori, S.~B.~Gudnason, K.~Konishi, T.~Nagashima, M.~Nitta, K.~Ohashi and W.~Vinci,
  ``Fractional Vortices and Lumps,''
  Phys.\ Rev.\ D {\bf 80}, 045018 (2009)
  [arXiv:0905.3540 [hep-th]].
  %%CITATION = ARXIV:0905.3540;%%
  %25 citations counted in INSPIRE as of 17 Feb 2015

\end{thebibliography}
\end{document}